\documentclass[longauth,twocolumn,traditabstract]{aa}

\def\setsymbol#1#2{\expandafter\def\csname #1\endcsname{#2}}
\def\getsymbol#1{\csname #1\endcsname}

\def\Planck{\textit{Planck}}

\def\allearlypapers{\nocite{planck2011-1.1, planck2011-1.3, planck2011-1.4, planck2011-1.5, planck2011-1.6, planck2011-1.7, planck2011-1.10, planck2011-1.10sup, planck2011-5.1a, planck2011-5.1b, planck2011-5.2a, planck2011-5.2b, planck2011-5.2c, planck2011-6.1, planck2011-6.2, planck2011-6.3a, planck2011-6.4a, planck2011-6.4b, planck2011-6.6, planck2011-7.0, planck2011-7.2, planck2011-7.3, planck2011-7.7a, planck2011-7.7b, planck2011-7.12, planck2011-7.13}}

\newbox\tablebox    \newdimen\tablewidth
\def\leaderfil{\leaders\hbox to 5pt{\hss.\hss}\hfil}
\def\tablenote#1 #2\par{\begingroup \parindent=0.8em
    \abovedisplayshortskip=0pt\belowdisplayshortskip=0pt
    \noindent
    $$\hss\vbox{\hsize\tablewidth \hangindent=\parindent \hangafter=1 \noindent
    \hbox to \parindent{$^#1$\hss}\strut#2\strut\par}\hss$$
    \endgroup}

\def\L2{\ifmmode L_2\else $L_2$\fi}

\def\DeltaT{\ifmmode \Delta T\else $\Delta T$\fi}
\def\deltat{\ifmmode \Delta t\else $\Delta t$\fi}
\def\fknee{\ifmmode f_{\rm knee}\else $f_{\rm knee}$\fi}
\def\Fmax{\ifmmode F_{\rm max}\else $F_{\rm max}$\fi}
\def\solar{\ifmmode{\rm M}_{\mathord\odot}\else${\rm M}_{\mathord\odot}$\fi}
\def\Msolar{\ifmmode{\rm M}_{\mathord\odot}\else${\rm M}_{\mathord\odot}$\fi}
\def\Lsolar{\ifmmode{\rm L}_{\mathord\odot}\else${\rm L}_{\mathord\odot}$\fi}
\def\inv{\ifmmode^{-1}\else$^{-1}$\fi}
\def\mo{\ifmmode^{-1}\else$^{-1}$\fi}
\def\sup#1{\ifmmode ^{\rm #1}\else $^{\rm #1}$\fi}
\def\expo#1{\ifmmode \times 10^{#1}\else $\times 10^{#1}$\fi}
\def\,{\thinspace}
\def\lsim{\mathrel{\raise .4ex\hbox{\rlap{$<$}\lower 1.2ex\hbox{$\sim$}}}}
\def\gsim{\mathrel{\raise .4ex\hbox{\rlap{$>$}\lower 1.2ex\hbox{$\sim$}}}}

\def\simprop{\mathrel{\raise .4ex\hbox{\rlap{$\propto$}\lower 1.2ex\hbox{$\sim$}}}}
\def\deg{\ifmmode^\circ\else$^\circ$\fi}
\def\pdeg{\ifmmode $\setbox0=\hbox{$^{\circ}$}\rlap{\hskip.11\wd0 .}$^{\circ}
          \else \setbox0=\hbox{$^{\circ}$}\rlap{\hskip.11\wd0 .}$^{\circ}$\fi}
\def\arcs{\ifmmode {^{\scriptstyle\prime\prime}}
          \else $^{\scriptstyle\prime\prime}$\fi}
\def\arcm{\ifmmode {^{\scriptstyle\prime}}
          \else $^{\scriptstyle\prime}$\fi}
\newdimen\sa  \newdimen\sb
\def\parcs{\sa=.07em \sb=.03em
     \ifmmode \hbox{\rlap{.}}^{\scriptstyle\prime\kern -\sb\prime}\hbox{\kern -\sa}
     \else \rlap{.}$^{\scriptstyle\prime\kern -\sb\prime}$\kern -\sa\fi}
\def\parcm{\sa=.08em \sb=.03em
     \ifmmode \hbox{\rlap{.}\kern\sa}^{\scriptstyle\prime}\hbox{\kern-\sb}
     \else \rlap{.}\kern\sa$^{\scriptstyle\prime}$\kern-\sb\fi}
\def\ra[#1 #2 #3.#4]{#1\sup{h}#2\sup{m}#3\sup{s}\llap.#4}
\def\dec[#1 #2 #3.#4]{#1\deg#2\arcm#3\arcs\llap.#4}
\def\deco[#1 #2 #3]{#1\deg#2\arcm#3\arcs}
\def\rra[#1 #2]{#1\sup{h}#2\sup{m}}

\def\dots{\relax\ifmmode \ldots\else $\ldots$\fi}
\def\WHzsr{\ifmmode $W\,Hz\mo\,sr\mo$\else W\,Hz\mo\,sr\mo\fi}
\def\mHz{\ifmmode $\,mHz$\else \,mHz\fi}
\def\GHz{\ifmmode $\,GHz$\else \,GHz\fi}
\def\mKs{\ifmmode $\,mK\,s$^{1/2}\else \,mK\,s$^{1/2}$\fi}
\def\muKs{\ifmmode \,\mu$K\,s$^{1/2}\else \,$\mu$K\,s$^{1/2}$\fi}
\def\muKRJs{\ifmmode \,\mu$K$_{\rm RJ}$\,s$^{1/2}\else \,$\mu$K$_{\rm RJ}$\,s$^{1/2}$\fi}
\def\muKHz{\ifmmode \,\mu$K\,Hz$^{-1/2}\else \,$\mu$K\,Hz$^{-1/2}$\fi}
\def\MJysr{\ifmmode \,$MJy\,sr\mo$\else \,MJy\,sr\mo\fi}
\def\MJysrmK{\ifmmode \,$MJy\,sr\mo$\,mK$_{\rm CMB}\mo\else \,MJy\,sr\mo\,mK$_{\rm CMB}\mo$\fi}
\def\microns{\ifmmode \,\mu$m$\else \,$\mu$m\fi}

\def\muK{\ifmmode \,\mu$K$\else \,$\mu$\hbox{K}\fi}
\def\microK{\ifmmode \,\mu$K$\else \,$\mu$\hbox{K}\fi}
\def\muW{\ifmmode \,\mu$W$\else \,$\mu$\hbox{W}\fi}
\def\kms{\ifmmode $\,km\,s$^{-1}\else \,km\,s$^{-1}$\fi}
\def\kmsMpc{\ifmmode $\,\kms\,Mpc\mo$\else \,\kms\,Mpc\mo\fi}
\providecommand{\sorthelp}[1]{}

\usepackage[breaklinks,colorlinks,citecolor=blue]{hyperref}
\usepackage{amsmath}
\usepackage{natbib}
\usepackage{graphicx}
\usepackage{txfonts}
\usepackage{natbib}
\usepackage{fixltx2e}
\usepackage{inputenc} 
\usepackage{color}
\usepackage[usenames,dvipsnames,svgnames,table]{xcolor}
\usepackage{ifthen}
\usepackage[flushleft]{threeparttable}

\bibpunct{(}{)}{;}{a}{}{,}

\def\deg{^{\circ}}

\def\all2103resultspapers{\nocite{planck2013-p01, planck2013-p02, planck2013-p02a, planck2013-p02d, planck2013-p02b, planck2013-p03, planck2013-p03c, planck2013-p03f, planck2013-p03d, planck2013-p03e, planck2013-p01a, planck2013-p06, planck2013-p03a, planck2013-pip88, planck2013-p08, planck2013-p11, planck2013-p12, planck2013-p13, planck2013-p14, planck2013-p15, planck2013-p05b, planck2013-p17, planck2013-p09, planck2013-p09a, planck2013-p20, planck2013-p19, planck2013-pipaberration, planck2013-p05, planck2013-p05a, planck2013-pip56, planck2013-p06b}}

\begin{document}

\title{\Planck{} 2015 results. IV. Low Frequency Instrument beams and window functions}

\authorrunning{Planck Collaboration}
\titlerunning{LFI beams and window functions}

\author{\small
Planck Collaboration: P.~A.~R.~Ade\inst{85}
\and
N.~Aghanim\inst{58}
\and
M.~Ashdown\inst{69, 5}
\and
J.~Aumont\inst{58}
\and
C.~Baccigalupi\inst{84}
\and
A.~J.~Banday\inst{93, 7}
\and
R.~B.~Barreiro\inst{64}
\and
N.~Bartolo\inst{28, 65}
\and
E.~Battaner\inst{94, 95}
\and
K.~Benabed\inst{59, 92}
% ...
}
\institute{\small
APC, AstroParticule et Cosmologie, Universit\'{e} Paris Diderot, CNRS/IN2P3, CEA/lrfu, Observatoire de Paris, Sorbonne Paris Cit\'{e}, 10, rue Alice Domon et L\'{e}onie Duquet, 75205 Paris Cedex 13, France\goodbreak
\and
Aalto University Mets\"{a}hovi Radio Observatory and Dept of Radio Science and Engineering, P.O. Box 13000, FI-00076 AALTO, Finland\goodbreak
\and
African Institute for Mathematical Sciences, 6-8 Melrose Road, Muizenberg, Cape Town, South Africa\goodbreak
\and
Agenzia Spaziale Italiana Science Data Center, Via del Politecnico snc, 00133, Roma, Italy\goodbreak
\and
Astrophysics Group, Cavendish Laboratory, University of Cambridge, J J Thomson Avenue, Cambridge CB3 0HE, U.K.\goodbreak
\and
CITA, University of Toronto, 60 St. George St., Toronto, ON M5S 3H8, Canada\goodbreak
\and
CNRS, IRAP, 9 Av. colonel Roche, BP 44346, F-31028 Toulouse cedex 4, France\goodbreak
\and
CRANN, Trinity College, Dublin, Ireland\goodbreak
\and
California Institute of Technology, Pasadena, California, U.S.A.\goodbreak
\and
Centre for Theoretical Cosmology, DAMTP, University of Cambridge, Wilberforce Road, Cambridge CB3 0WA, U.K.\goodbreak
\and
Computational Cosmology Center, Lawrence Berkeley National Laboratory, Berkeley, California, U.S.A.\goodbreak
\and
Consejo Superior de Investigaciones Cient\'{\i}ficas (CSIC), Madrid, Spain\goodbreak
\and
DSM/Irfu/SPP, CEA-Saclay, F-91191 Gif-sur-Yvette Cedex, France\goodbreak
\and
DTU Space, National Space Institute, Technical University of Denmark, Elektrovej 327, DK-2800 Kgs. Lyngby, Denmark\goodbreak
\and
D\'{e}partement de Physique Th\'{e}orique, Universit\'{e} de Gen\`{e}ve, 24, Quai E. Ansermet,1211 Gen\`{e}ve 4, Switzerland\goodbreak
\and
Departamento de Astrof\'{i}sica, Universidad de La Laguna (ULL), E-38206 La Laguna, Tenerife, Spain\goodbreak
\and
Departamento de F\'{\i}sica, Universidad de Oviedo, Avda. Calvo Sotelo s/n, Oviedo, Spain\goodbreak
\and
Department of Astronomy and Astrophysics, University of Toronto, 50 Saint George Street, Toronto, Ontario, Canada\goodbreak
\and
Department of Astrophysics/IMAPP, Radboud University Nijmegen, P.O. Box 9010, 6500 GL Nijmegen, The Netherlands\goodbreak
\and
Department of Physics \& Astronomy, University of British Columbia, 6224 Agricultural Road, Vancouver, British Columbia, Canada\goodbreak
\and
Department of Physics and Astronomy, Dana and David Dornsife College of Letter, Arts and Sciences, University of Southern California, Los Angeles, CA 90089, U.S.A.\goodbreak
\and
Department of Physics and Astronomy, University College London, London WC1E 6BT, U.K.\goodbreak
\and
Department of Physics, Florida State University, Keen Physics Building, 77 Chieftan Way, Tallahassee, Florida, U.S.A.\goodbreak
\and
Department of Physics, Gustaf H\"{a}llstr\"{o}min katu 2a, University of Helsinki, Helsinki, Finland\goodbreak
\and
Department of Physics, Princeton University, Princeton, New Jersey, U.S.A.\goodbreak
\and
Department of Physics, University of California, Santa Barbara, California, U.S.A.\goodbreak
\and
Department of Physics, University of Illinois at Urbana-Champaign, 1110 West Green Street, Urbana, Illinois, U.S.A.\goodbreak
\and
Dipartimento di Fisica e Astronomia G. Galilei, Universit\`{a} degli Studi di Padova, via Marzolo 8, 35131 Padova, Italy\goodbreak
\and
Dipartimento di Fisica e Scienze della Terra, Universit\`{a} di Ferrara, Via Saragat 1, 44122 Ferrara, Italy\goodbreak
\and
Dipartimento di Fisica, Universit\`{a} La Sapienza, P. le A. Moro 2, Roma, Italy\goodbreak
\and
Dipartimento di Fisica, Universit\`{a} degli Studi di Milano, Via Celoria, 16, Milano, Italy\goodbreak
\and
Dipartimento di Fisica, Universit\`{a} degli Studi di Trieste, via A. Valerio 2, Trieste, Italy\goodbreak
\and
Dipartimento di Fisica, Universit\`{a} di Roma Tor Vergata, Via della Ricerca Scientifica, 1, Roma, Italy\goodbreak
\and
Dipartimento di Matematica, Universit\`{a} di Roma Tor Vergata, Via della Ricerca Scientifica, 1, Roma, Italy\goodbreak
\and
Discovery Center, Niels Bohr Institute, Blegdamsvej 17, Copenhagen, Denmark\goodbreak
\and
European Space Agency, ESAC, Planck Science Office, Camino bajo del Castillo, s/n, Urbanizaci\'{o}n Villafranca del Castillo, Villanueva de la Ca\~{n}ada, Madrid, Spain\goodbreak
\and
European Space Agency, ESTEC, Keplerlaan 1, 2201 AZ Noordwijk, The Netherlands\goodbreak
\and
Facolt\`{a} di Ingegneria, Universit\`{a} degli Studi e-Campus, Via Isimbardi 10, Novedrate (CO), 22060, Italy\goodbreak
\and
Gran Sasso Science Institute, INFN, viale F. Crispi 7, 67100 L'Aquila, Italy\goodbreak
\and
HGSFP and University of Heidelberg, Theoretical Physics Department, Philosophenweg 16, 69120, Heidelberg, Germany\goodbreak
\and
Haverford College Astronomy Department, 370 Lancaster Avenue, Haverford, Pennsylvania, U.S.A.\goodbreak
\and
Helsinki Institute of Physics, Gustaf H\"{a}llstr\"{o}min katu 2, University of Helsinki, Helsinki, Finland\goodbreak
\and
INAF - Osservatorio Astrofisico di Catania, Via S. Sofia 78, Catania, Italy\goodbreak
\and
INAF - Osservatorio Astronomico di Padova, Vicolo dell'Osservatorio 5, Padova, Italy\goodbreak
\and
INAF - Osservatorio Astronomico di Roma, via di Frascati 33, Monte Porzio Catone, Italy\goodbreak
\and
INAF - Osservatorio Astronomico di Trieste, Via G.B. Tiepolo 11, Trieste, Italy\goodbreak
\and
INAF/IASF Bologna, Via Gobetti 101, Bologna, Italy\goodbreak
\and
INAF/IASF Milano, Via E. Bassini 15, Milano, Italy\goodbreak
\and
INFN, Sezione di Bologna, Via Irnerio 46, I-40126, Bologna, Italy\goodbreak
\and
INFN, Sezione di Roma 1, Universit\`{a} di Roma Sapienza, Piazzale Aldo Moro 2, 00185, Roma, Italy\goodbreak
\and
INFN, Sezione di Roma 2, Universit\`{a} di Roma Tor Vergata, Via della Ricerca Scientifica, 1, Roma, Italy\goodbreak
\and
INFN/National Institute for Nuclear Physics, Via Valerio 2, I-34127 Trieste, Italy\goodbreak
\and
IUCAA, Post Bag 4, Ganeshkhind, Pune University Campus, Pune 411 007, India\goodbreak
\and
Imperial College London, Astrophysics group, Blackett Laboratory, Prince Consort Road, London, SW7 2AZ, U.K.\goodbreak
\and
Infrared Processing and Analysis Center, California Institute of Technology, Pasadena, CA 91125, U.S.A.\goodbreak
\and
Institut N\'{e}el, CNRS, Universit\'{e} Joseph Fourier Grenoble I, 25 rue des Martyrs, Grenoble, France\goodbreak
\and
Institut Universitaire de France, 103, bd Saint-Michel, 75005, Paris, France\goodbreak
\and
Institut d'Astrophysique Spatiale, CNRS (UMR8617) Universit\'{e} Paris-Sud 11, B\^{a}timent 121, Orsay, France\goodbreak
\and
Institut d'Astrophysique de Paris, CNRS (UMR7095), 98 bis Boulevard Arago, F-75014, Paris, France\goodbreak
\and
Institut f\"ur Theoretische Teilchenphysik und Kosmologie, RWTH Aachen University, D-52056 Aachen, Germany\goodbreak
\and
Institute of Astronomy, University of Cambridge, Madingley Road, Cambridge CB3 0HA, U.K.\goodbreak
\and
Institute of Theoretical Astrophysics, University of Oslo, Blindern, Oslo, Norway\goodbreak
\and
Instituto de Astrof\'{\i}sica de Canarias, C/V\'{\i}a L\'{a}ctea s/n, La Laguna, Tenerife, Spain\goodbreak
\and
Instituto de F\'{\i}sica de Cantabria (CSIC-Universidad de Cantabria), Avda. de los Castros s/n, Santander, Spain\goodbreak
\and
Istituto Nazionale di Fisica Nucleare, Sezione di Padova, via Marzolo 8, I-35131 Padova, Italy\goodbreak
\and
Jet Propulsion Laboratory, California Institute of Technology, 4800 Oak Grove Drive, Pasadena, California, U.S.A.\goodbreak
\and
Jodrell Bank Centre for Astrophysics, Alan Turing Building, School of Physics and Astronomy, The University of Manchester, Oxford Road, Manchester, M13 9PL, U.K.\goodbreak
\and
Kavli Institute for Cosmological Physics, University of Chicago, Chicago, IL 60637, USA\goodbreak
\and
Kavli Institute for Cosmology Cambridge, Madingley Road, Cambridge, CB3 0HA, U.K.\goodbreak
\and
Kazan Federal University, 18 Kremlyovskaya St., Kazan, 420008, Russia\goodbreak
\and
LAL, Universit\'{e} Paris-Sud, CNRS/IN2P3, Orsay, France\goodbreak
\and
LERMA, CNRS, Observatoire de Paris, 61 Avenue de l'Observatoire, Paris, France\goodbreak
\and
Laboratoire AIM, IRFU/Service d'Astrophysique - CEA/DSM - CNRS - Universit\'{e} Paris Diderot, B\^{a}t. 709, CEA-Saclay, F-91191 Gif-sur-Yvette Cedex, France\goodbreak
\and
Laboratoire Traitement et Communication de l'Information, CNRS (UMR 5141) and T\'{e}l\'{e}com ParisTech, 46 rue Barrault F-75634 Paris Cedex 13, France\goodbreak
\and
Laboratoire de Physique Subatomique et Cosmologie, Universit\'{e} Grenoble-Alpes, CNRS/IN2P3, 53, rue des Martyrs, 38026 Grenoble Cedex, France\goodbreak
\and
Laboratoire de Physique Th\'{e}orique, Universit\'{e} Paris-Sud 11 \& CNRS, B\^{a}timent 210, 91405 Orsay, France\goodbreak
\and
Lawrence Berkeley National Laboratory, Berkeley, California, U.S.A.\goodbreak
\and
Lebedev Physical Institute of the Russian Academy of Sciences, Astro Space Centre, 84/32 Profsoyuznaya st., Moscow, GSP-7, 117997, Russia\goodbreak
\and
Max-Planck-Institut f\"{u}r Astrophysik, Karl-Schwarzschild-Str. 1, 85741 Garching, Germany\goodbreak
\and
McGill Physics, Ernest Rutherford Physics Building, McGill University, 3600 rue University, Montr\'{e}al, QC, H3A 2T8, Canada\goodbreak
\and
National University of Ireland, Department of Experimental Physics, Maynooth, Co. Kildare, Ireland\goodbreak
\and
Nicolaus Copernicus Astronomical Center, Bartycka 18, 00-716 Warsaw, Poland\goodbreak
\and
Niels Bohr Institute, Blegdamsvej 17, Copenhagen, Denmark\goodbreak
\and
SISSA, Astrophysics Sector, via Bonomea 265, 34136, Trieste, Italy\goodbreak
\and
School of Physics and Astronomy, Cardiff University, Queens Buildings, The Parade, Cardiff, CF24 3AA, U.K.\goodbreak
\and
School of Physics and Astronomy, University of Nottingham, Nottingham NG7 2RD, U.K.\goodbreak
\and
Sorbonne Universit\'{e}-UPMC, UMR7095, Institut d'Astrophysique de Paris, 98 bis Boulevard Arago, F-75014, Paris, France\goodbreak
\and
Space Sciences Laboratory, University of California, Berkeley, California, U.S.A.\goodbreak
\and
Special Astrophysical Observatory, Russian Academy of Sciences, Nizhnij Arkhyz, Zelenchukskiy region, Karachai-Cherkessian Republic, 369167, Russia\goodbreak
\and
Sub-Department of Astrophysics, University of Oxford, Keble Road, Oxford OX1 3RH, U.K.\goodbreak
\and
Theory Division, PH-TH, CERN, CH-1211, Geneva 23, Switzerland\goodbreak
\and
UPMC Univ Paris 06, UMR7095, 98 bis Boulevard Arago, F-75014, Paris, France\goodbreak
\and
Universit\'{e} de Toulouse, UPS-OMP, IRAP, F-31028 Toulouse cedex 4, France\goodbreak
\and
University of Granada, Departamento de F\'{\i}sica Te\'{o}rica y del Cosmos, Facultad de Ciencias, Granada, Spain\goodbreak
\and
University of Granada, Instituto Carlos I de F\'{\i}sica Te\'{o}rica y Computacional, Granada, Spain\goodbreak
% ...
}


\abstract{

This paper presents the characterization of the in-flight beams, the beam window functions, and the associated uncertainties for the \Planck\ Low Frequency Instrument (LFI).
The structure of the paper is similar to that presented in the 2013 $Planck$ release; the main differences concern the beam normalization and the delivery of the window functions to be used for polarization analysis.
The in-flight assessment of the LFI main beams relies on measurements performed during observations of Jupiter. 
By stacking data from seven Jupiter transits, the main beam profiles are measured down to --25 dB at 30 and 44\,GHz, and down to --30 dB at 70\,GHz.
It has been confirmed that the agreement between the simulated beams and the measured beams is better than 1\% at each LFI frequency band (within the 20 dB contour from the peak, the rms values are 0.1\% at 30 and 70 GHz; 0.2\% at 44 GHz).
Simulated polarized beams are used for the computation of the effective beam window functions.
The error budget for the window functions is estimated from both main beam and sidelobe contributions, and accounts for the radiometer band shapes.
The total uncertainties in the effective beam window functions are 0.7\% and 1\% at 30 and 44\,GHz, respectively (at $\ell \approx 600$); and 0.5\,\% at 70\,GHz (at $\ell \approx 1000$).

}

\maketitle

\allearlypapers

\section{Introduction}
\label{introduction}
This paper, one of a set associated with the 2015 release of data from the \Planck\footnote{\Planck\ (\url{http://www.esa.int/Planck}) is a project of the European Space Agency  (ESA) with instruments provided by two scientific consortia funded by ESA member states and led by Principal Investigators from France and Italy, telescope reflectors provided through a collaboration between ESA and a scientific consortium led and funded by Denmark, and additional contributions from NASA (USA).} mission, describes the beams and window functions of the Low Frequency Instrument (LFI). The structure of the paper is similar to that presented in \cite{planck2013-p02d}; the main differences concern the beam normalization and the delivery of the window functions to be used for polarization analysis.

We summarize here the general framework and the nomenclature adopted, which is the same as that used in \cite{planck2013-p02d}.
The LFI optical layout is composed of an array of 11 corrugated feed horns, each coupled to an orthomode transducer (OMT), which splits the incoming electromagnetic wave into two orthogonal, linearly polarized components.
Thus, the LFI observed the sky with 11 pairs of beams, associated with 22 pseudo-correlation radiometers. 
Each beam in a pair is named \texttt{LFIXXM} or \texttt{LFIXXS} for the two polarization states (``Main'' Arm and ``Side'' Arm of the orthomode transducer, respectively). 
Here \texttt{XX} is the radiometer chain assembly number, ranging from \texttt{18} to \texttt{28}. 
The beams from \texttt{LFI18} to \texttt{LFI23} are in the V--band (nominally from 63 to 77\,GHz); we refer to them as 70\,GHz.
The beams from \texttt{LFI24} to \texttt{LFI26} are in the Q--band (from 39.6 to 48.4\,GHz); we refer to them as 44\,GHz.
The beams \texttt{LFI27} and \texttt{LFI28} are in the Ka--band (from 27 to 33\,GHz); we refer to them as 30\,GHz.
The fundamental definitions introduced in \cite{planck2013-p02d}, i.e., optical beams, scanning beams, and effective beams, can be found in Appendix~\ref{regions}.

In the framework of this paper, and the \Planck\ LFI companion papers, we considered three regions defined with respect to the beam boresight:
\begin{enumerate}
\item the main beam, which is defined as extending to 1.9$^\circ$, 1.3$^\circ$, and 0.9$^\circ$ at 30, 44, and 70\,GHz, respectively;
\item the near sidelobes, which are defined as extending between the main beam angular limit and 5$^\circ$;
\item the far sidelobes, which are defined as the beam response greater than 5$^\circ$ from the boresight.
\end{enumerate}

The scanning beams used in the LFI pipeline (affecting calibration, effective beams, and beam window functions) are very similar to those presented in \cite{planck2013-p02d}: they are beams computed with \texttt{GRASP}\footnote{The \texttt{GRASP} software was de\-vel\-oped by TICRA (Co\-pen\-hagen, DK) for analysing gen\-eral re\-flec\-tor antennas (\url{http://www.ticra.it}).}, properly smeared to take into account the satellite motion. 
Simulations have been performed using the optical model described in \cite{planck2013-p02d}, which was derived from the $Planck$ Radio Frequency Flight Model \citep{tauber2010b} by va\-ry\-i\-n\-g some optical parameters (e.g., the relative distance between the two mirrors and the focal plane unit, the feed horn locations and orientations) within the nominal tolerances expected from the thermo\-elastic model, in order to reproduce the me\-a\-su\-re\-men\-ts of the LFI main beams from seven Jupiter transits. 
This tuned optical model is able to represent all the measured LFI main beams with an accuracy of about 0.1\% at 30 and 70 GHz, and 0.2\% at 44 GHz\footnote{These values represent the rms value of the difference between measurements and simulations, computed within the 20 dB contour.}.
 
Unlike the case in \cite{planck2013-p02d}, a different beam normalization is introduced here to properly take into account the actual power entering the main beam (typically about 99\% of the total power). This is discussed in Sect.~\ref{normalization}.

In Sect.~\ref{scanning_beams} the details of the main beam reconstruction from the Jupiter transits are presented. 
The comparison between the measured scanning beam and \texttt{GRASP} scanning beams is also shown.
Section \ref{effective_beams} presents the descriptive parameters of the effective beams, needed for the evaluation of the flux densities of the point sources from the maps.  
In Sect.~\ref{window_function} we present the beam window functions for temperature and polarization analysis.
In the computation of the effective beams and their related window functions, we have significantly increased the outer radius (for 70\,GHz this means a change from 2.5 FWHM to 4 FWHM) to minimize the effect of the cut-off radius. 
The effect of near and far sidelobes on the window functions is described in the same section.
The normalization of the window function reflects the main beam efficiencies presented in Sect.~\ref{normalization}. 
The main parameter that affects the polarization (\textit{EE}) beam window functions was confirmed to be the beam ellipticity, which leads to a temperature-to-polarization leakage of about 15\,\% at multipole $\ell$ equal to 1000 (at 70 GHz) compared to an ideal case of a symmetrical Gaussian beam. 
The error budget on the window functions is presented in Sect.~\ref{error_propagation}. 

\section{Beam normalization}
\label{normalization}
\begin{figure}[hptb]
\centering
\includegraphics[width=9cm]{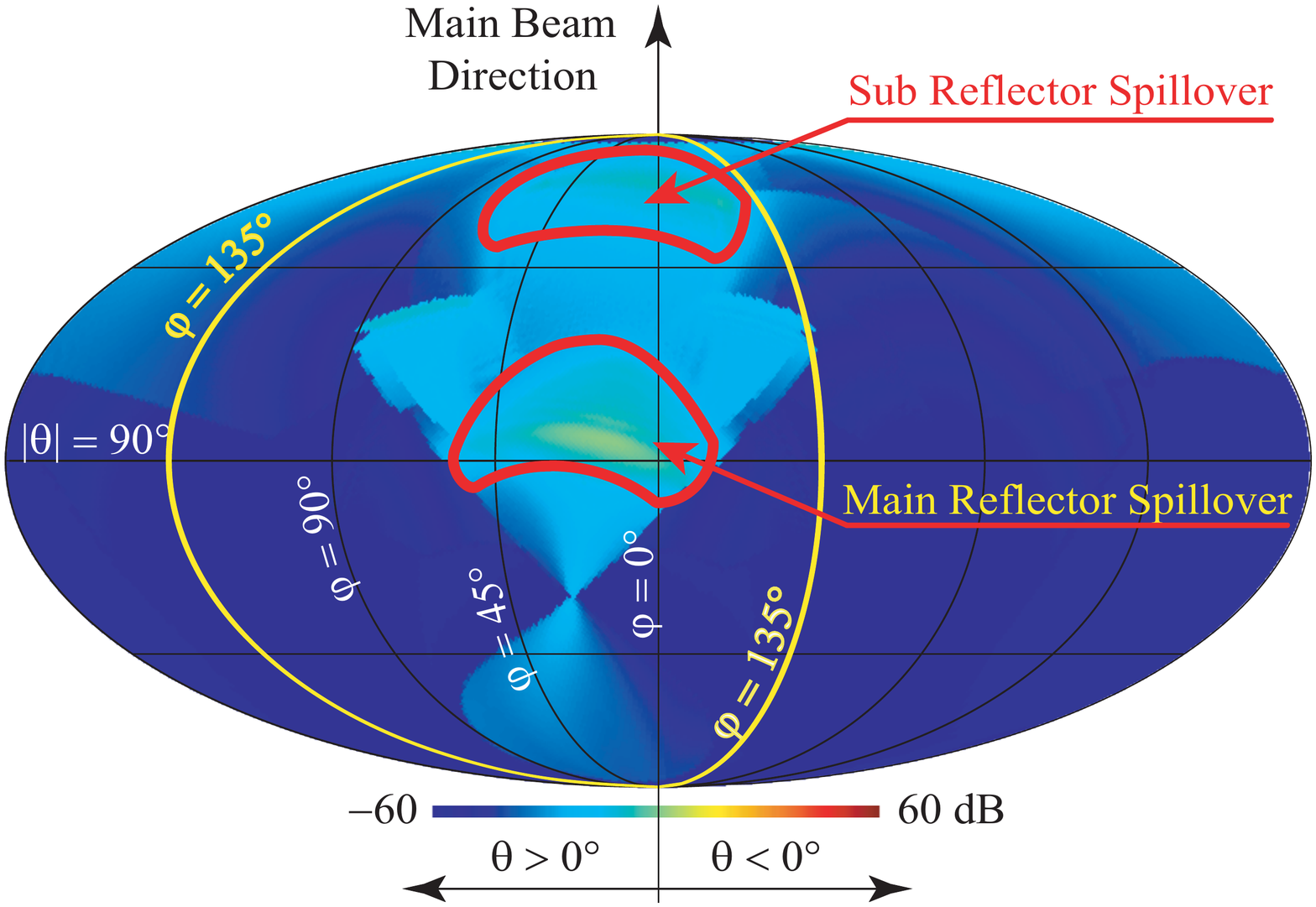} 
\caption{Far sidelobes at 30 GHz. The main beam points to the top of the map ($\theta$ = 0$^\circ$). The main and sub-reflector spillover regions are highlighted with red contours. The main-reflector spillover points at about 85$^\circ$ from the main beam pointing direction and it is peaked at about 2 dBi. The sub\-reflector spillover is mainly due to the feed sidelobe and peaks at about --8 dBi.}
\label{fig:sl_legend}
\end{figure}

In previous work \citep{planck2013-p02d}, the main beam used in the calculation of the effective beams (and effective beam window functions) was a full-power main beam (i.e., unrealistically set to 100\,\% efficiency). 
The resulting beam window function was normalized to unity because the calibration was performed assuming a pencil beam. 
This assumption considers that all the power entering the feed horn comes from the beam line of sight. 
We know that this assumption is not realistic, since up to 1\% of the solid angle of the LFI beams falls into the sidelobes, unevenly distributed and concentrated mainly in two areas, namely the main-reflector and sub-reflector spillover (see Fig.~\ref{fig:sl_legend}). 
The main-reflector spillover is primarily due to the rays reflected by the lower part of the sub-reflector and those diffracted by the two reflectors; it peaks at about 90$^\circ$ from the telescope line of sight, along the direction of the satellite spin axis, and it has an intensity below --50 dB from the main beam power peak. The sub-reflector spillover (whose intensity is lower than the main-reflector spillover) is generated by the rays entering the feed without any interaction with the reflectors; its shape aligns roughly with the feed, pointing at about 20$^\circ$ from the line of sight of the telescope. 
They are both extended structures whose shape and power change significantly across the band.

Because we accurately model the dipole signal, by convolving the sky dipole with the full $4\pi$ beam response of each radiometer, our calibration procedure correctly converts the time-ordered data into received antenna temperature in kelvin, where that temperature represents the full-sky temperature weighted by the $4\pi$ beam \citep{planck2014-a06}. 
Our mapping procedure assumes a pencil beam \citep{planck2014-a07}, which in the ideal case of a circularly-symmetric beam would yield a map of the beam-convolved sky; 
however, a fraction of the signal from any source appears in the far sidelobes, and would be missed by integration of the map over the main beam alone. 
By the same token, bright resolved features in the map have temperatures fractionally lower than in the sky, due to signal lost to the sidelobes. 
In essence this description remains true even given the highly asymmetric sidelobes of the \Planck\ beam: the main difference is that the far sidelobe contribution to a given pixel
varies according to the orientation of the satellite at the time of observation.  
This is handled by explicitly subtracting a model of the Galactic straylight and treating the remaining effect as a noise term. 
Important to note is that the roughly 1\% of the signal found in the sidelobes is missing from the vicinity of the main beam, so the main beam efficiency $\eta \approx 99$\,\%; and this must be accounted for in any analysis of the maps. In particular, the window function used to correct the power spectra extracted from the maps (which is based on the main beam only) allows for this efficiency. 
Likewise, to calculate the flux densities of compact sources from LFI maps, we should correct for the main beam efficiency or, alternatively, deconvolve the beam from the map before calculating the flux densities (the latter approach takes into account the true beam shape, not just its angular resolution and/or solid angle).
In other words, the source flux densities must be scaled up by a correction factor, as presented in \cite{planck2014-a03}. 
In particular, the scaling factors are 1.00808, 1.00117, and 1.00646 at 30, 44, and 70 GHz, respectively.
 
The efficiency values listed in Table \ref{mbe} were calculated by taking into account the variation across the band of the optical response (coupling between feed horn pattern and telescope) and the radiometric response. 
The bandpass of each radiometer is unique, since it depends very sensitively on the manufacturing process. 
Therefore two beams that are optically similar (e.g., mirror-symmetric, such as 18S and 23S) have different main beam efficiencies; and two beams that might be expected to have different efficiencies (because the off-axis telescope responds differently to the two polarizations) are characterized by very similar efficiencies (for example, 22M and 22S). 
The impact of this imbalance in the efficiencies is not negligible for the 30 GHz window functions; it generates a bump at low $\ell$ compared to the previous release which treated an ideal case in which the beams were normalized to unity (see Fig.~\ref{fig:bump} in Sect.~\ref{window_function}). 
At higher frequencies this impact is negligible.

Table \ref{mbe} reports the main beam efficiency of each LFI beam as well as the percentage of the power entering the near and far sidelobes.
We note that there is a small fraction of $missing$ power in LFI sidelobes, resulting from the first-order approximation adopted in the computation carried out with the \texttt{GRASP} Multi-reflector Geometrical Theory of Diffraction (\texttt{MrGTD}) \citep{graspmrgtd}.
In \cite{planck2013-p02d} we expressed the hope that we could include in the current paper the higher order contributions, but we found that the computational cost of such an analysis, performed across the band, was prohibitive.
However, we performed some tests in collaboration with HFI, comparing the straylight evaluated with sidelobes computed at the 1st and 7th orders, and found that the resulting differences are negligible, both at map level (0.2\,$\mu$K) and power spectrum level (lower than 10$^{-15}$ K$^2$). 
In other words, it does seem that the missing power is broadly distributed at a low power level and does not have a significant impact on the straylight contamination, which is clearly dominated by the main-reflector spillover.

\begin{table}[htpb]
\centering
\caption{Beam efficiency computed from \texttt{GRASP} beams. In the first column the main beam efficiency, $\eta$, is presented. The second and third columns report the percentage of the power entering the near and far sidelobes, respectively ($n_{{\rm sl}}$ and $f_{{\rm sl}}$): these values are directly calculated as the integral of the electric field computed with \texttt{GRASP}. The sum of the three beam components is presented in the fourth column. The three regions considered (main beam, near, and far sidelobes) are those defined in Sect.~\ref{introduction}.}
\begin{tabular}{c c c c c} 
\hline
\hline
\noalign{\vskip 2pt}
Beam &  $\eta$ & $n_{{\rm sl}}$ & $f_{{\rm sl}}$ & Total \\ 
\hline
\noalign{\vskip 2pt}
\multicolumn{5}{l}{ 70 GHz} \\
\noalign{\vskip 4pt}
\texttt{18S} 	 & 98.87	&	0.12	&	0.62	&	99.60	\\
\texttt{18M} 	 & 99.21	&	0.09	&	0.38	&	99.68	\\
\texttt{19S} 	 & 98.98	&	0.11	&	0.58	&	99.66	\\
\texttt{19M} 	 & 98.83	&	0.13	&	0.60	&	99.56	\\
\texttt{20S} 	 & 98.81	&	0.13	&	0.70	&	99.64	\\
\texttt{20M} 	 & 98.85	&	0.13	&	0.63	&	99.61	\\
\texttt{21S} 	 & 98.82	&	0.13	&	0.70	&	99.65	\\
\texttt{21M} 	 & 98.94	&	0.11	&	0.59	&	99.64	\\
\texttt{22S} 	 & 99.15	&	0.08	&	0.50	&	99.73	\\
\texttt{22M} 	 & 99.16	&	0.08	&	0.44	&	99.69	\\
\texttt{23S} 	 & 99.19	&	0.09	&	0.43	&	99.71	\\
\texttt{23M} 	 & 99.26	&	0.08	&	0.35	&	99.69	\\
\hline                                                                                                           
\noalign{\vskip 2pt}                                                                                             
\multicolumn{5}{l}{ 44 GHz} \\                                                                                
\noalign{\vskip 4pt}                                                                                             
\texttt{24S}   & 99.73	&	0.03	&	0.15	&	99.91	\\  
\texttt{24M} 	 & 99.72	&	0.03	&	0.15	&	99.90	\\
\texttt{25S} 	 & 99.76	&	0.02	&	0.06	&	99.84	\\
\texttt{25M} 	 & 99.75	&	0.03	&	0.08	&	99.86	\\
\texttt{26S} 	 & 99.77	&	0.02	&	0.05	&	99.84	\\
\texttt{26M} 	 & 99.74	&	0.03	&	0.08	&	99.85	\\
\hline\noalign{\vskip 2pt}                                                                                       
\multicolumn{5}{l}{ 30 GHz} \\                                                                                
\noalign{\vskip 4pt}                                                                                             
\texttt{27S} 	 & 98.89	&	0.09	&	0.76	&	99.75	\\
\texttt{27M} 	 & 99.04	&	0.08	&	0.64	&	99.76	\\
\texttt{28S} 	 & 98.79	&	0.10	&	0.83	&	99.73	\\
\texttt{28M} 	 & 99.07	&	0.07	&	0.62	&	99.76	\\
\hline
\end{tabular}
\label{mbe}
\end{table}

\section{Scanning beams}
\label{scanning_beams}
\subsection{Planet data}
\label{planet_data}

\begin{figure*}[hptb]
\centering
\begin{tabular}{c c}
\includegraphics[width=8.5cm]{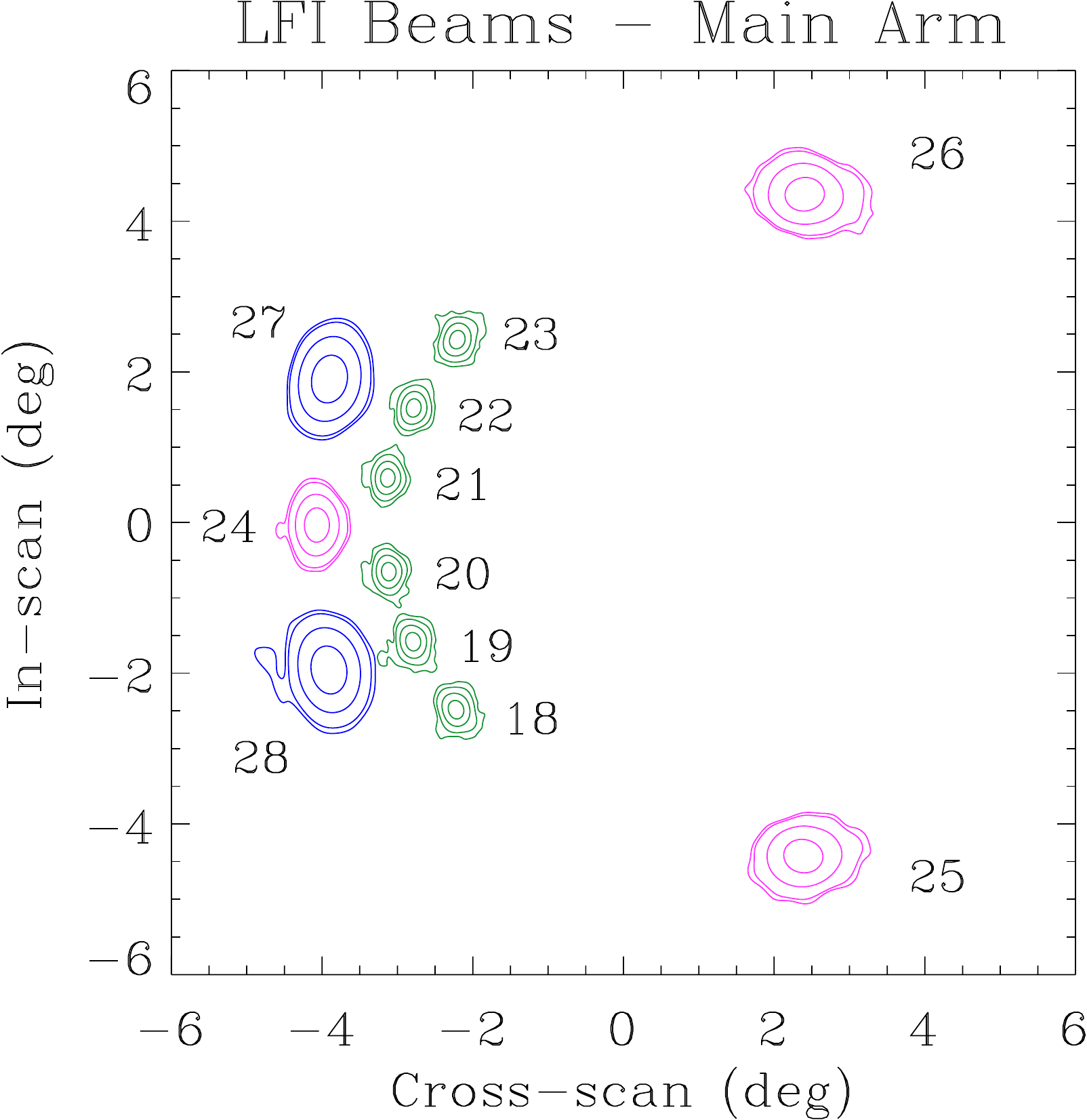} &
\includegraphics[width=8.5cm]{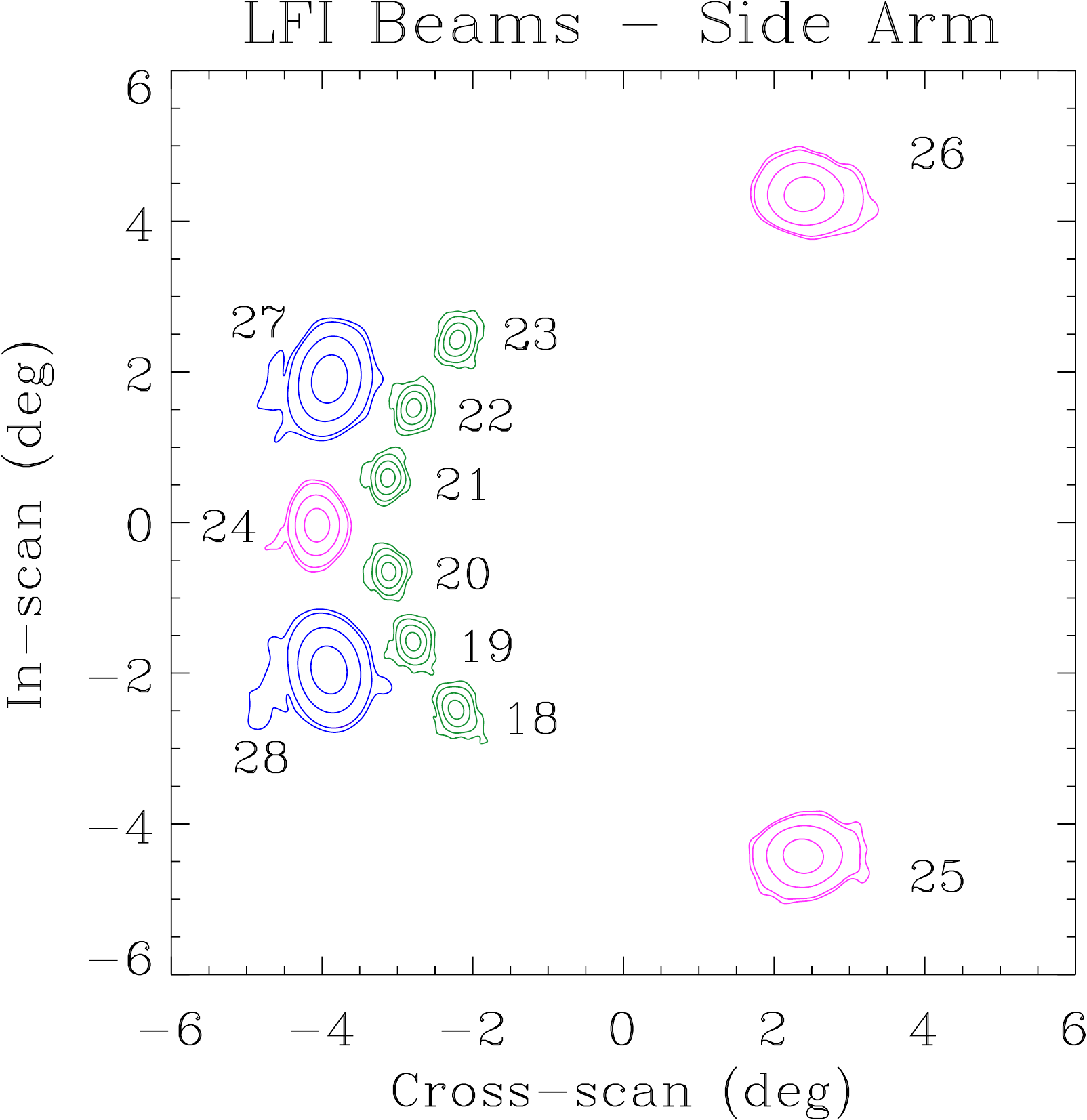} \\
\end{tabular}
\caption{Scanning beam profiles for both polarization arms, reconstructed from seven Jupiter transits. The beams are plotted in contours of --3, --10, --20, and --30 dB from the peak at 70\,GHz (green), and --3, --10, --20, --25 dB at 30\,GHz (blue) and 44\,GHz (pink).}
\label{fig:uv}
\end{figure*}

The LFI in-flight main beam reconstruction is based on the same method adopted in the past release \citep{planck2013-p02d}.
In Fig.~\ref{fig:uv} the LFI footprint on the sky is shown for both polarization arms.
In contrast to the analogous figure reported in \cite{planck2013-p02d}, here the beams are plotted down to --30 dB at 70 GHz, and --25 dB at 30 and 44 GHz.

\begin{figure}[hptb]
\centering
\includegraphics[width=8.5cm]{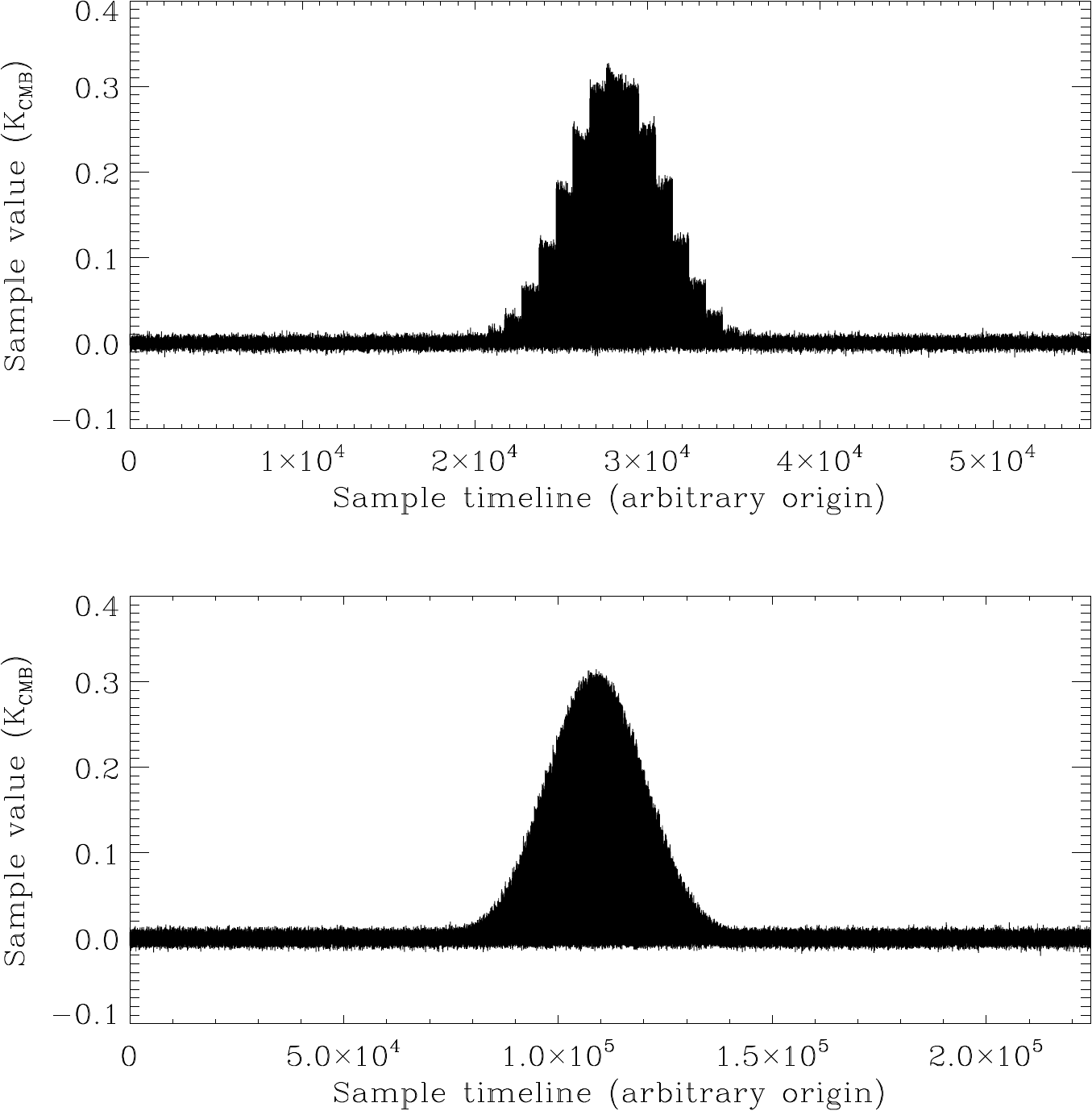} 
\caption{Timelines corresponding to radiometer \texttt{LFI18S} at 70\,GHz: first nominal scan (upper panel) and seventh deep scan (bottom panel).}
\label{fig:scan-18s}
\end{figure}

To assess the beam properties, we used seven Jupiter transits \citep{planck2014-a03}. The first four transits (J1 to J4) occurred in nominal scan mode (spin shift 2 arcmin, 1 degree per day), and the last three scans (J5 to J7) in deep mode (shift of the spin axis between rings of 0.5 arcmin, 15 arcmin per day). 
Figure \ref{fig:scan-18s} shows two Jupiter scans at 70 GHz: the first one, in nominal mode; and the seventh, in deep mode. 
Some data from the first deep scan have been discarded and, for this reason we used only the last two deep scans at the lower frequencies (30 and 44 GHz). 
For the 70 GHz channel, the resulting sampling of the uv-plane is about 3.4 times better than in the earlier paper and, consequently, the signal-to-noise ratio is about 1.8 times better. 
At 44 and 30 GHz the improvement is slightly lower (1.3 and 1.5 times better, respectively), since data from the first deep scan could not be used due to spacecraft manoeuvrements.
As a result of the deeper sampling, the error on the reconstructed beam parameters is lower with respect to the previous release, as can be seen by comparing Table~\ref{tab:imo} with table~2 of \citet{planck2013-p02d}, and the error envelope on the window functions is lower as well.


\begin{figure}[htpb]
\centering
\includegraphics[width=9cm]{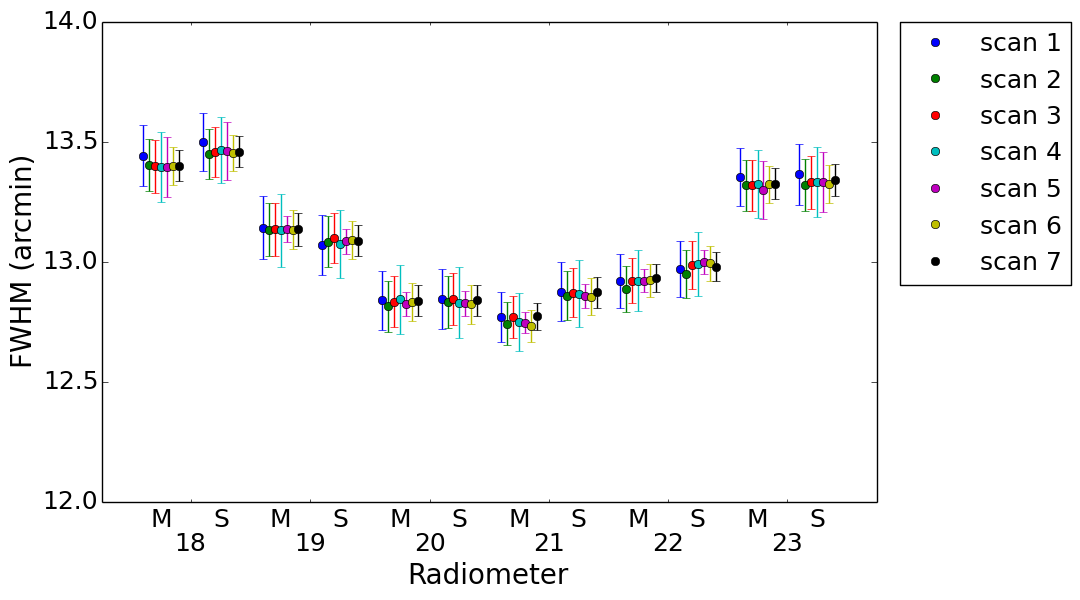} 
\includegraphics[width=9cm]{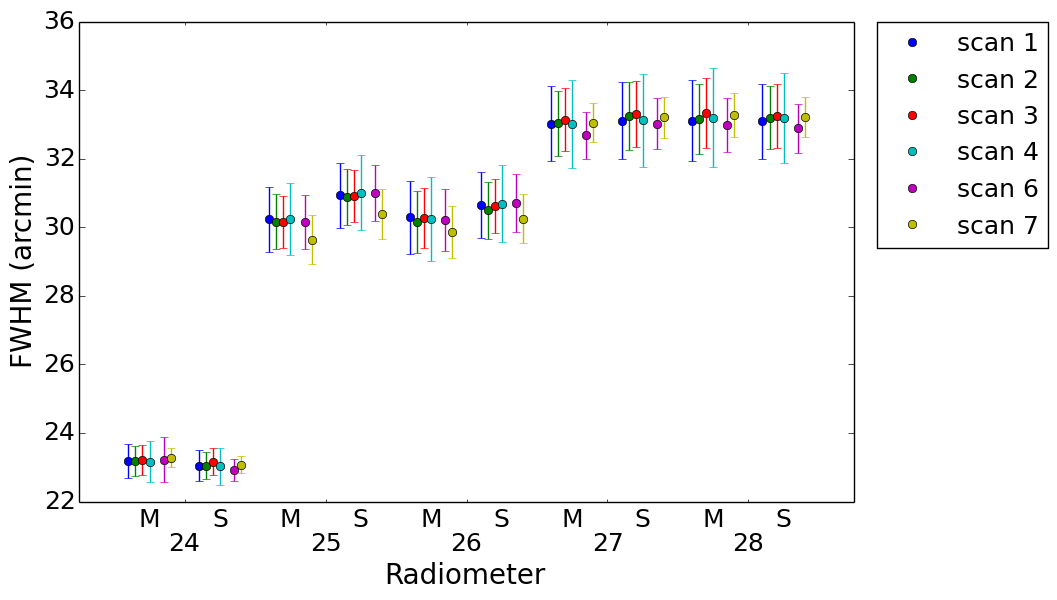}
\caption{FWHM at 70 GHz (upper panel) and 30/44\,GHz (bottom panel) for the seven Jupiter scans.}
\label{fig:histofwhm}
\end{figure}

\begin{figure}[htpb]
\centering
\includegraphics[width=9cm]{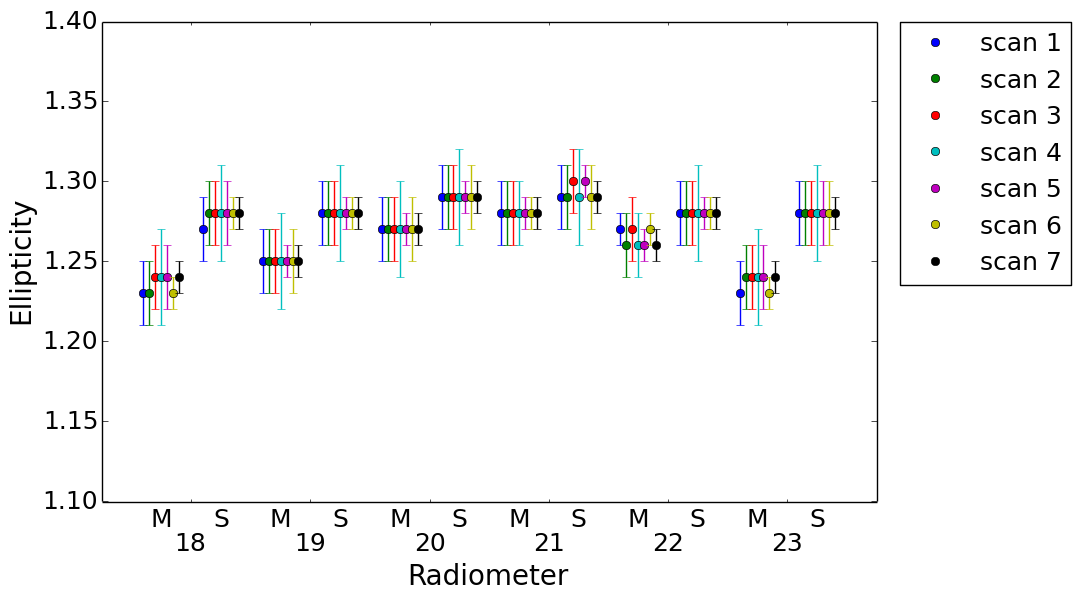} 
\includegraphics[width=9cm]{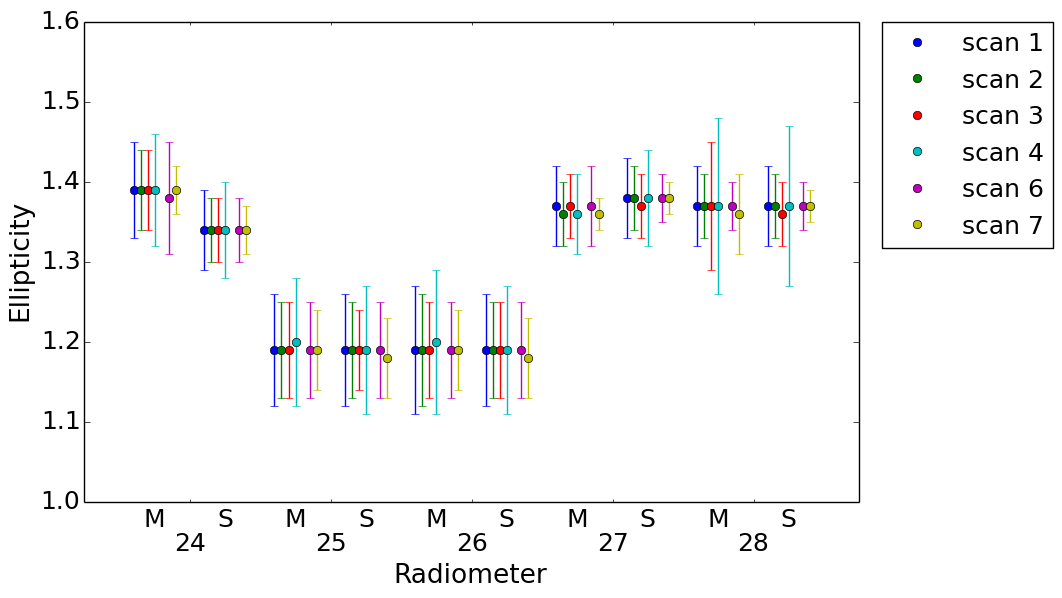}
\caption{Ellipticity at 70 GHz (upper panel) and 30/44\,GHz (bottom panel) for the seven Jupiter scans.}
\label{fig:histoe}
\end{figure}

\begin{figure}[htpb]
\centering
\includegraphics[width=9cm]{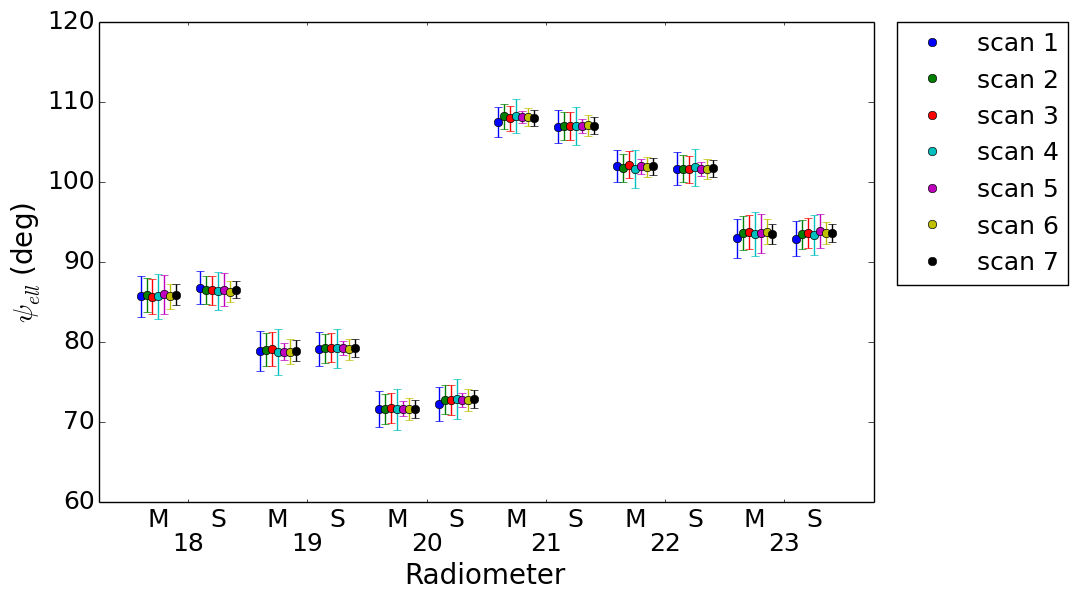}
\includegraphics[width=9cm]{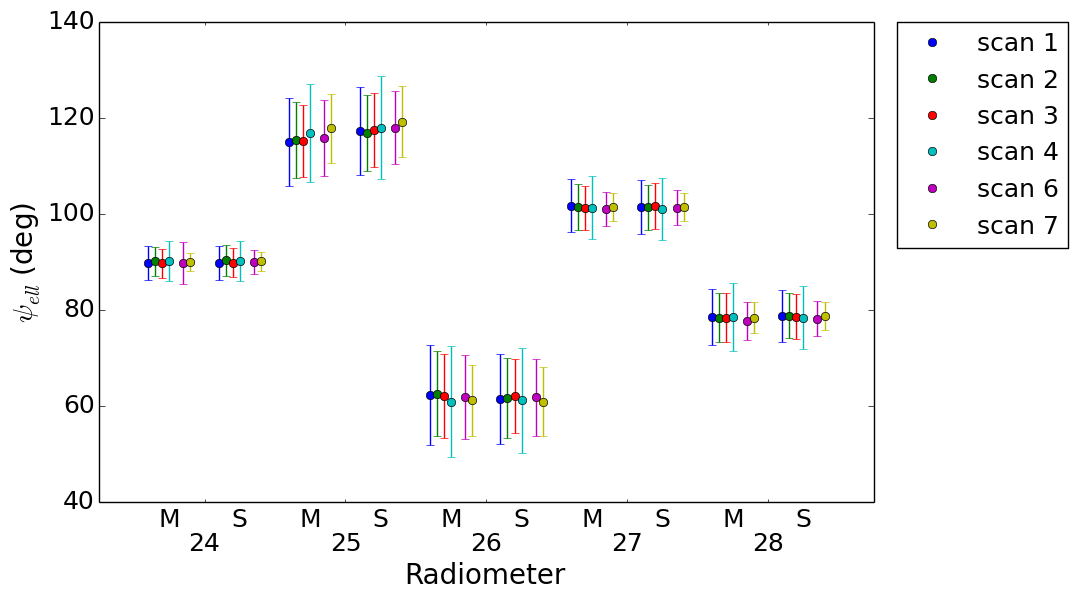} 
\caption{Beam orientation at 70 GHz (upper panel) and 30/44 GHz (bottom panel) for the seven Jupiter scans. $\psi_\textrm{ell}$ is defined in \cite{planck2013-p02d}.}
\label{fig:histopsi}
\end{figure}

Table \ref{tab:imo} reports the main beam descriptive parameters with the estimated uncertainties evaluated from the stacked beams obtained using all seven Jupiter transits.
Figs.~\ref{fig:histofwhm}, \ref{fig:histoe}, and \ref{fig:histopsi} show the values of FWHM, ellipticity, and beam orientation derived from each Jupiter transit. 
The scatter among the values reconstructed from different transits is much smaller than that expected from the errors quoted for each transit, which conservatively includes any possible systematic effects. The main uncertainty comes from the fact that the elliptical Gaussian representation of the beam shape, adopted only in this fit, although accurate at a level of $\sim$ few $\mu$K for characterizing the power entering the main beam and the signal convolved with sky diffuse emissions \citep{burigana2001}, shows a small point-to-point difference with the reconstructed beam at a level of a few percentage points. If we consider only the statistical properties of the noise and sky fluctuations, as measured from the analysis of the signal variance just outside the main beam, the resulting error bars would be about ten times smaller, but in this release we adopted a conservative approach to define the uncertainties in the beam window functions.
It is evident that the seven measurements give basically the same results.
Thus, no time-dependent optical effects are evident in these data, which were taken from October 2009 to February 2013. 

With respect to the previous main beam reconstruction using four Jupiter transits \citep{planck2013-p02d}, there is an improvement in the uncertainties on the FWHM, ellipticity, and $\psi_{{\rm ell}}$, respectively, by factors of about 1.8, 3.1, and 1.8 at 70 GHz; 1.5, 2.1, and 1.5 at 44 GHz; and 1.6, 1.9, and 1.6 at 30 GHz.
These numbers reflect the improvement in the coverage of the uv-plane of the stacked beams: the number of samples including the three deep scans is about 3.4 times higher at 70 GHz, 1.7 at 44 GHz, and 2.2 at 30 GHz.   
For completeness, in Appendix~\ref{beam_fit_results} the fitted parameters are reported for each scan. 

\begin{table}[hptb]
\footnotesize
\centering
\caption{Main beam de\-scrip\-tive pa\-ram\-e\-ters of the scanning beams, with $\pm 1\,\sigma$ uncertainties.}
\begin{tabular}{c c c r@{.}l}
\hline
\hline
\noalign{\vskip 2pt}
Beam & FWHM & Ellipticity & \multicolumn{2}{c}{$\psi_{{\rm ell}}$} \\
 & (arcmin) &  & \multicolumn{2}{c}{(degrees)} \\
\hline
\noalign{\vskip 2pt}
\multicolumn{5}{l}{ 70\,GHz} \\
\noalign{\vskip 4pt}
\texttt{18M} 	& 13.40	$\pm$ 0.02  &	1.235 $\pm$ 0.004	& 85&74	  $\pm$ 0.41 \\  
\texttt{18S} 	& 13.46	$\pm$ 0.02	& 1.278	$\pm$ 0.004	& 86&41	  $\pm$ 0.33 \\  
\texttt{19M} 	& 13.14	$\pm$ 0.02	& 1.249	$\pm$ 0.003	& 78&82	  $\pm$ 0.35 \\  
\texttt{19S} 	& 13.09	$\pm$ 0.02	& 1.281	$\pm$ 0.002	& 79&15	  $\pm$ 0.30 \\  
\texttt{20M}	& 12.83	$\pm$ 0.02	& 1.270	$\pm$ 0.003	& 71&59	  $\pm$ 0.32 \\  
\texttt{20S}	& 12.83	$\pm$ 0.02	& 1.289	$\pm$ 0.004	& 72&69	  $\pm$ 0.31 \\  
\texttt{21M}	& 12.75	$\pm$ 0.02	& 1.280	$\pm$ 0.003	& 107&99	$\pm$ 0.27 \\  
\texttt{21S}	& 12.86	$\pm$ 0.02	& 1.294	$\pm$ 0.003	& 106&96	$\pm$ 0.29 \\  
\texttt{22M}	& 12.92	$\pm$ 0.02	& 1.264	$\pm$ 0.003	& 101&87	$\pm$ 0.30 \\  
\texttt{22S}	& 12.99	$\pm$ 0.02	& 1.279	$\pm$ 0.003	& 101&61	$\pm$ 0.30 \\  
\texttt{23M}	& 13.32	$\pm$ 0.02	& 1.235	$\pm$ 0.004	& 93&53	  $\pm$ 0.40 \\  
\texttt{23S}	& 13.33	$\pm$ 0.02	& 1.279	$\pm$ 0.004	& 93&49  	$\pm$ 0.36 \\ 
\hline
\noalign{\vskip 2pt}
\multicolumn{5}{l}{ 44\,GHz} \\
\noalign{\vskip 4pt} 
\texttt{24M}	& 23.18	$\pm$ 0.05	& 1.388	$\pm$ 0.005	& 89&82	  $\pm$ 0.33 \\  
\texttt{24S}	& 23.03	$\pm$ 0.04	& 1.344	$\pm$ 0.003	& 89&97	  $\pm$ 0.34 \\  
\texttt{25M}	& 30.02	$\pm$ 0.07	& 1.191	$\pm$ 0.005	& 115&95	$\pm$ 0.75 \\  
\texttt{25S}	& 30.79	$\pm$ 0.07	& 1.188	$\pm$ 0.005	& 117&70	$\pm$ 0.74 \\  
\texttt{26M}	& 30.13	$\pm$ 0.08	& 1.191	$\pm$ 0.006	& 61&89  	$\pm$ 0.84 \\  
\texttt{26S}	& 30.52	$\pm$ 0.08	& 1.189	$\pm$ 0.006	& 61&53	  $\pm$ 0.77 \\
\hline
\noalign{\vskip 2pt}
\multicolumn{5}{l}{ 30\,GHz} \\
\noalign{\vskip 4pt}  
\texttt{27M}	& 32.96	$\pm$ 0.06	& 1.364	$\pm$ 0.005	& 101&20	$\pm$ 0.34 \\  
\texttt{27S}	& 33.16	$\pm$ 0.07	& 1.379	$\pm$ 0.005	& 101&29	$\pm$ 0.34 \\  
\texttt{28M}	& 33.17	$\pm$ 0.07	& 1.366	$\pm$ 0.006	& 78&17  	$\pm$ 0.36 \\  
\texttt{28S}	& 33.12	$\pm$ 0.07	& 1.367	$\pm$ 0.005	& 78&47	  $\pm$ 0.33 \\  
\hline
\end{tabular}
\label{tab:imo}
\end{table}                                                                  

\subsection{Polarized scanning beams}  
\label{polarized_beams} 

The polarized scanning beams have been evaluated from optical simulations carried out by the application of physical optics and physical theory of diffraction using {\tt GRASP}.
As reported in \cite{planck2013-p02d}, these beams came from a dedicated optical study that has been carried out with the goal of fitting the simulated beams to the in-flight measurements. Of course, to take into account the satellite motion, the optical beams have been properly smeared.
The impact of the polarization of Jupiter is negligible because it is well below the level of beam measurements (see Appendix B of \cite{planck2013-p02d} for a detailed description and evaluation of such effect).

The Low Frequency Instrument performs polarization measurements of the cosmic microwave background (CMB) anisotropies by combining the signal received by the feed horns appropriately aligned in the focal plane \citep{leahy2010}.
All LFI feed horns are off-axis and the respective main beams, located at 3 to 5$^\circ$ from the telescope line of sight (LOS), suffer some aberration.
The LFI main beams can be considered linearly polarized, to first order, but we are conscious of the impact of a non-null cross-polarization close to the main beam pointing direction.  Knowledge of the polarization properties of each main beam (i.e., co- and cross-polar components) and of the spacecraft pointing direction are required to perform polarization measurements.
Since we were not able to measure the cross-polar beam in flight, we have relied on simulations validated by far more accurate beam measurements than those reported earlier in \cite{planck2013-p02d}.
The strength of the model adopted is twofold: (i) we have a description of the beams at levels lower than the instrumental noise; and (ii) the beam cross-polar component is fully characterized.

The GRASP main beams were computed in uv-spherical polar grids (see Appendix \ref{regions} for the definition of the main beam region).
In each point of the uv-grid, the far field was computed in the co- and cross-polar basis according to Ludwig's third definition \citep{ludwig}.

Although the GRASP beams are computed as the far-field angular transmission function of a highly polarized radiating element in the focal plane, the far-field pattern is in general no longer exactly linearly polarized: a spurious component, induced by the optics, is present \citep{sandri2010}.
The co-polar pattern is interpreted as the response of the linearly polarized detector to radiation from the sky that is linearly polarized in the direction defined as co-polar, and the same is true for the cross-polar pattern, where the cross-polar direction is orthogonal to the co-polar one.
The Jupiter scans allow us to measure only the total field, that is, the co- and cross-polar components combined in quadrature.
The total field of GRASP beams fits the Jupiter data, but these beams also have the co- and cross-polar pattern defined separately.
The adopted beam reference frame, in which each main beam was computed, implies that the power peak of the co-polar component lies in the centre of the uv-grid, and a minimum in the cross-polar component appears at the same point.
In particular, the major axis of the polarization ellipse is along the u-axis for the radiometer side arm and it is aligned with v- for the radiometer main arm.
This means that, very close to the beam pointing direction, the main beam can be assumed to be linearly polarized; the x-axis of the main beam frame can be assumed to be the main beam polarization direction for the radiometers S; and the y-axis of the main beam frame can be assumed to be the main beam polarization direction for the radiometers M.

We have evaluated the effect of cross-polarization on the window functions, and find that it is roughly 1\% at 70 GHz for $\ell$ equal to 1000.
The GRASP beams are normalized to have an integrated solid angle of $4\pi$\,sr.
The integral over the main beam region (the summed co- and cross-polar power) is representative of the main beam efficiency. 

\subsection{Hybrid beams}

\begin{figure}[htpb]
\centering
\includegraphics[width=8cm]{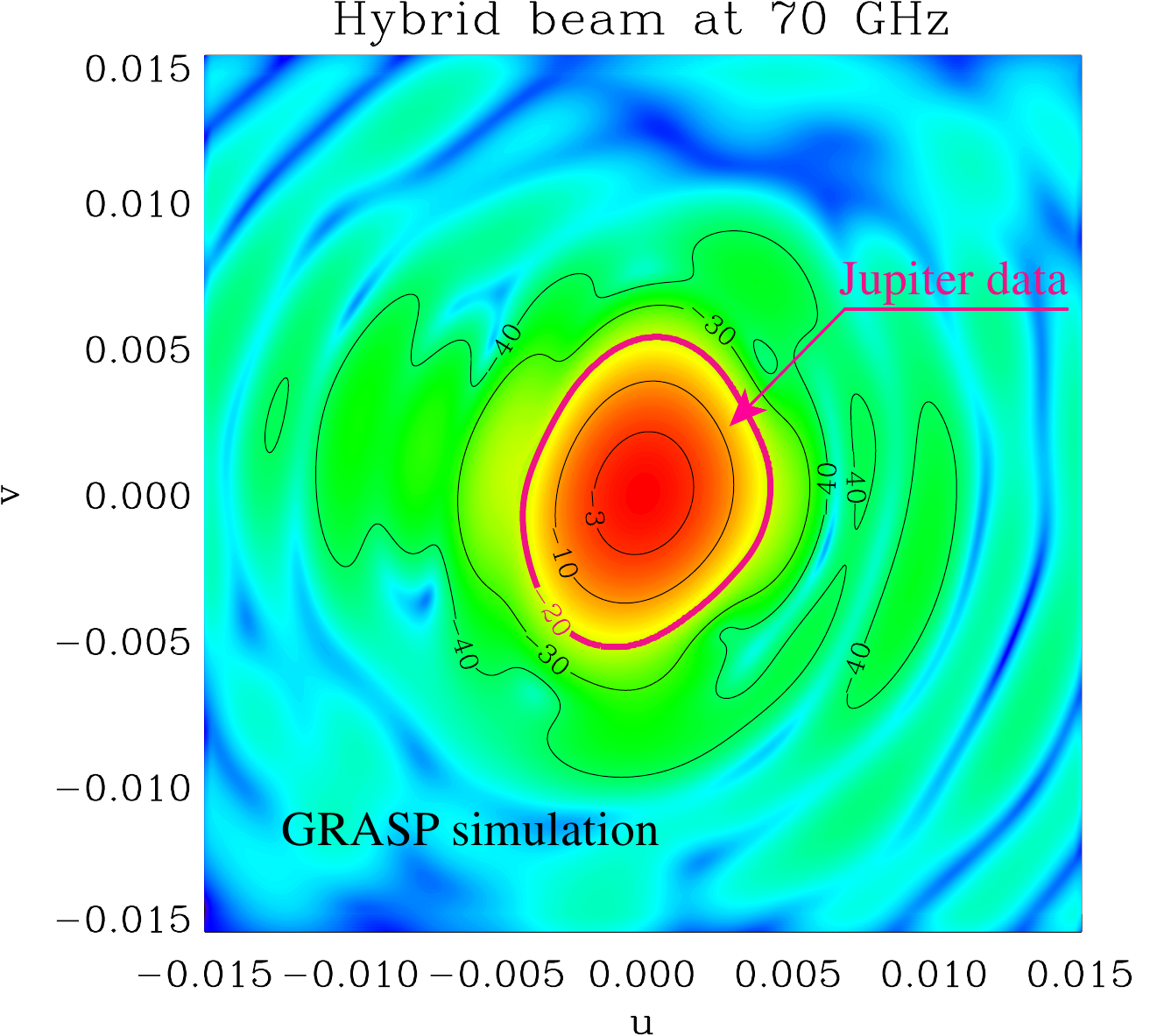} 
\caption{Hybrid beam at 70 GHz. The data within the 20 dB contour are measurements (i.e., Jupiter data), filtered and interpolated on a regular grid. The data at lower levels are GRASP si\-mu\-lations, smeared to take into account the satellite motion.}
\label{fig:hybrid}
\end{figure}

\begin{figure}[htpb]
\centering
\includegraphics[width=8cm]{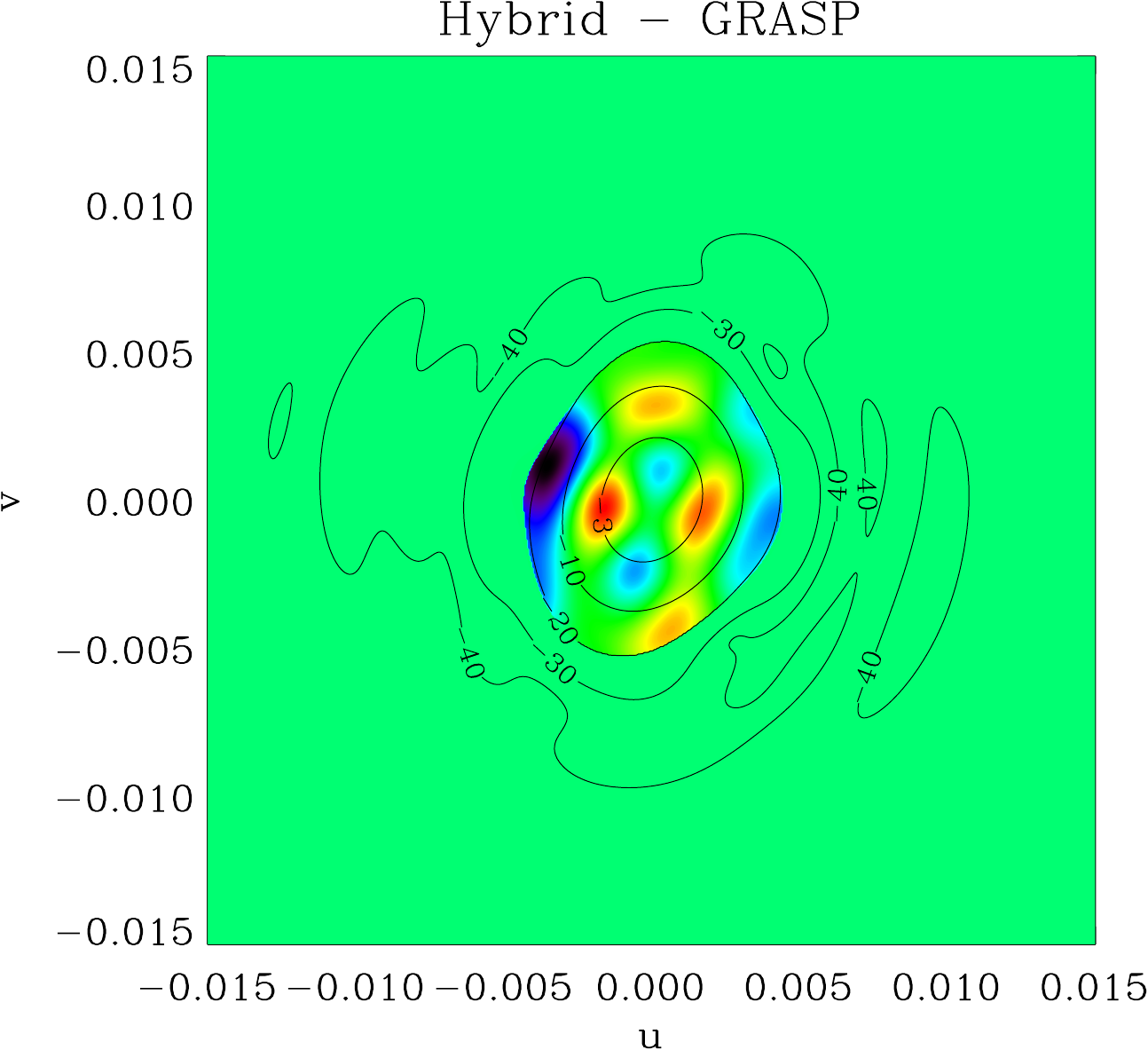} 
\caption{Difference between hybrid beam and GRASP simulation. The colour scale spans 2.25 times the rms of the beam difference, i.e., 0.1\% of the beam maximum.}
\label{fig:grasp}
\end{figure}

\begin{figure*}[htpb]
\centering
\begin{tabular}{ccc}
\includegraphics[width=6cm]{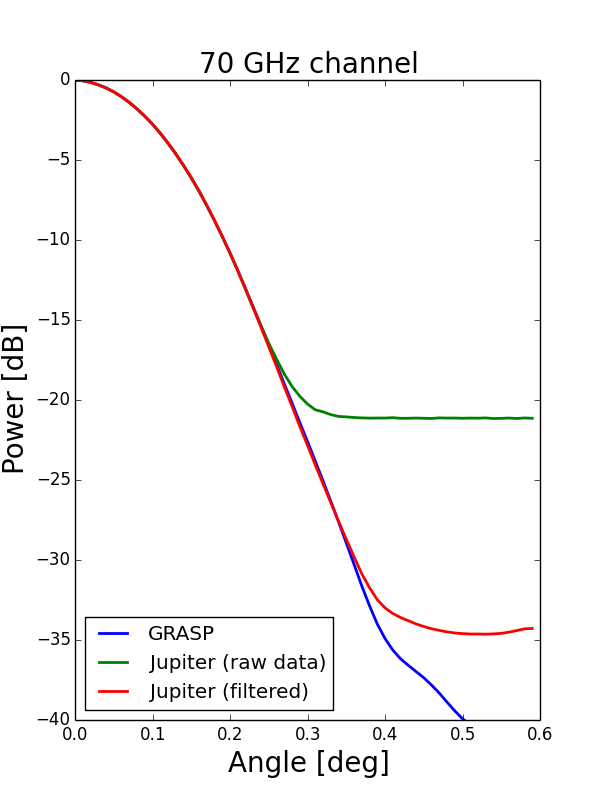} 
\includegraphics[width=6cm]{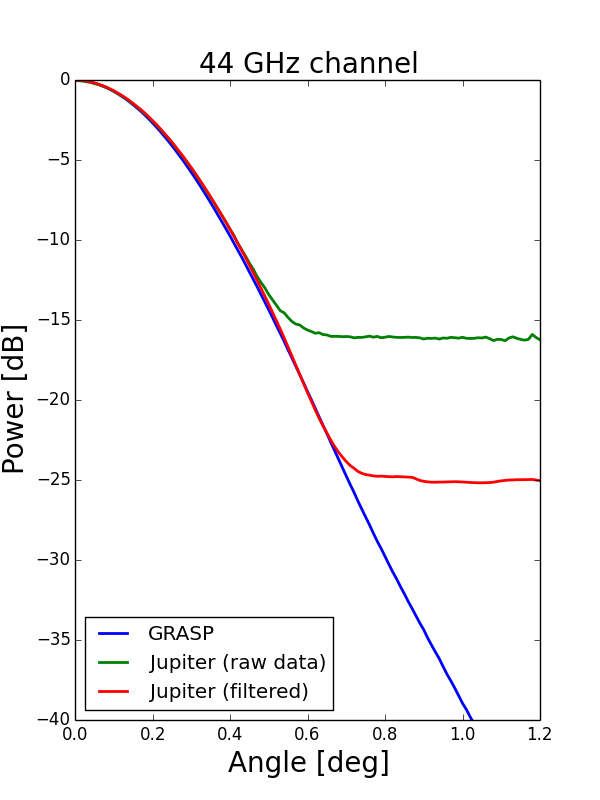} 
\includegraphics[width=6cm]{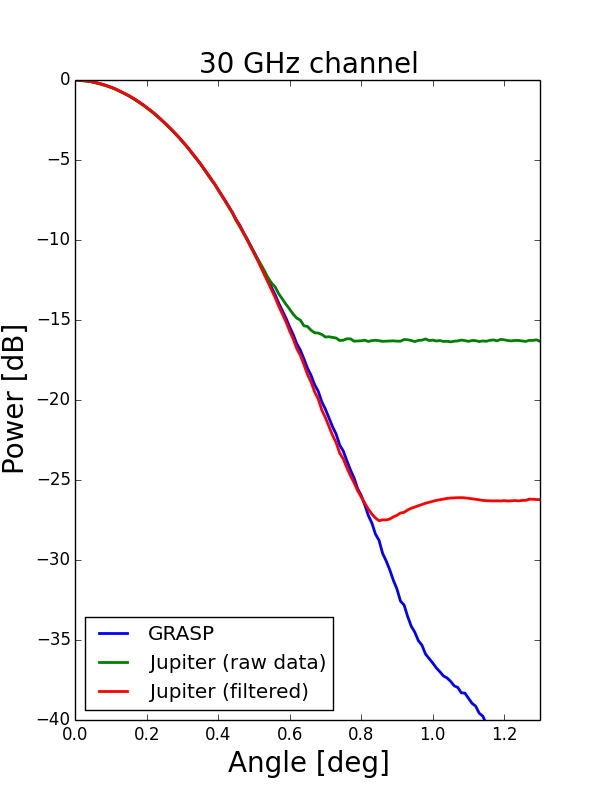} 
\end{tabular}
\caption{Comparison between the (symmetrized) \texttt{GRASP} beams and the (symmetrized) beam profile measured with Jupiter (raw data and filtered data) for the three channels, 30, 44, and 70 GHz. The symmetrized beams have been computed by averaging all the beams of each channel, and then averaging over radial angle to be circularly symmetric.}
\label{fig:comp_sym}
\end{figure*}

Unlike in the previous release, this time we have produced a new main beam model named the ``hybrid beam''.
Hybrid beams have been created using planet measurements above 20 dB below the main beam power peak and GRASP beams below this threshold (see Figs.~\ref{fig:hybrid} and \ref{fig:grasp}).
The planet data have been filtered using a maximally flat magnitude filter (Butterworth filter) to reduce the noise.
The hybrid beams have been normalized according to the GRASP beams (i.e., the main beam efficiency is set to the same value).  
We used the hybrid beams to perform a further check on the consistency between the GRASP model and the planet data, in terms of window functions.
Figure \ref{fig:comp_sym} shows the comparison between the symmetrized GRASP beams and the planet data.
The polarized beams provide the best fit to the available measurements of the LFI main beams from Jupiter; this model represents all the LFI beams with an accuracy of about 0.1\% at 30 and 70 GHz, and 0.2\% at 44 GHz (rms value of the difference between measurements and simulations, computed within the 20~dB contour).
Figure \ref{fig:wf} shows for all channels the comparison between the window functions computed using GRASP beams and the window functions computed using hybrid beams.

\section{Effective beams} 
\label{effective_beams}
The effective beam is defined in the map domain, and is obtained by averaging the scanning beams that are pointed at a given pixel of the sky map, while taking into account both the scanning strategy and the orientation of the scanning beams when they point at that pixel.   
The effective beams thus capture information about the difference between the true and observed images of the sky. 
They are, by definition, the objects whose convolution with the true CMB sky produces the observed sky map, at least in the absence of sidelobes.
Similarly, the effective beam window functions capture the ratio between the true and observed angular power spectra.
As in \cite{planck2013-p02d}, we compute in this paper the effective beam at each sky pixel for each LFI frequency scanning beam, and scan history using the {\tt FEBeCoP} method.
For a detailed account of the algebra involving the effective beams for temperature and polarization see \citet{mitra2010}.

The main beam solid angle of the effective beam, $\Omega_{{\rm eff}}$, is estimated as the integral over the full extent of the effective beam. 
A larger cut-off radius has been applied to the main beams: 113.6 arcmin at 30 GHz; 79 arcmin at 44 GHz; and 52 arcmin at 70 GHz.
From the effective beam solid angle, we can estimate the effective full width half maximum (FWHM$_{{\rm eff}}$), assuming a Gaussian of equivalent solid angle.
These values have been averaged across the map to obtain the band (quadruplets) averaged effective beam solid angles listed in Table \ref{tab:stat1}. 
The spatial variation is the 1$\sigma$ uncertainty associated with the band (quadruplets) averaged beams.

In Table \ref{tab:stat2}, we report the FWHM computed in a different way, by forming the averages of the FWHM evaluated evaluated from a Gaussian fit to the effective beam maps. 
The former is best used for flux determination, the latter for source identification.

\begin{table*}[ht]
\centering
\caption{Band averaged effective beam solid angles under a Gaussian approximation. $\Omega_{{\rm eff}}$ is the beam solid angle estimated up to a radius equal to the main beam radius.
FWHM$_{{\rm eff}}$ is the effective FWHM estimated from $\Omega_{{\rm eff}}$.  
$\Omega_{{\rm eff}}^{(1)}$  is the beam solid angle estimated up to a radius equal to the FWHM$_{{\rm eff}}$, while $\Omega_{{\rm eff}}^{(2)}$ indicates the beam solid angle estimated up to a radius = 2 $\times$ FWHM$_{{\rm eff}}$.}
\begin{tabular}{c r@{.}l r@{.}l r@{.}l r@{.}l r@{.}l r@{.}l r@{.}l r@{.}l}
\hline 
\hline
\noalign{\vskip 2pt}
\noalign{\vskip 2pt}
Band    & \multicolumn{2}{c}{$\Omega_{{{\rm eff}}}$} & \multicolumn{2}{c}{spatial variation} & 
          \multicolumn{2}{c}{$\Omega_{{{\rm eff}}}^{(1)}$} & \multicolumn{2}{c}{spatial variation} & 
          \multicolumn{2}{c}{$\Omega_{{{\rm eff}}}^{(2)}$} & \multicolumn{2}{c}{spatial variation} & 
          \multicolumn{2}{c}{FWHM$_{\rm eff}$} & \multicolumn{2}{c}{spatial variation} \\
           
        & \multicolumn{2}{c}{(arcmin$^{2}$)} & \multicolumn{2}{c}{(arcmin$^{2}$)} &
          \multicolumn{2}{c}{(arcmin$^{2}$)} & \multicolumn{2}{c}{(arcmin$^{2}$)} &
          \multicolumn{2}{c}{(arcmin$^{2}$)} & \multicolumn{2}{c}{(arcmin$^{2}$)} &
          \multicolumn{2}{c}{(arcmin)} & \multicolumn{2}{c}{(arcmin)} \\
\hline 
\noalign{\vskip 2pt}
30	    & 1190&06	&  \hskip 20pt 0&69 &	1117&3 & \hskip 20pt 1&8	& 1188&93	& \hskip 20pt 0&70	& 32&408 & \hskip 20pt 0&009 \\
44	    &  832&00 & 34&00$^{\dag}$ &  758&0 &	32&0$^{\dag}$  &	832&00	& 35&00$^{\dag}$	& 27&100	 &  0&570$^{\dag}$ \\                 
70	    &  200&90	&  0&99 &	 186&1 &   1&8	& 200&59	&  0&99	& 13&315 &  0&033 \\ 
18/23	&  210&13	&  0&63 &	 194&2 &   2&6	& 209&82	&  0&64	& 13&618 &  0&020 \\
19/22	&  199&19	&  0&64 &	 185&0 &   1&6	& 198&90	&  0&64	& 13&259 &  0&021 \\
20/21	&  192&58	&  0&67 &	 179&1 &   1&9	& 192&27	&  0&67	& 13&037 &  0&023 \\
25/26	& 1019&63	&  0&65 &	 942&2 &   2&4	& 1019&05	&  0&64	& 29&998 &  0&009 \\
\hline 
\label{tab:stat1}
\end{tabular} 
\begin{tablenotes}
\item $^{\dag}$ The large spatial variation associated with the 44 channel is due to the combination of beams with very different shapes and orientations, due to the different location of horn 24 with respect to horns 25 and 26 in the focal plane \citep{sandri2010}. 
Indeed, the value associated with the quadruplet 25/26 (the spatial variation of the 24 is about 0.78 arcmin$^2$) is in line with other quadruplets.
\end{tablenotes}
\end{table*}

\begin{table*}[ht]
\centering
\caption{Statistics of the {\tt FEBeCoP} effective beams computed with the \texttt{GRASP} scanning beams.}
\begin{tabular}{c r@{.}l r@{.}l r@{.}l r@{.}l c c}
\hline 
\hline
\noalign{\vskip 2pt}
\noalign{\vskip 2pt}
        & \multicolumn{4}{c}{FWHM} &  
          \multicolumn{4}{c}{Ellipticity} &  
          \multicolumn{2}{c}{$\psi$} \\
           
Band    & \multicolumn{2}{c}{mean} & \multicolumn{2}{c}{stdev} & \multicolumn{2}{c}{mean} & \multicolumn{2}{c}{stdev} & mean & stdev \\
        & \multicolumn{2}{c}{(arcmin)} & \multicolumn{2}{c}{(arcmin)} & \multicolumn{2}{c}{} & \multicolumn{2}{c}{} & (degree) & (degree) \\
\hline 
\noalign{\vskip 2pt}
30	    &  32&293 & 0&024 & 1&318 & 0&037 & 0 & 54 \\
44	    &  27&000 & 0&590 & 1&035 & 0&035 & 0 & 50 \\                 
70	    &  13&213 & 0&034 & 1&223 & 0&026 & 3 & 54 \\ 
18/23	&  13&525 & 0&021 & 1&188 & 0&021 & 3 & 54 \\
19/22	&  13&154 & 0&037 & 1&230 & 0&027 & 2 & 54 \\
20/21	&  12&910 & 0&037 & 1&256 & 0&036 & 3 & 54 \\
25/26	&  29&975 & 0&013 & 1&177 & 0&030 & --2 & 47\\
\hline 
\label{tab:stat2}
\end{tabular} 
\end{table*}

\section{Beam window function}
\label{window_function}
\subsection{LFI window functions based on FEBeCoP}
\label{febecop_wf}

\begin{figure*}[htpb]
\centering
\begin{tabular}{c c}
\includegraphics[width=9cm]{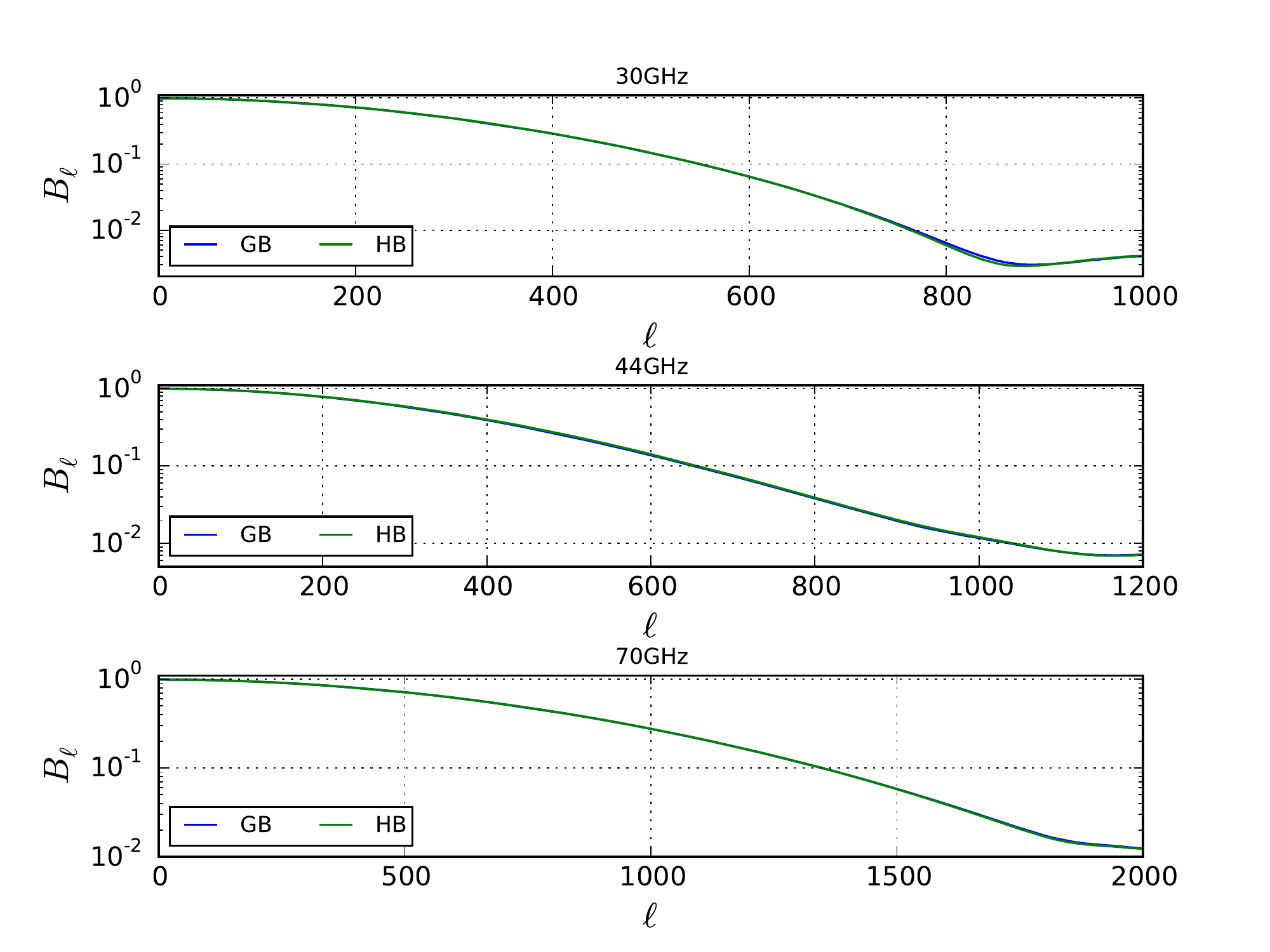} &
\includegraphics[width=9cm]{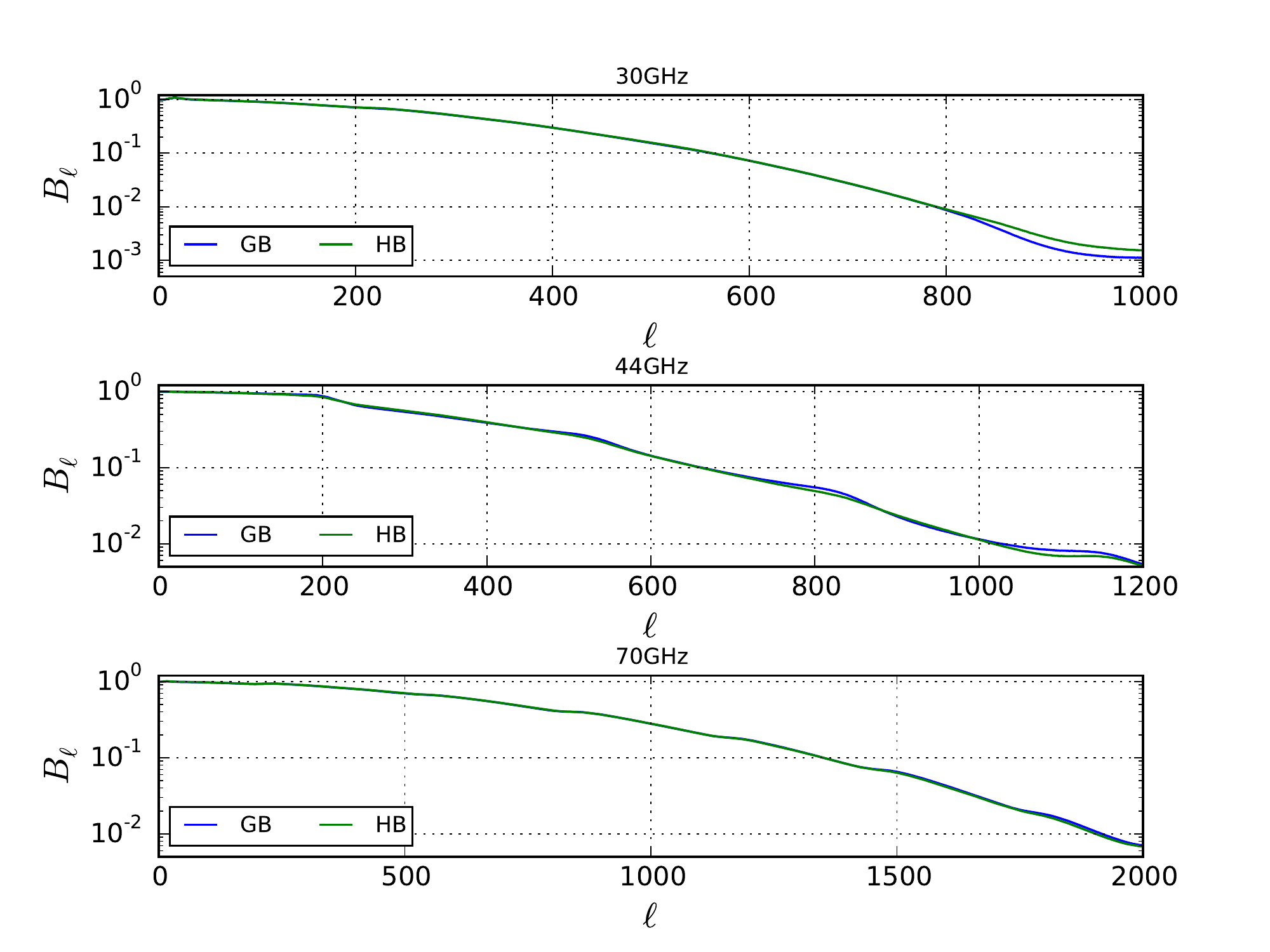}
\end{tabular}
\caption{{\tt FEBeCoP} beam window functions for \Planck\ 30, 44, and 70\,GHz frequency maps: temperature (left panels) and polarization (right panels) computed from \texttt{GRASP} beams (GB) and hybrid beams (HB).}
\label{fig:wf}
\end{figure*}

\begin{figure}[htpb]
\centering
\includegraphics[width=9.5cm]{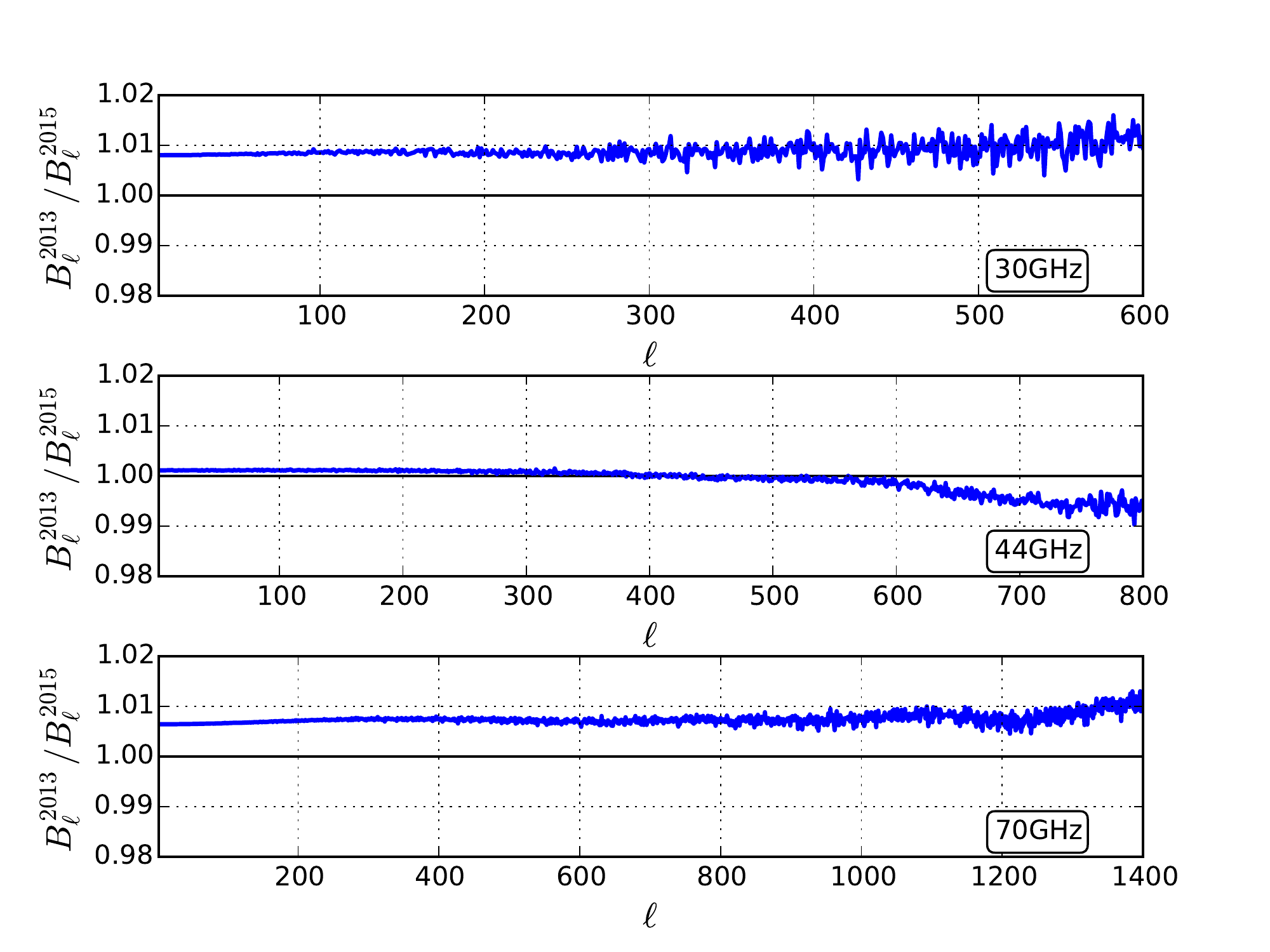} 
\caption{Comparison between the LFI window functions delivered in the previous release \citep{planck2013-p02d} and the current LFI window functions. 
The curves are slightly biased above unity due to the different normalization adopted in the two releases. In the 2013 release we assumed a full-power main beam whereas in the current release we are considering that not all the power falls into the main beam (see Sect.~\ref{normalization}).}
\label{fig:comp1}
\end{figure}

{\tt FEBeCoP} beam window functions have been computed as presented in \cite{planck2013-p02d}. 
In the current release we deliver both $TT$ and $EE$ window functions defined as
\begin{equation}
\label{eq:wlfullsky}
W^{TT,EE}_{\ell} = \langle {\tilde C}_{\ell }^{TT,EE} \rangle / C^{TT,EE}_{\ell} ,
\end{equation}
where the ensemble average is taken over the Monte Carlo (MC) simulations of the CMB observations, ${\tilde C}_{\ell }$ is the power spectrum of the CMB-only maps simulated by {\tt FEBeCoP} as described in \citep{mitra2010}, and $C_{\ell }$ is the fiducial model used as input. These are shown in Fig.~\ref{fig:wf} for 30, 44, and 70\,GHz frequency maps (temperature and polarization), using two different beam models (\texttt{GRASP} beams and hybrid beams).
Figure \ref{fig:comp1} shows the difference between the current window functions and the old ones, delivered in 2013.
The main difference is in the normalization, with the current window functions taking into account the power missed by the main beams, whereas the old ones were computed using full-power main beams.
Naturally, the actual pointing solution is different with respect to that used in the past release \citep{planck2014-a03}.

As done in 2013, we verified that for the Galactic mask used for power spectrum estimation \citep{planck2014-a03,planck2014-a13} the differences between full-sky and cut-sky window functions are marginal with respect to the error envelopes discussed in Sect.~\ref{error_propagation}, therefore the full-sky approximation has been used.

\begin{figure*}[htpb]
\centering
\begin{tabular}{c c}
\includegraphics[width=9cm]{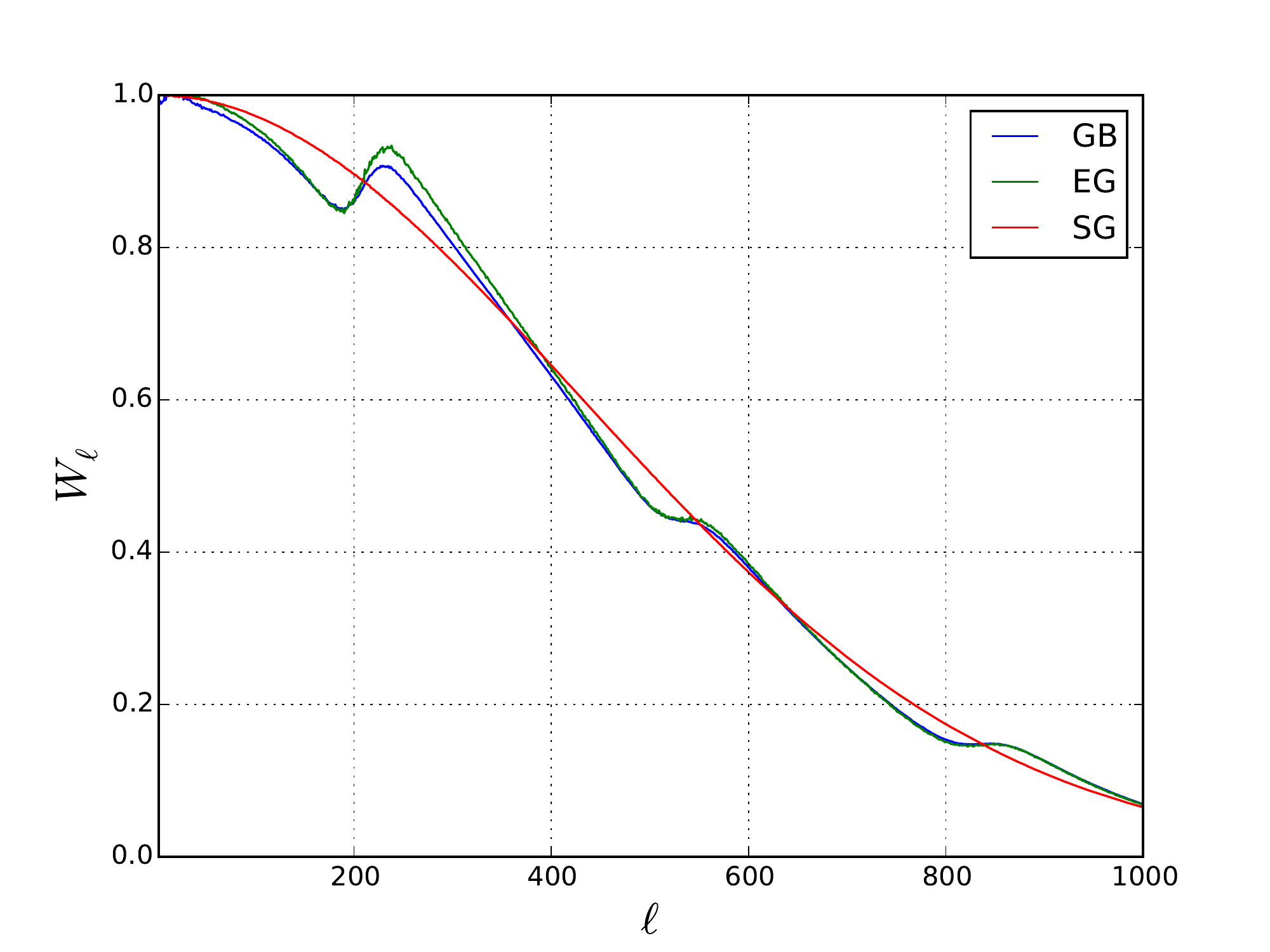} &
\includegraphics[width=9cm]{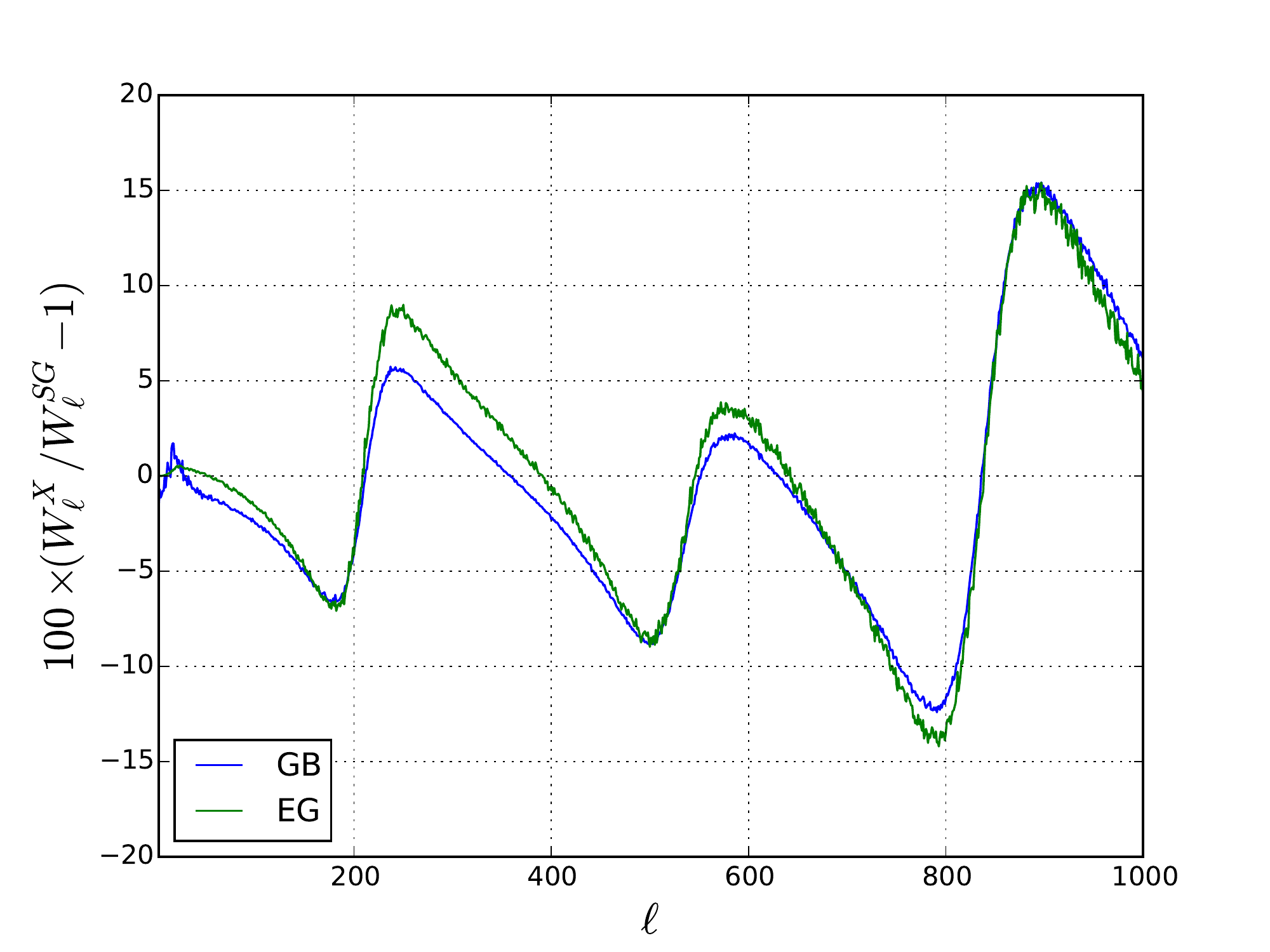}
\end{tabular}
\caption{$Left$: beam window functions computed using \texttt{GRASP} beams (blue curve), elliptical Gaussian beams (with the same descriptive parameters of the \texttt{GRASP} beams, green curve), and circular Gaussian beams (with the same FWHM of the \texttt{GRASP} beams, red curve). $Right$: per cent difference between the window function computed using \texttt{GRASP} beams and circular Gaussian beams (blue curve). The same ratio using elliptical Gaussian beams instead of \texttt{GRASP} beams is also shown (green curve).}
\label{fig:leakage}
\end{figure*}

The oscillations in the $EE$ window functions, located at values of ${\ell}$ corresponding to the $C_{\ell}^{TT}$ acoustic peaks, hint at the presence of temperature-to-polarization leakage, likely caused by the coupling of the scanning strategy with the particular shape of scanning beams.
To demonstrate this, we compare the window functions of the 18/23 quadruplet computed using a circular Gaussian, an elliptical Gaussian and a more realistic \texttt{GRASP} scanning beam. 
In Fig.~\ref{fig:leakage} the $EE$ window function for the 18/23 quadruplet (at 70\,GHz) is shown. 
It is noteworthy that for the circular Gaussian no oscillations are present, while the main contribution to the leakage (15\,\% at $\ell$ = 900) is due to the beam ellipticity.
The actual beam shape also has an effect (see the right panel of the Fig.~\ref{fig:leakage}, comparison between blue and green curves), but it is minor with respect to the ellipticity effect (blue curve).

\begin{figure*}[htpb]
\centering
\begin{tabular}{c c}
\includegraphics[width=9cm]{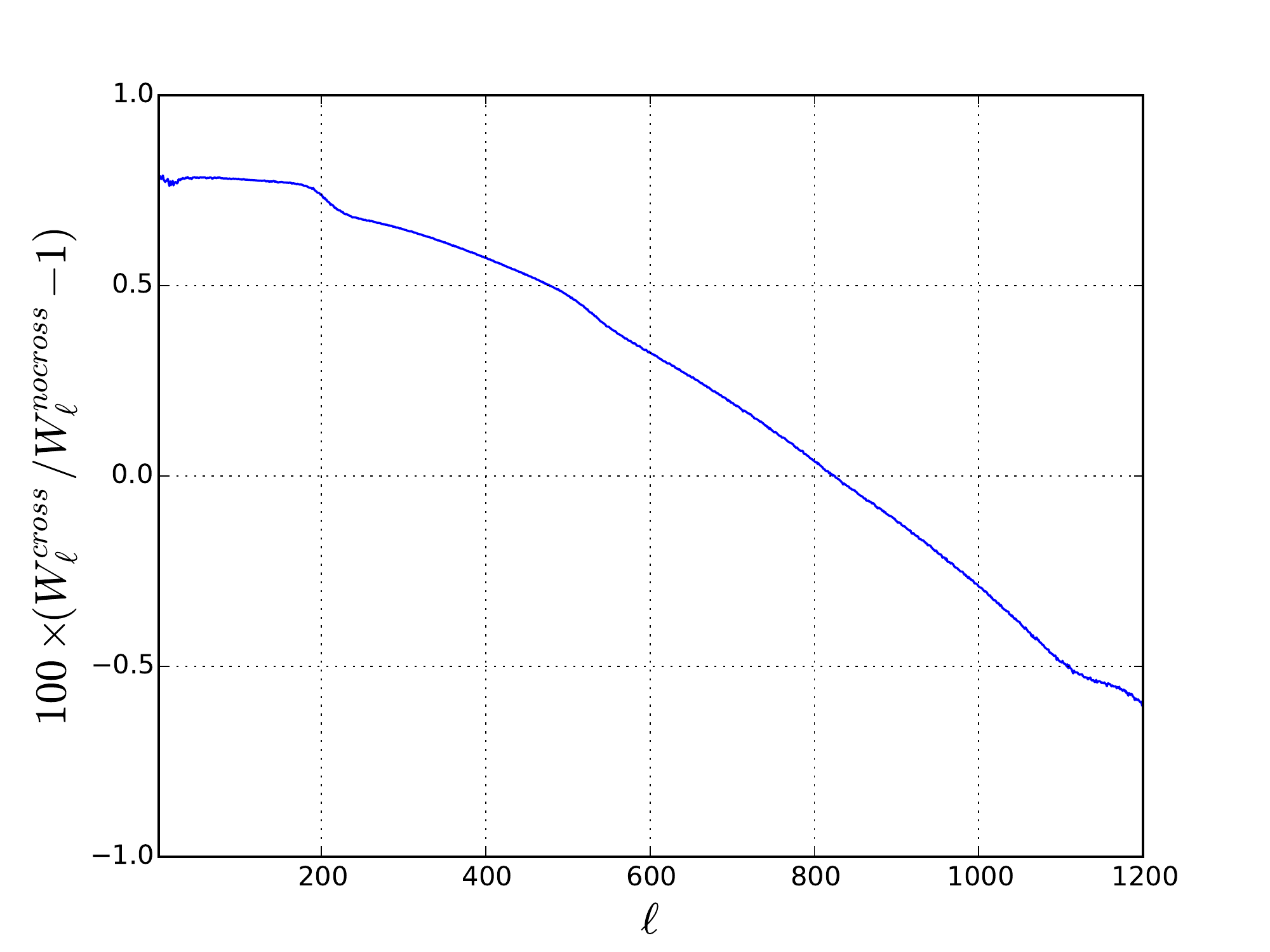} &
\includegraphics[width=9cm]{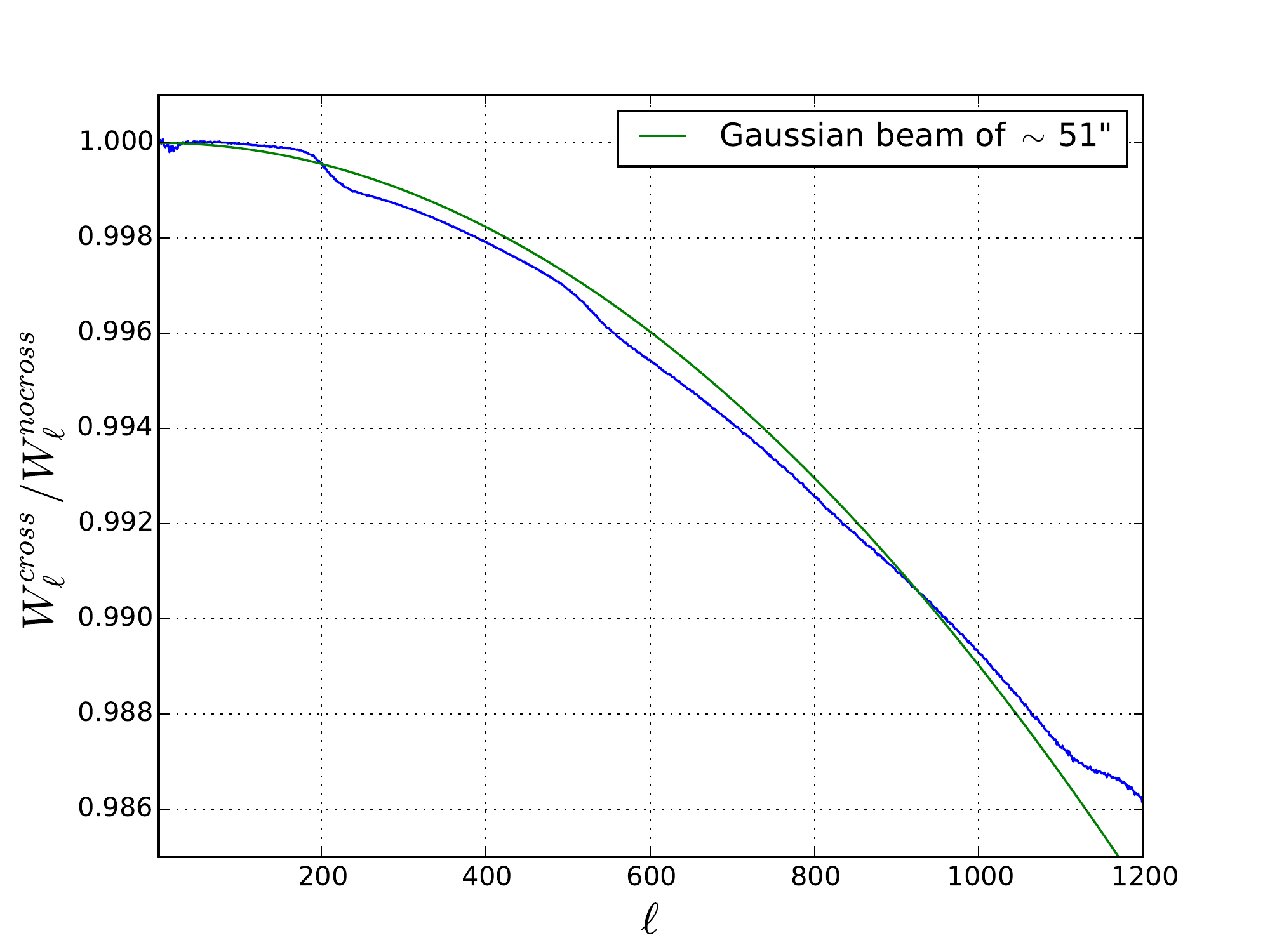} 
\end{tabular}
\caption{Per cent difference between the two $EE$ window functions obtained with \texttt{GRASP} beams with and without considering the cross-polar component (left panel). The ratio between the two $EE$ window functions forcing the normalization is also shown (right panel). The window function corresponding to a Gaussian beam with FWHM = 51 arcseconds is overplotted in green colour.}
\label{fig:crossnocross}
\end{figure*}

Regarding the beam cross-polarization, since the delivered window functions have been obtained from \texttt{GRASP} beams, where the cross-polarization is properly taken into account, no approximation is required.
Nevertheless we evaluated the effect of the beam cross-polarization by computing the window functions, including and not including the cross-polar beam component, as described in \citep{jones2007}. 
The results are presented in Fig.~\ref{fig:crossnocross}. 
The effect of including the cross-polar beam in the window function computation for 70~GHz is roughly 1\% at $\ell$ = 1000. 
Including the cross-polar term can approximately be described by an overall smoothing effect. 
A Gaussian beam of about 51 arcseconds accurately describes the extra smoothing effect up to $\ell$ = 1000 for 70~GHz, and deviates from the real effect for large multipoles, overpredicting the amount of smoothing.

\begin{figure}[htpb]
\centering
\includegraphics[width=9.5cm]{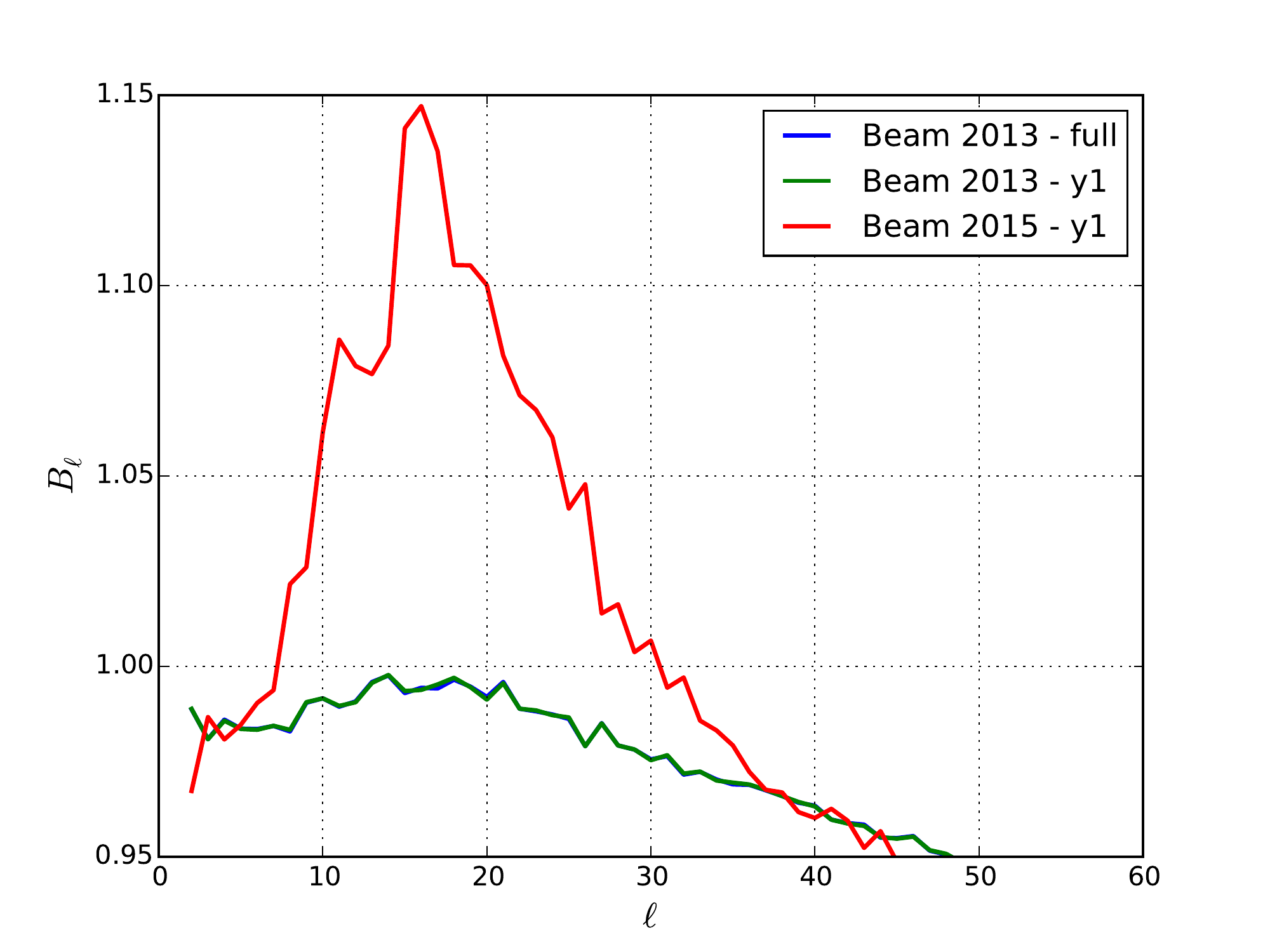} 
\caption{$EE$ beam window functions, $B_\ell$ computed from the 30\,GHz beams associated with the 2013 delivery (blue and green curves, which nearly overlap) and $B_\ell$ computed from the beams of the current release (red and light blue curves). The only difference here is in the beam efficiencies. In the legend, y1 indicates the first year.}
\label{fig:bump}
\end{figure}

Another interesting effect on the polarized window function is related to the different main beam efficiencies, mainly at 30~GHz.
This effect is shown in Fig.~\ref{fig:bump}.
From this figure it is evident that there is a bump at low multipoles with respect to the window function delivered in 2013.
This bump is not due to the different beam shape or the different pointing solution, but rather to the fact that the beams have, reasonably, different efficiencies due to mechanical issues.
 
Since the beam window functions are computed using CMB-only Monte Carlo simulations, the oscillations we see in the polarization $B_{\ell}$s only account for the leakage of the CMB signal itself and not for foreground-induced leakage.

\subsection{Simulated timeline-to-map Monte Carlo window functions}
\label{sec:timeline-to-map}

To see the effect of sidelobes and to provide a consistency check for the {\tt FEBeCoP} window functions, we also calculated the window functions via simulated timelines.
This is more suitable for including the sidelobes, although costly, limiting us to a small number of realizations and thus leaving a large simulation variance in the results.

Signal-only timeline-to-map Monte Carlo simulations were produced using Level-S \citep{reinecke2006} and {\tt HEALPix} \citep{gorski2005} subroutines and the {\tt Madam} map-maker \citep{kurki-suonio2009, keihanen2010} on the \emph{Sisu} supercomputer at the \emph{CSC-IT Center for Science} in Finland, as described in \cite{planck2013-p02d}. In the 2013 analysis, only the main beam was simulated; now we simulated all three parts of the beam, i.e., the main beam, near sidelobes, and far sidelobes.

We started from the simulated input CMB sky $a_{\ell m}$ realizations of the FFP8 CMB Monte Carlo simulation set \citep{planck2014-a14}. 
Given the high computational cost of the timelime-to-map simulation, we used only the first 50 realizations. These sky $a_{\ell m}$ were then convolved, using the Level-S code {\tt conviqt\_v4}, with the beam $a_{\ell m}$ (called here $b_{\ell m}$).  The $a_{\ell m}$ and $b_{\ell m}$ both have three components: $T$ for intensity and $E$, $B$ for polarization.  Here the $a^B_{\ell m}$ represent just the $B$-mode polarization due to gravitational lensing of the $E$-mode polarization, i.e., there was no primordial $B$-mode in these simulations, so they were much smaller than the $a^E_{\ell m}$.
In order to evaluate the sidelobe effect on the window function, and also for practical computational reasons, the three contributions corresponding to the three beam regions presented in Sect.~\ref{introduction} (i.e., main beam, near sidelobes, and far sidelobes) were considered separately. The CMB timelines for each realization and beam component were produced using {\tt multimod}, according to the detector pointing for each radiometer. 

Because of the very different extent of the different beam parts, different Level-S parameters were used for each (see Table~\ref{tab:LevelS}). 
The significance of these parameters is that only multipoles $\ell$ up to {\tt conv\_lmax} are modelled, but the accuracy falls off near {\tt conv\_lmax} and can be improved by increasing {\tt lmax\_out} and {\tt interpol\_order}. The parameter {\tt beammmax} controls how accurately the azimuthal structure of the beam is modelled. Increasing the values of these parameters increases the computational cost.

\begin{table*}[ht]
\centering
\caption{Parameters for the Level-S codes {\tt conviqt\_v4} and {\tt multimod} for the different beam parts.}
\begin{tabular}{l c c c c c c}
\hline 
\hline
\noalign{\vskip 2pt}
\noalign{\vskip 2pt}
        & \multicolumn{2}{c}{Main beam} &  
          \multicolumn{2}{c}{Near sidelobes} &  
          \multicolumn{2}{c}{Far sidelobes} \\
           
parameter    & 30\&44 GHz & 70 GHz & 30\&44 GHz & 70 GHz & 30\&44 GHz & 70 GHz \\
        \hline 
\noalign{\vskip 2pt}
{\tt conv\_lmax}	    &  2048 & 2048 & 1000 & 1500 & 180 & 180 \\
{\tt lmax\_out}	    &  4096 & 4096 & 2000 & 3000 & 360 & 360 \\
{\tt beammmax}	    &  9 & 9 & 18 & 18 & 180 & 180 \\
{\tt interpol\_order}	    &  5 & 9 & 5 & 5 & 5 & 5 \\
\hline 
\label{tab:LevelS}
\end{tabular} 
\end{table*}

Maps were then made with {\tt Madam}, separately from just the main beam timelines, from the sum of the main beam and near sidelobe timelines, and the sum of all three beam component timelines, using the same {\tt Madam} parameter settings as were used for the flight maps \citep{planck2014-a07}.
In this way, we produced 30 GHz, 44 GHz, and 70 GHz frequency maps ({\tt HEALPix} resolution $N_{\rm side}$ = 1024) and the quadruplet maps for 44~GHz 25/26 and 70 GHz 18/23, 19/22, and 20/21, for the 4-year full mission LFI survey. The angular power spectra $C_\ell$ were then calculated with {\tt anafast} (from full-sky maps).

We calculated the scalar beam window function $B_\ell$ as 
\begin{equation}
	B_\ell = \frac{1}{B^\mathrm{pix}_\ell}\sqrt{\frac{\langle C^{TT}_\ell(\mbox{out})\rangle}{\langle C^{TT}_\ell(\mbox{sky})\rangle}} \,,
\label{eq:timelinewf}
\end{equation}
where $C^{TT}_\ell(\mbox{sky})$ is the temperature angular power spectrum of the input $a_{\ell m}$ of the simulation, 
$C^{TT}_\ell(\mbox{out})$ is the temperature angular power spectrum of the map produced by the simulation pipeline, and $\langle\cdot
\rangle$ represents the mean over the first 49 realizations (the 50th realization was used as a test case for applying the window 
function). The quantity $B^\mathrm{pix}_\ell$ is the {\tt HEALPix} $N_{\rm side}$ 1024 pixel window function, which we divided out in order not to include the pixel window that comes from using pixelized output maps. The proper definition of the beam window function would refer to the model $C_\ell$ used to produce the different $a_{\ell m}$ realizations (see Eq.~\ref{eq:wlfullsky}), instead of the mean of the input realizations $\langle C^{TT}_\ell(\mbox{sky})\rangle$, but we do not have enough statistics for this formulation, and instead we use Eq.~(\ref{eq:timelinewf}) to reduce the simulation variance in the obtained window function. 

No filter function appears in Eq.~(\ref{eq:timelinewf}) since the LFI analysis uses no filtering (see \cite{poutanen2004} for destriping and filter functions).
We actually obtain two output maps from {\tt Madam}, the binned output map and the destriped output map. Destriping is a process that aims to remove correlated noise. It happens in the time domain and its effect does not properly belong to the beam window function, so we have used the binned output maps for calculating the window functions. For the main beam and near sidelobe timelines destriping has a rather small effect for our noiseless simulation. There is a small noiselike contribution due to pixelization noise \citep{kurki-suonio2009}. However, for the far sidelobe timelines, a given location of the sky appears completely different with different beam orientations, and the {\tt Madam} destriper interprets this difference as due to noise and tries to remove it. Therefore the contribution of far sidelobes to the destriped maps is very different from their contribution to the binned maps. To show this effect we have also calculated the beam window functions for the full beam using the destriped maps.

To summarize, for each frequency channel and horn pair, the beam window functions have been computed for:

\begin{itemize}
\item[(a)] just the main beam; 
\item[(b)] main beam + near sidelobes;  
\item[(c)] main beam + near sidelobes + far sidelobes.  
\end{itemize}
For the last case we calculated both a ``binned'' and ``destriped'' window function.

\begin{figure*}[htpb]
\centering
\begin{tabular}{c c}
\includegraphics[width=9cm]{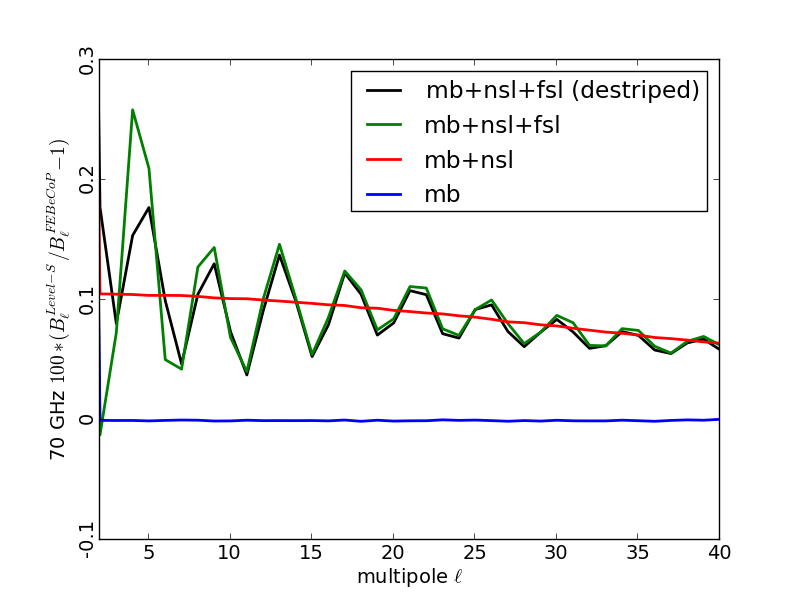} &
\includegraphics[width=9cm]{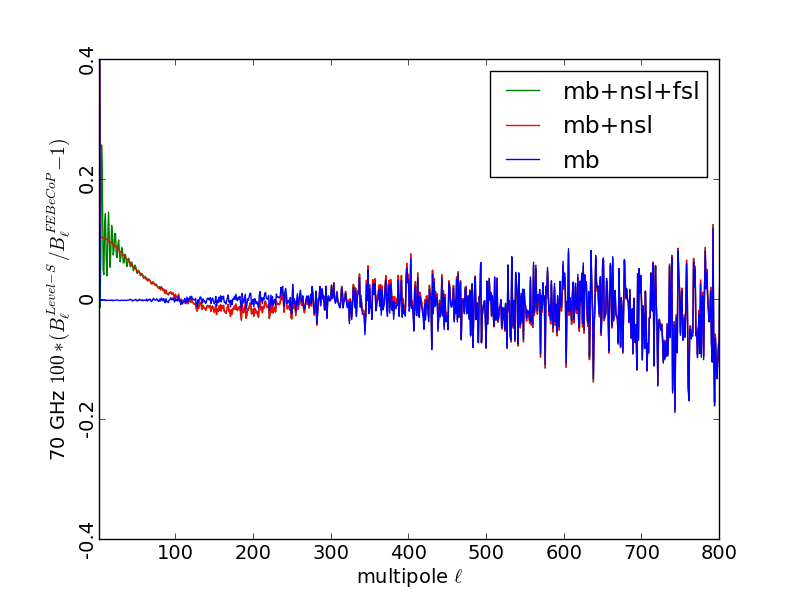}
\end{tabular}
\caption{Per cent difference between the window function obtained using Level-S (see Sect.~\ref{sec:timeline-to-map}) and the {\tt FEBeCoP} window function (see Sect.~\ref{febecop_wf}), at 70 GHz. The left panel is an enlargement of the right panel, concentrating on the low multipoles. The contribution from the main beam (mb), near sidelobes (nsl), and far sidelobes (fsl) is shown. The agreement between the Level-S window function computed using the main beams and the {\tt FEBeCoP} window function is evident, as presented also in \cite{planck2013-p02d}. The effect, $0.1 - 0.2 \%$, of near and far sidelobes is clearly visible at low $\ell$. At high $\ell$ the difference is mainly simulation variance, due to the small number (49) of CMB realizations.}
\label{fig:sleffect}
\end{figure*}

A comparison of the resulting window functions to the {\tt FEBeCoP} window functions is shown in Fig.~\ref{fig:sleffect} for the 70 GHz channel.
The sidelobe impact on the low multipoles for the 30 GHz channels is of nearly the same magnitude, whereas it is much lower at 44 GHz, since the main beam efficiency is higher. 

The increment at the quadrupole of the window functions computed considering the near sidelobes reflects the efficiencies  in Table \ref{mbe}, i.e., about 0.1\%.
Since the far sidelobes are very wide structures that are strongest in a direction almost orthogonal to the line of sight, they add power incoherently to the signal entering the main beam at scales of $\ell$ = 2 or higher multipoles.

This timeline-to-map Monte-Carlo approach is quite resource intensive, and since the timelines are communicated from Level-S to {\tt Madam} by writing them on disk from where {\tt Madam} reads them, there is an I/O bottleneck that limits massive parallelization of the simulations. The {\tt FEBeCoP} algorithm is much faster, hence it allows for a significantly larger number of simulations, resulting in a more accurate estimation of the window functions. Since {\tt FEBeCoP} cannot handle the sidelobes, the sidelobe effect is included in the error budget, as done in the previous release.

\subsection{Matrix window functions}
\label{matrix_window_functions}

The scalar window functions of the previous subsections depend on the assumed CMB angular power spectra, in addition to the instrument beam and scanning, because they contain contributions from the leakage between the temperature and polarization signals. 
This gives a large contribution for the $EE$ window function particularly. 
This is because the $EE$ window function is obtained from the ratio between the (simulated) output $EE$ spectrum and the input $EE$ spectrum; however, because of $T$ to $E$ leakage, the output $EE$ spectrum also depends on the input $TT$ spectrum.
We can isolate the leakage effect by introducing the matrix beam window function.

Assume that the spherical harmonic coefficients $\tilde{a}^X_{\ell m}$ of the output map are related to those of the sky, $a^X_{\ell m}$, by
\begin{equation}
   \tilde{a}^X_{\ell m} = \sum_{m'X'}K_{\ell mm'}^{XX'}a^{X'}_{\ell m'} \,,
\end{equation}
where $X = T, E, B$.  For the expectation value of the angular power spectrum of the map we then get
\begin{equation}
   \langle\tilde{C}^{XY}_\ell\rangle = \sum_{X'Y'}W^{XY,X'Y'}_\ell C^{X'Y'}_\ell \,,
\end{equation}
where
\begin{equation}
   W^{XY,X'Y'}_\ell \equiv \frac{1}{2\ell+1}\sum_{mm'}K_{\ell mm'}^{XX'}K_{\ell mm'}^{YY'\ast}
\end{equation}
and 
\begin{equation}
   C^{X'Y'}_\ell = \langle a^{X'}_{\ell m'}a^{Y'}_{\ell m'} \rangle
\end{equation}
is the expectation value of the angular power spectrum of the sky. 

Assuming the only effect is that of the instrument beams, the $W^{XY,X'Y'}_\ell$ is the matrix beam window function (as formulated here, it also includes the pixel window). It should be a combined property of the beams and the scanning strategy that determines the beam orientations at different times, and independent of the angular power spectrum.  We evaluate it by timeline-to-map simulations where the input skies are realizations of constant (as a function of $\ell$) ${\cal D}^{XY}_\ell \equiv \ell(\ell+1)C^{XY}_\ell/2\pi$. (Using realistic ${\cal D}^{XY}_\ell$ here would result in poor accuracy due to some ${\cal D}^{XY}_\ell$ being very small, especially near where the cross-correlations change sign.) We used ${\cal D}^{TT}_\ell = {\cal D}^{EE}_\ell = {\cal D}^{BB}_\ell = 1000 \,\mu\mathrm{K}^2$ and ${\cal D}^{TE}_\ell = {\cal D}^{TB}_\ell = {\cal D}^{EB}_\ell = 500\,\mu\mathrm{K}^2$ to produce a set of 25 realizations of $a^T_{\ell m}$, $a^E_{\ell m}$, $a^B_{\ell m}$.  

To evaluate the individual matrix elements, we need three separate simulations where the input sky contains only $T$, only $E$, and only $B$. We carried out such timeline-to-map simulations using the prescription of Sect.~\ref{sec:timeline-to-map}, but always setting two of the input $a^T_{\ell m}$, $a^E_{\ell m}$, $a^B_{\ell m}$ to zero. 

From the output maps of these simulations we get directly those matrix elements that represent leakage from $C^{TT}_\ell$, $C^{EE}_\ell$ and $C^{BB}_\ell$ as, e.g., 
\begin{equation}
   W^{XY,TT}_\ell = \frac{\langle\tilde{C}^{XY}_\ell(T\mathrm{-only})\rangle}
   {\langle C^{TT}_\ell\rangle} \,.
\end{equation}
We note that ``$T$-only'' represents here the output map obtained from a $T$-only input sky, not that the output map itself would have only $T$ -- the presence of $E$ and $B$ in this output map is precisely due to the leakage we want to measure.

To get those matrix elements that represent the additional effect of the correlations, e.g., $C^{TE}_\ell$, we also need output maps where both $T$ and $E$ were present in the input sky. Because of the linearity of the map-making process we obtain these directly by taking the sum of the ``$T$-only" and ``$E$-only'' maps.  We then get
\begin{equation}
   W^{XY,TE}_\ell = \frac{\langle \tilde{C}^{XY}_\ell(TE) - \tilde{C}^{XY}_\ell(T\mathrm{-only}) - \tilde{C}^{XY}_\ell(E\mathrm{-only})\rangle}
   {\langle C^{TE}_\ell\rangle} \,.
\end{equation}



The evaluation of the matrix beam window functions is three times as costly as evaluating the scalar window functions. 
Therefore we only used 25 realizations to estimate them. This took 500\,000 core-hours, the sidelobes being more costly than the main beam. Of this {\tt multimod} used about 60\,\% and {\tt Madam} about 40\,\%; all other steps, including {\tt conviqt\_v4} and {\tt anafast}, took less than 1\,\% taken together. (The 50 realizations for the scalar window functions took 300\,000 core-hours.)
The simulations were run on the CSC \emph{Sisu} Cray-XC30 (Intel Haswell 2.6 GHz) computer, using 1728 cores (72 nodes), which allowed running four 70\,GHz {\tt Madam} map making tasks simultaneously. 
A larger number of simultaneous map making tasks would have led to I/O congestion. Therefore these simulations took several weeks to run.  

\begin{figure*}[htpb]
\centering
\begin{tabular}{c c}
\includegraphics[width=9cm]{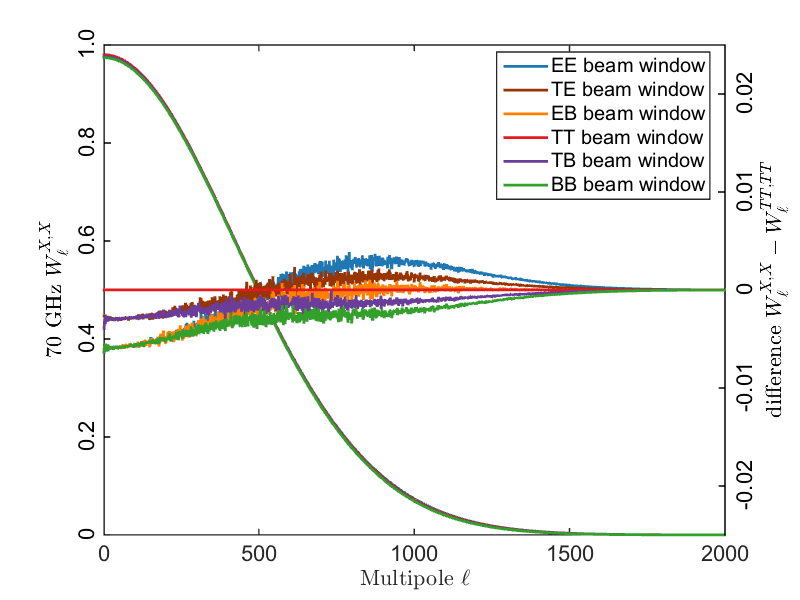} &
\includegraphics[width=9cm]{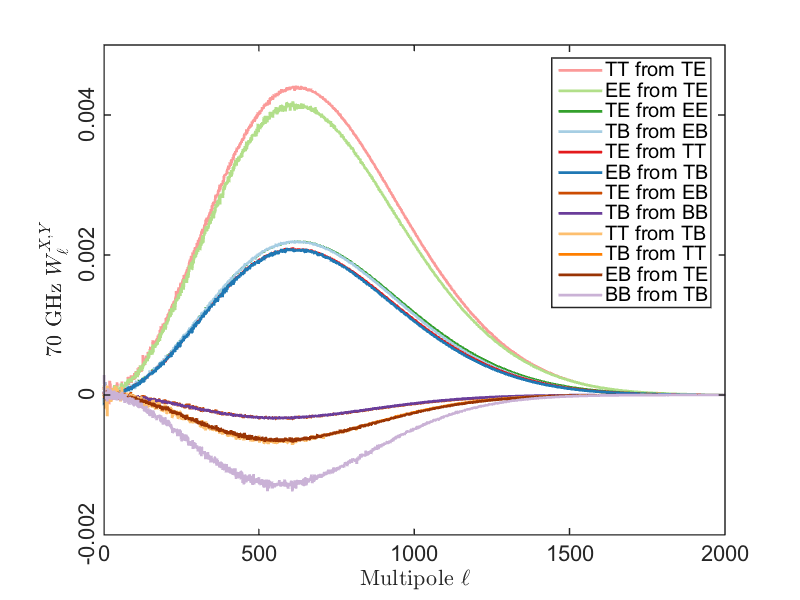} \cr
\includegraphics[width=9cm]{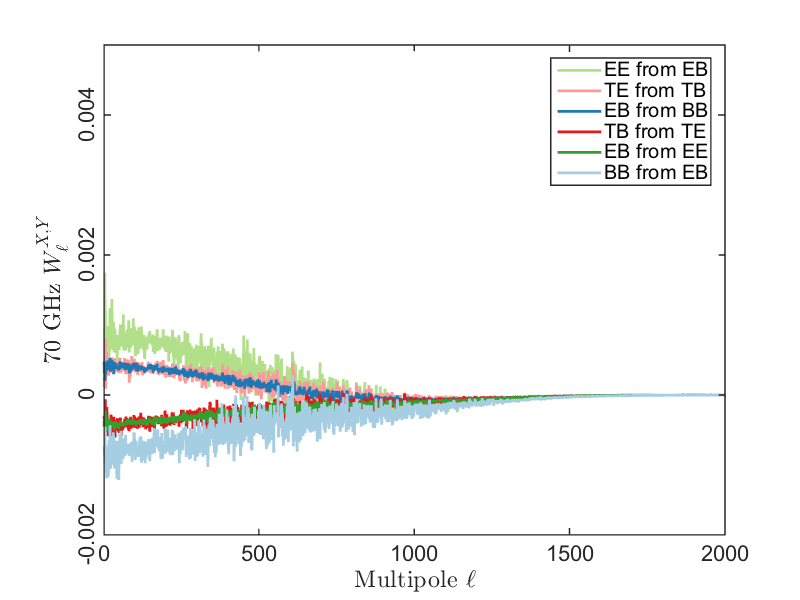} &
\includegraphics[width=9cm]{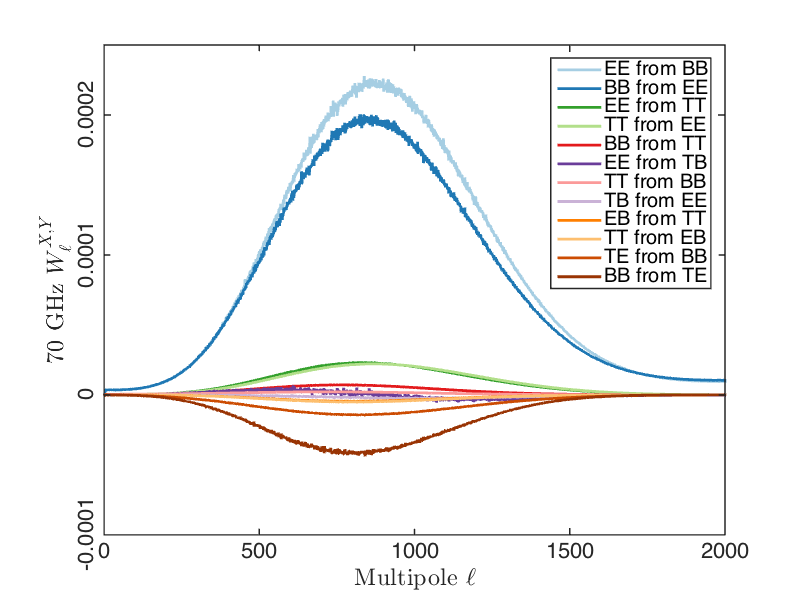} 
\end{tabular}
\caption{Diagonal (top left panel) and off-diagonal components of the 70 GHz main beam matrix window function. ``X from Y'' stands for $W^{X,Y}_\ell$ and ``X beam window" for the diagonal components $W^{X,X}_\ell$. To bring out the difference between the six diagonal components, we also show the difference $W^{X,X}_\ell-W^{TT,TT}_\ell$ (the near-horizontal lines in the top left panel, with the scale indicated at the right side of the panel). The 30 off-diagonal components are divided into three panels.  The bottom left shows the six components that are largest at low $\ell$. The two right panels shows those components that peak at intermediate $\ell$.  These were further divided according to whether the leakage involves the same components (of $T$, $E$, and $B$) in the `from' and `to' sides (top right) or not (bottom right) -- the latter tend to be smaller (we note the difference in scale). There are too many curves for all of them to be clearly visible, so to indicate where each of them lies, the legend lists the components in approximately the same order as they appear in the plot. In the top right panel $TE,EE$ lies under $TB,EB$; $TE,TT$ under $EB,TB$; $TE,EB$ under $TB,BB$; and both $TT,TB$ and $TB,TT$ under $EB,TE$. In the bottom right panel $EE,TT$ lies under $TT,EE$ and $EB,TT$ under $TT,EB$.}
\label{fig:matrix_window}
\end{figure*}

We show all 36 components of the 70 GHz main beam matrix window function in Fig.~\ref{fig:matrix_window}.  
The relative effect of the near sidelobes on the diagonal components is less than or (at the lowest multipoles) roughly equal to $\pm 2\times10^{-3}$. The relative effect of the near sidelobes to the off-diagonal components is largest where the off-diagonal components are small; the absolute effect is less than or (at the lowest multipoles) roughly equal $\pm 6\times10^{-5}$. 

\begin{figure*}[htpb]
\centering
\begin{tabular}{ccc}
\includegraphics[width=6cm]{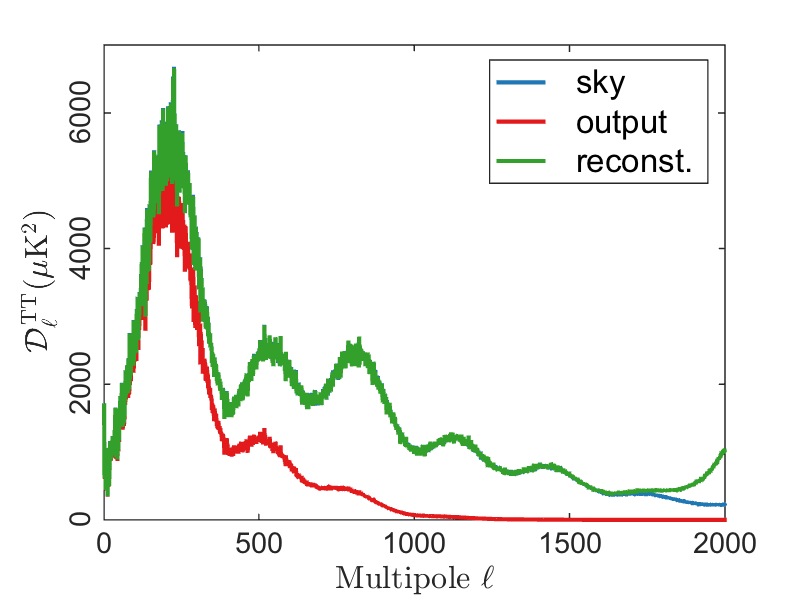}
\includegraphics[width=6cm]{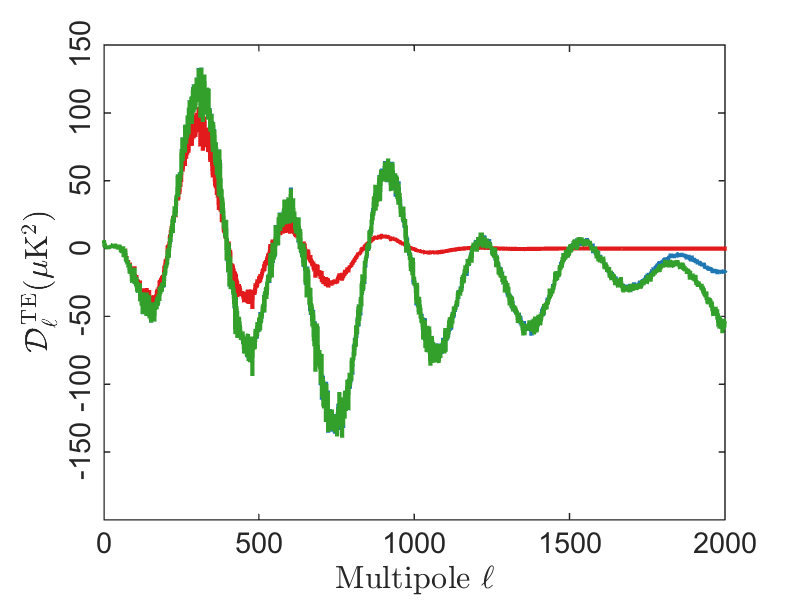}
\includegraphics[width=6cm]{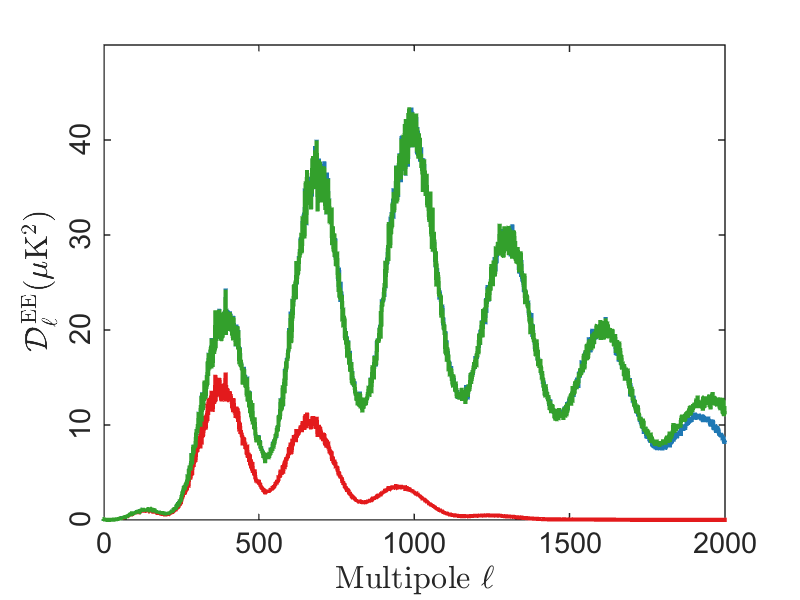} 
\end{tabular}
\caption{Reconstruction of the sky angular power spectrum using the matrix beam window function.  The blue curves show the 70 GHz input ``sky'' angular power spectrum of the 50th CMB realization.  The red curves show the angular power spectrum of the corresponding output map. The green curves show the result of applying the inverse of the matrix beam window function to it. We note that for $\ell \lesssim 1700$ the blue curve is not visible, since it lies under the green curve, showing that the reconstruction was successful. The left panel shows the $TT$ spectrum, the middle panel the $TE$ spectrum, and the right panel the $EE$ spectrum.}
\label{fig:matrix_window_applied}
\end{figure*}

In Fig.~\ref{fig:matrix_window_applied} we apply the obtained inverse window function to the output $C_\ell$ of the 50th realization of our CMB simulation (Sect.~\ref{sec:timeline-to-map}) to reconstruct the input $C_\ell$. We see that the reconstruction works for $C^{TT}_\ell$, $C^{TE}_\ell$, and $C^{EE}_\ell$ to the accuracy of simulation variance, except at the highest multipoles, where the window function is very small and not calculated as accurately as for the lower multipoles. 
For 30 GHz and 44 GHz the performance is similar, except that the accuracy falls at lower $\ell$ reflecting the wider beams.

The matrix window function approach presented in this section is work in progress; it was not yet mature enough to be used later in the analysis due to the lack of further testing in the pipeline. 
We are working on this approach to consolidate it for future data releases.

\section{Error budget}
\label{error_propagation}
The propagation of the uncertainties in the beam knowledge to the window function has been evaluated using the simulated beams derived from the MC pipeline on the \Planck\ optics performed for the last release \citep{planck2013-p02d}.  
Of course, the selected sample is smaller because the uncertainties in the main beam parameters are smaller (see Table \ref{tab:imo}) with respect to those presented in the 2013 paper.
Since the difference between the window functions obtained with {\tt FEBeCoP} and those obtained with the simple harmonic transformation is very small (less than the error on the window function calculated in 2013), it was decided to calculate the error budget using the harmonic transform approach instead of {\tt FEBeCoP} because it is much faster. This assumption is conservative (the errors calculated in this way are slightly higher than those calculated with {\tt FEBeCoP}).

Using the set of simulated beam window functions, we have built the covariance matrix $\tens{C}$ in $\ell$-space computing
\begin{equation}
{C}_{\ell\ell'} = \left\langle (W_{\ell}- \left\langle W_{\ell} \right\rangle)(W_{\ell'}-\left\langle W_{\ell'} \right\rangle ) \right\rangle ,
\end{equation}
where the 65 simulations are averaged.
Then we have decomposed the covariance matrix into eigenvalues ($\Lambda_k$) and eigenvectors (${ V}_k$). 
The error content is substantially encompassed in the first two eigenvalues, which account for the cutoff radius and main beam uncertainties, respectively.

\begin{figure}[htpb]
\centering
\includegraphics[width=9cm]{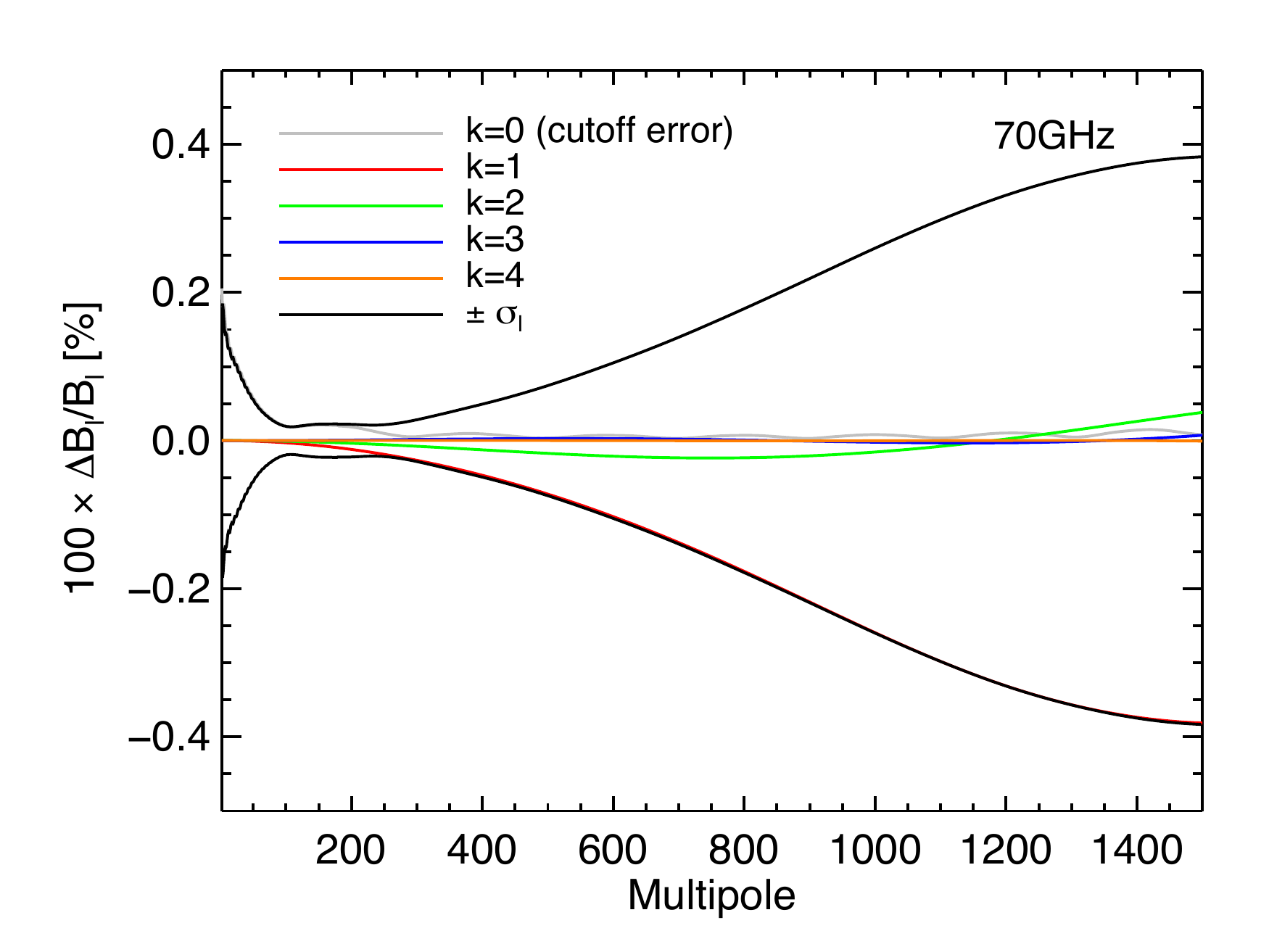} 
\caption{Eigenmodes of the covariance matrix of the 70\,GHz channel.}
\label{fig:ploteigen70}
\end{figure}

\begin{figure}[htpb]
\centering
\includegraphics[width=9cm]{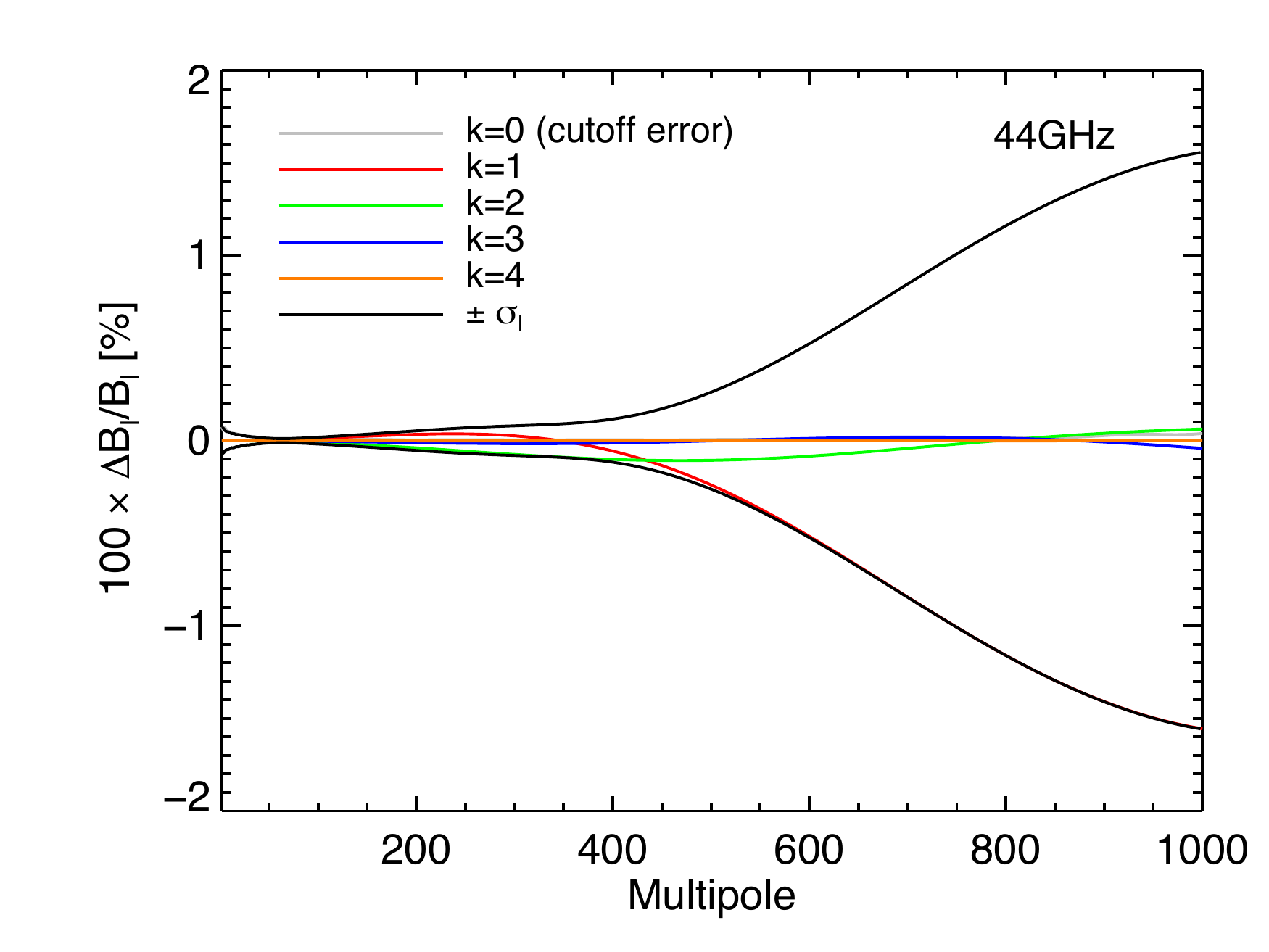} 
\caption{Eigenmodes of the covariance matrix of the 44\,GHz channel.}
\label{fig:ploteigen44}
\end{figure}

\begin{figure}[htpb]
\centering
\includegraphics[width=9cm]{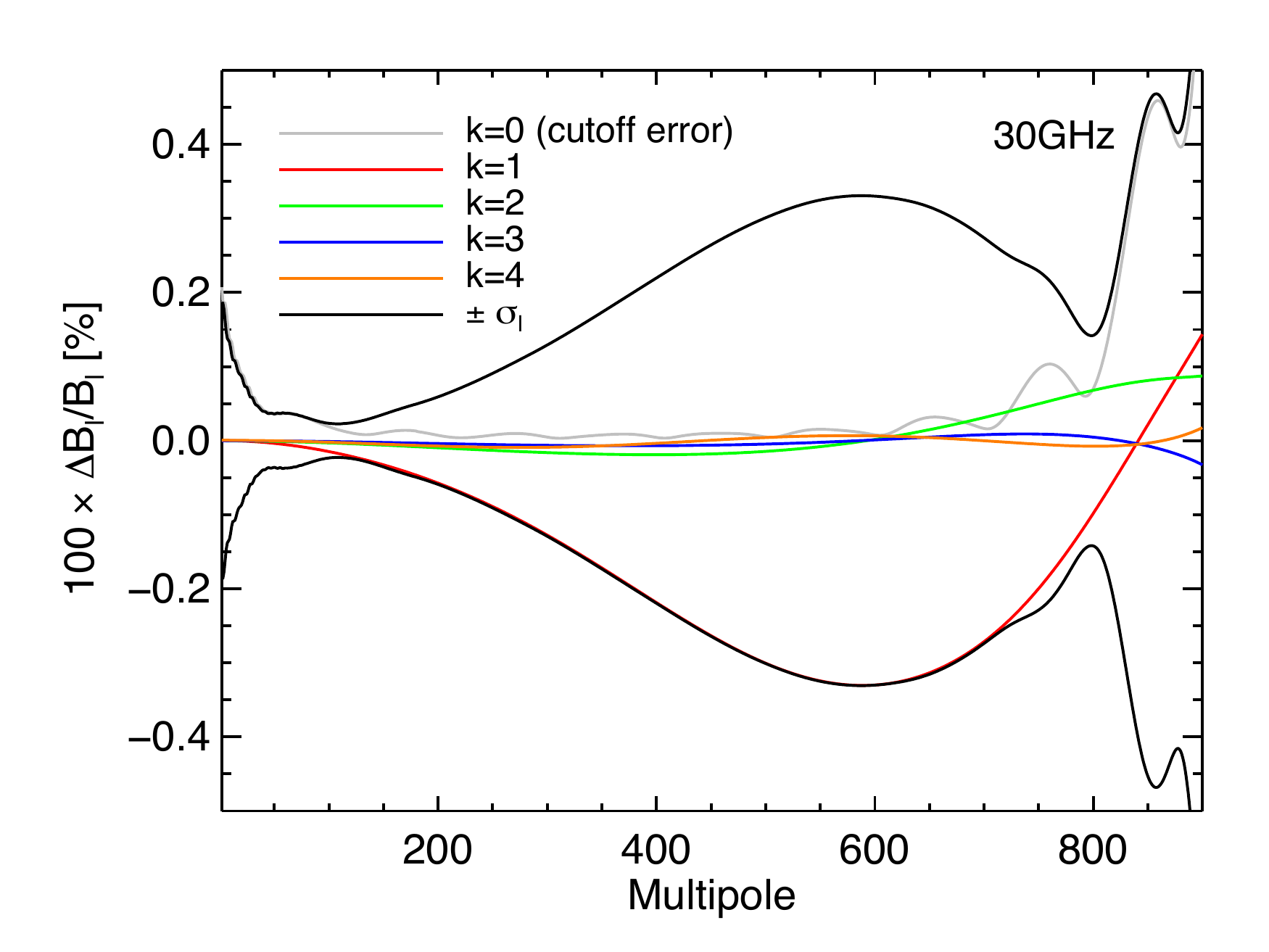} 
\caption{Eigenmodes of the covariance matrix of the 30\,GHz channel.}
\label{fig:ploteigen30}
\end{figure}

The {\tt FEBeCoP} window functions are computed using only the main beam. 
In Sect.~\ref{sec:timeline-to-map} we evaluated the impact on the beam window functions of neglecting near and far sidelobes. 
To evaluate the total error budget, we added this term as the first eigenmode in error decomposition described above and we show the total error budget in Figs.~\ref{fig:ploteigen70}, \ref{fig:ploteigen44}, and \ref{fig:ploteigen30} for the 70, 44, and 30\,GHz, respectively.
The grey line (eigenvector $k=0$) represents the cutoff radius term. 
The widening of the error at low $\ell$ accounts for the uncertainty introduced neglecting the near and far sidelobe contribution. 
Since for this release the new window functions are not normalized, the errors themselves are not normalized to zero. 

Whereas the main beam shape has been verified via the Jupiter observations, we have no direct measurement of the near and far sidelobes.
The LFI sidelobes have been computed using {\tt GRASP} and taking into account the nominal radiometer bandshapes.
The impact on sidelobes of the uncertainty in the knowledge of the radiometer bandshape is under investigation and will be introduced in the next release.
  
\begin{figure}[htpb]
\centering
\includegraphics[width=9cm]{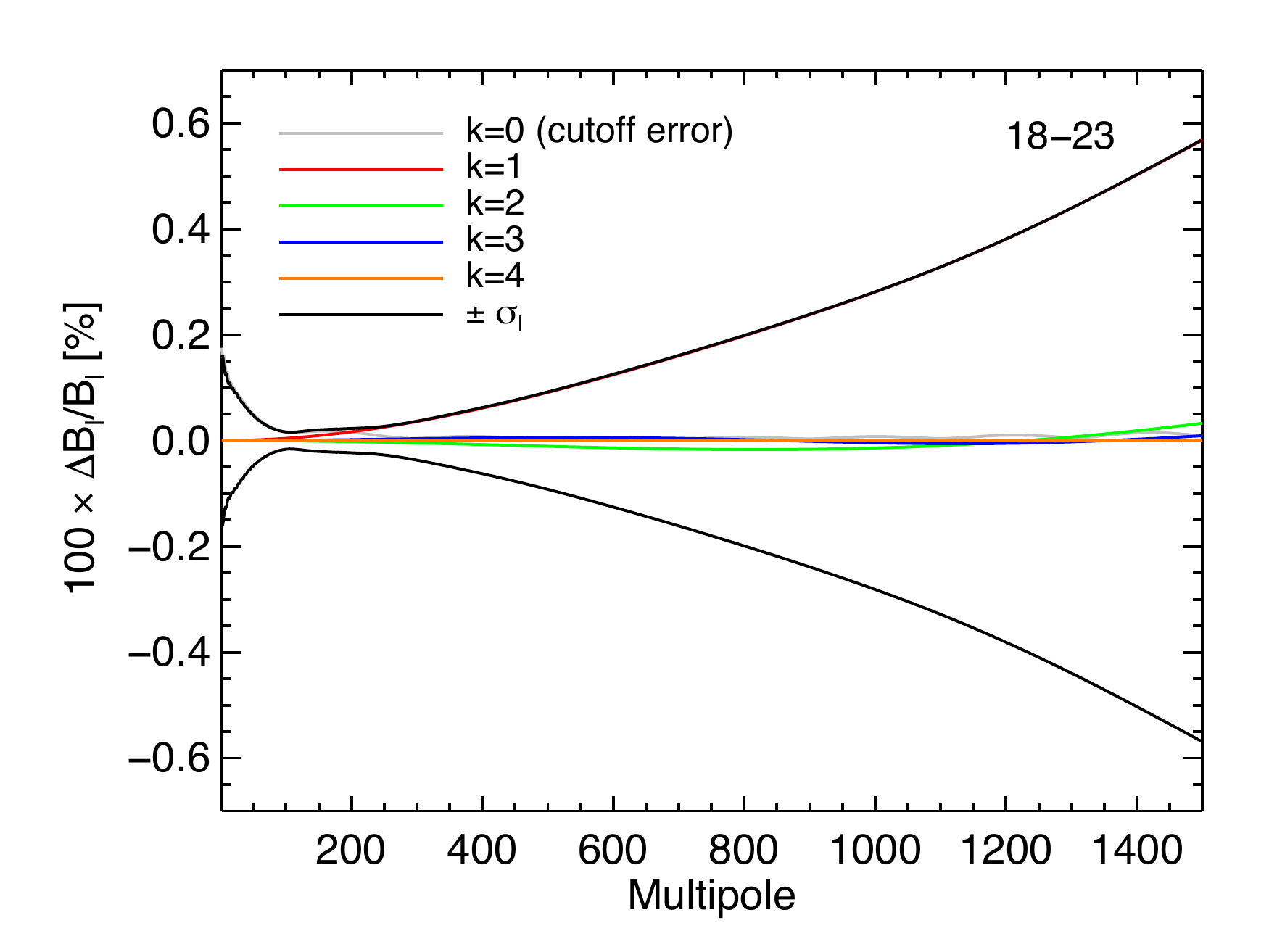} 
\caption{Eigenmodes of the covariance matrix of the quadruplet 18/23 at 70\,GHz.}
\label{fig:ploteigen1823}
\end{figure}

\begin{figure}[htpb]
\centering
\includegraphics[width=9cm]{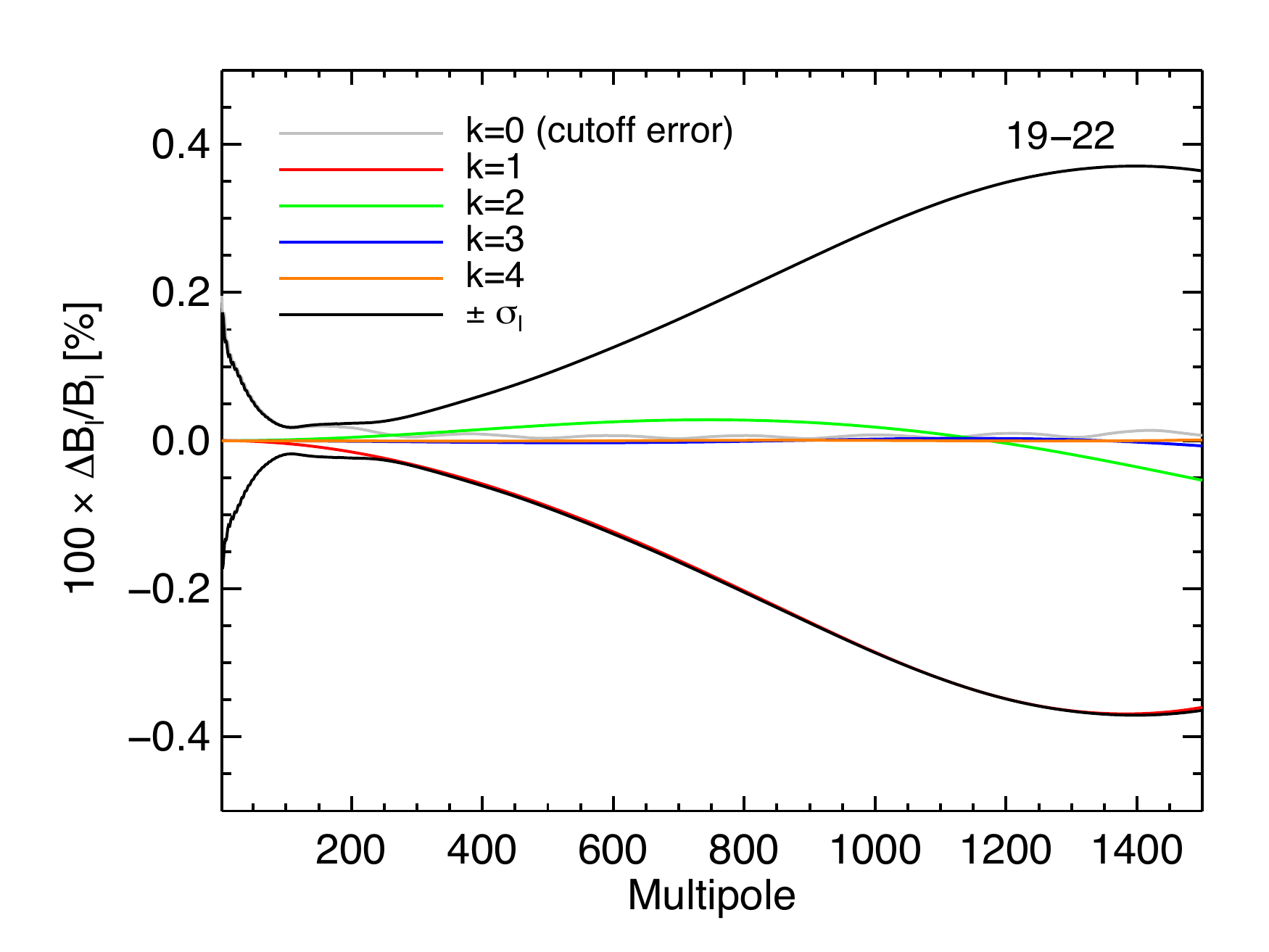} 
\caption{Eigenmodes of the covariance matrix of the quadruplet 19/22 at 70\,GHz.}
\label{fig:ploteigen1922}
\end{figure}

\begin{figure}[htpb]
\centering
\includegraphics[width=9cm]{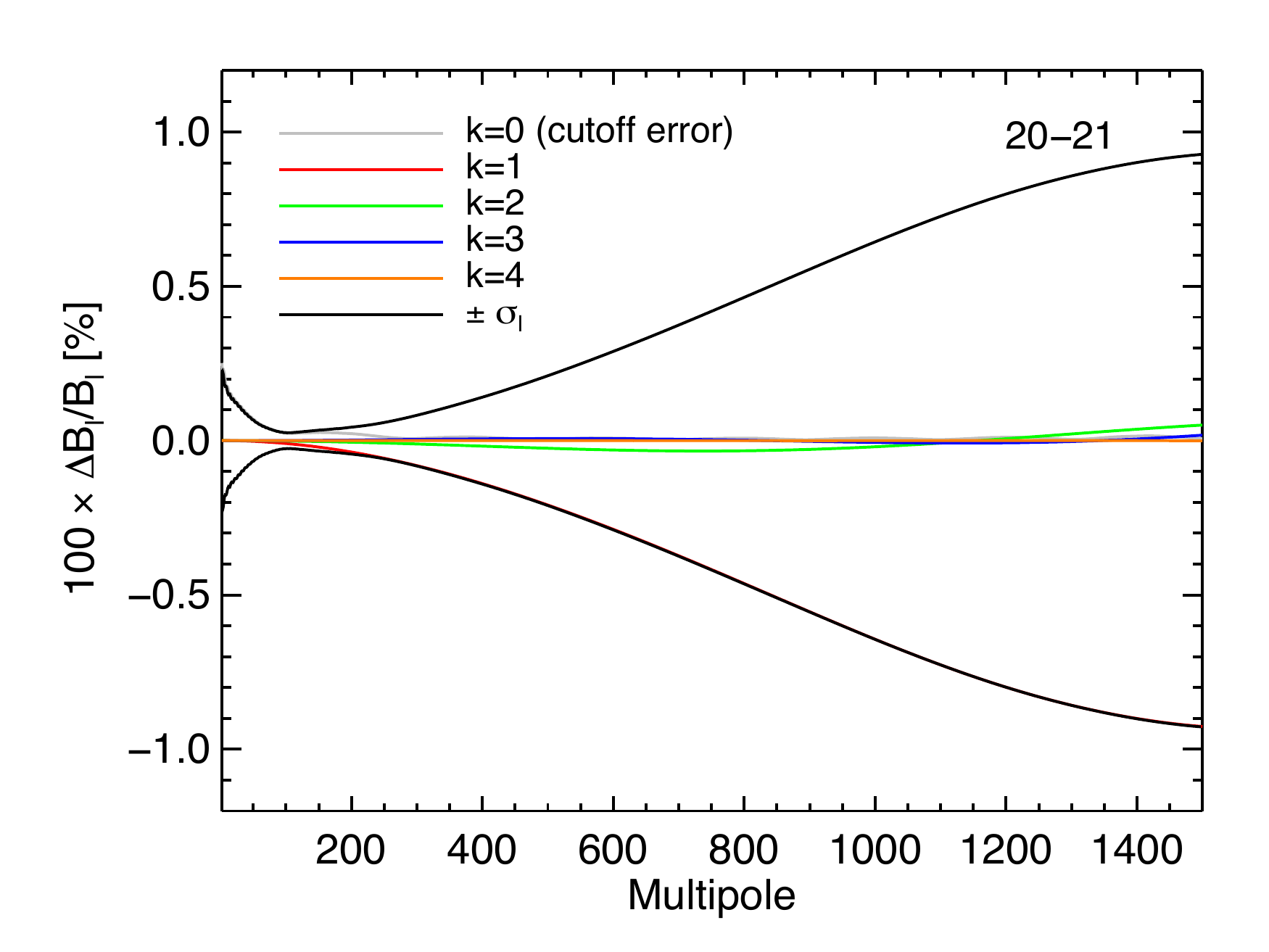} 
\caption{Eigenmodes of the covariance matrix of the quadruplet 20/21 at 70\,GHz.}
\label{fig:ploteigen2021}
\end{figure}

The LFI beams are not used in the $Planck$ 2015 likelihood at high $\ell$ \citep{planck2014-a13}, nevertheless we decide to estimate the impact of the beam error on the cosmological parameters. As done in 2013 release we apply the Markov Chain Beam Randomization (\cite{rocha2010a}) procedure to a simulated 70\,GHz dataset. We found that the impact for all the $\Lambda$CDM parameters is well below $10\%$ of $\sigma$ confirming that the uncertainty on beam knowledge is negligible in the cosmological parameter estimation.

\section{Conclusions}
\label{conclusions}
In this paper we discussed: (i) the improvement in the LFI main beam reconstruction with respect to the 2013 release; (ii) the beam normalization convention adopted in the LFI pipeline; (iii) the temperature and polarized beam window functions; and (iv) the error budgets on the beam parameters and window functions. 
The in-flight assessment of the LFI main beams relied mainly on the measurements performed during seven Jupiter crossings, the first four transits occurring in nominal scan mode and the last three scans in deep mode.
The calibrated data from the Jupiter scans were used to determine the \textit{scanning beams}: the signal-to-noise ratio for these data makes it possible to follow the LFI beams profile down to --30 dB.
These measurements have been used to further validate the beam model presented in 2013 (\texttt{GRASP} beams properly smeared to take into account the satellite motion).
Fitting the main beam shapes with an elliptical Gaussian, we expressed the uncertainties of the measured scanning beam in terms of statistical errors for the Gaussian parameters: ellipticity; orientation; and FWHM. 
The polarized beams, described in Sect.~\ref{polarized_beams}, provide the best fit to the available measurements of the LFI main beams from Jupiter.
We found that this model represents all the LFI beams with an accuracy of about 0.1\% at 30 and 70 GHz, and 0.2\% at 44 GHz (rms value of the difference between measurements and simulations, computed within the 20 dB contour), which has been considered in the propagation of the uncertainties at the window function level.
The corresponding simulated sidelobes have been used in the calibration pipeline to evaluate the gains and to subtract Galactic straylight from the calibrated timelines \citep{planck2014-a03}.
This model, together with the pointing information derived from the focal plane geometry reconstruction, gives the most advanced and precise noise-free representation of the LFI beams.  
The polarized beams were the input to calculate the effective beams, which take into account the specific scanning strategy to include any smearing and orientation effects on the beams themselves. 

To evaluate the beam window function, we adopted two independent approaches, both based on Monte Carlo simulations. In one case, we convolved a fiducial model $C_\ell$ with realistic scanning beams in harmonic space to generate the corresponding timelines and maps; in the other case, we convolved the maps derived from the fiducial model $C_\ell$ with effective beams in pixel space. Using the first approach, we have also evaluated the contribution of the near and far sidelobes on the window functions: it is seen that the impact of sidelobes on the low multipole region is at about the 0.1\,\% level.

The error budget comes from two contributions: the propagation of the main beam uncertainties through the analysis; and the contribution of near and far sidelobes.
As found in the past release, the two error sources have different relevance, depending on the angular scale. Ignoring the near and far sidelobes is the dominant error at low multipoles, while the main beam uncertainties dominate the total error budget at $\ell\geq 600$. 
The total uncertainties in the effective beam window functions are: 0.7\% and 1\% at 30 and 44\,GHz, respectively (at $\ell \approx 600$); and 0.5\% at 70\,GHz at $\ell \approx 1000$.

The results presented in this paper, and in the LFI companion papers, prove the extraordinary capabilities of LFI in achieving the expected objectives in terms of sensitivity, angular resolution, and control of systematic effects. In particular, we found an impressive consistency between main beam simulations and measurements, which demonstrates the reliability and the accuracy of the optical model. The methods used to evaluate the beam window functions and the corresponding error budget have proved to be very well consolidated. In addition, a new promising approach -- the matrix beam window function -- has been presented and it will be consolidated for future data releases.

\begin{acknowledgements}
  \Planck\ is too large a project to allow full acknowledgement of all
  contributions by individuals, institutions, industries, and funding
  agencies. The main entities involved in the mission operations are
  as follows. The European Space Agency (ESA) operates the satellite via its
  Mission Operations Centre located at ESOC (Darmstadt, Germany) and
  coordinates scientific operations via the Planck Science Office
  located at ESAC (Madrid, Spain). Two Consortia, comprising around 50
  scientific institutes within Europe, the USA, and Canada, and funded
  by agencies from the participating countries, developed the
  scientific instruments LFI and HFI, and continue to operate them via
  Instrument Operations Teams located in Trieste (Italy) and Orsay
  (France). The Consortia are also responsible for scientific
  processing of the acquired data. The Consortia are led by the
  Principal Investigators: J.L. Puget in France for HFI (funded
  principally by CNES and CNRS/INSU-IN2P3-INP) and N. Mandolesi in Italy
  for LFI (funded principally via ASI). NASA US Planck Project,
  based at JPL and involving scientists at many US institutions,
  contributes significantly to the efforts of these two Consortia. The
  author list for this paper has been selected by the Planck Science
  Team, and is composed of individuals from all of the above entities
  who have made multi-year contributions to the development of the
  mission. It does not pretend to be inclusive of all contributions.
  The \Planck -LFI project is developed by an International Consortium
  led by Italy and involving Canada, Finland, Germany, Norway, Spain,
  Switzerland, UK, and USA. The Italian contribution to \Planck\ is
  supported by the Italian Space Agency (ASI) and INAF. 
  This work was supported by the Academy of Finland grants 253204, 256265, 257989, and 283497. 
  We acknowledge that the results of this research have been achieved
  using the PRACE-3IP project (FP7 RI-312763) resource Sisu
  based in Finland at CSC.
  We thank CSC -- IT Center for Science Ltd (Finland) for computational resources.  
  We acknowledge financial support provided by the Spanish Ministerio
  de Ciencia e Innovaci{\~o}n through the Plan Nacional del Espacio y
  Plan Nacional de Astronomia y Astrofisica.  
  We acknowledge the Max Planck Institute for Astrophysics Planck Analysis Centre (MPAC),
  funded by the Space Agency of the German Aerospace Center (DLR)
  under grant 50OP0901 with resources of the German Federal Ministry
  of Economics and Technology, and by the Max Planck Society. This
  work has made use of the Planck satellite simulation package
  (Level-S), which is assembled by the Max Planck Institute for
  Astrophysics Planck Analysis Centre (MPAC). We
  acknowledge financial support provided by the National Energy
  Research Scientific Computing Center, which is supported by the
  Office of Science of the U.S. Department of Energy under Contract
  No. DE-AC02-05CH11231. Some of the results in this paper have been
  derived using the HEALPix package.
The Planck Collaboration acknowledges the support of: ESA; CNES and CNRS/INSU-IN2P3-INP (France); ASI, CNR, and INAF (Italy); NASA and DoE (USA); STFC and UKSA (UK); CSIC, MINECO, JA, and RES (Spain); Tekes, AoF, and CSC (Finland); DLR and MPG (Germany); CSA (Canada); DTU Space (Denmark); SER/SSO (Switzerland); RCN (Norway); SFI (Ireland); FCT/MCTES (Portugal); ERC and PRACE (EU). A description of the Planck Collaboration and a list of its members, indicating which technical or scientific activities they have been involved in, can be found at \url{http://www.cosmos.esa.int/web/planck/planck-collaboration}.
\end{acknowledgements}

\appendix
\section{Useful definitions}
\label{regions}

\subsection{\texttt{GRASP} simulations}

The far field pattern in the three regions reported above has been computed with {\tt GRASP} using different computational methods and different field storage.

Main beams have been computed in two-dimensional grids over the spherical surface, defined by the variables u and v, related to the spherical angles by u = $\sin\theta \times \cos\phi$ and v = $\sin\theta \times \sin\phi$.
The variables u and v range from --0.033 to 0.033 ($\theta \leq 1.9^\circ$) for the 30 GHz channel, from --0.023 to 0.023 ($\theta \leq 1.3^\circ$) for the 44 GHz channels, and from --0.015 to 0.015 ($\theta \leq 0.9^\circ$) for the 70 GHz channel.
Each grid is sampled with 601 $\times$ 601 points, therefore the spatial resolution is about 23 arcsec for the 30 GHz channel, 16 arcsec for the 44 GHz channel, and 10 arcsec for the 70 GHz channel.

Near and far sidelobes have been computed in spherical polar cuts, for which $\phi$ is constant and $\theta$ is varying. 
These cuts pass through the pole of the sphere (i.e., the beam pointing direction) at $\theta = 0$.
Near sidelobes have been computed with a spatial resolution of 1' in $\theta$ and 0.5$^\circ$ in $\phi$.
Far sidelobes have been computed with a spatial resolution of 0.5$^\circ$, both in $\theta$ and $\phi$.

Main beams and near sidelobes have been computed using physical optics and physical theory of diffraction \citep{grasptd}, whereas far sidelobes have been computed using Multi-reflector Geometrical Theory of Diffraction (MrGTD) \citep{graspmrgtd}.

\begin{figure}
\centering
\includegraphics[width=8.5cm]{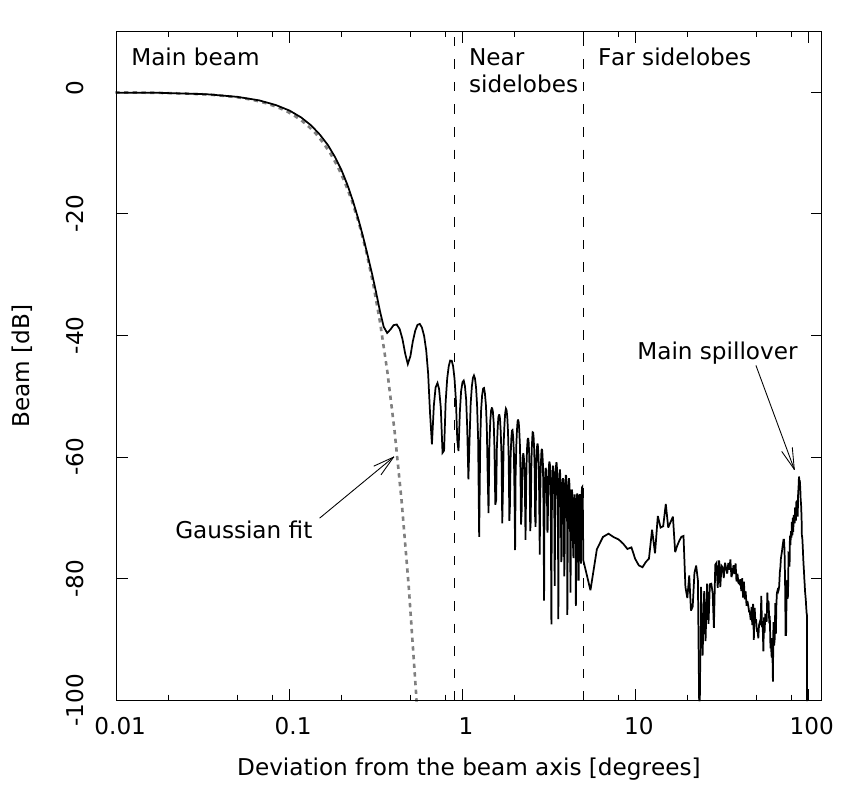}
\caption{Typical shape of a 70\,GHz beam (\texttt{LFI18S}). The plot shows the distinction between the main beam, near sidelobes, and far sidelobes. The distinction between ``near'' and ``far'' sidelobes is of course arbitrary, and here their boundary is marked at 5$^\circ$. The peak of the spillover of the primary mirror is clearly visible, at an angle of roughly 90$^\circ$.}
\label{fig:beamplot}
\end{figure}

\subsection{Nomenclature for beams}

In the present paper, and in the \Planck\ companion papers, we used three relevant definitions:
\begin{enumerate}
\item The ``optical beam'' is the optical response of the feed horn coupled to the telescope. It is independent of both the radiometer response (bandshape and non-linearity) and of the satellite motion (spinning and scanning strategy). It represents the pure optical transfer function. The main beam properties of the optical beams can be evaluated using optical simulations performed with methods largely validated by ground measurements.
\item The ``scanning beam'' is the beam that can be directly measured in-flight using planet observations. It stems from the optical beam, coupled with the radiometer response, and smeared by the satellite motion. So, with respect to the optical beams, the scanning beams have slightly higher values of angular size and ellipticity.
\item The ``effective beam'' is a beam defined in the map-domain, and is obtained by averaging the scanning beams pointing at a given pixel of the map, taking into account the scanning strategy and the orientation of the beams themselves when they point along the direction to that pixel. Therefore, whereas for each radiometer there is one corresponding optical and scanning beam, the same radiometer has the same number of effective beams as there are pixels in the observed sky map. The importance of the effective beams is twofold: they are used in the window function computation; and their solid angles are needed for the estimation of the flux density of point sources. 
\end{enumerate}

%

\section{Beam fit results}
\label{beam_fit_results}

As described in Sect.~\ref{planet_data}, the code used to fit the beam shape to an elliptical Gaussian function returns the full width half maximum (FWHM), the beam ellipticity ($e$), and the beam orientation ($\psi_{{\rm ell}}$).
Moreover, the fit procedure returns the main beam pointing directions in the \Planck\ field of view, centred along the nominal line of sight.
In Tables from \ref{Jupiter 1} to \ref{Jupiter 7}, the fitted parameters are reported for each scan, with their error at 68\,\%CL.
These values are those plotted in Figs.~\ref{fig:histofwhm}, \ref{fig:histoe}, and \ref{fig:histopsi}.  
The main beam descriptive parameters fitted from the stacked scans are those reported in Table \ref{tab:imo}, whereas the main beam pointing directions ($\theta_{{\rm uv}}$ and $\phi_{{\rm uv}}$) have been computed from $X_0$ and $Y_0$, and reported in \cite{planck2014-a03}, using these simple formulae:

\begin{equation}
\theta_{\rm uv} = \arcsin \sqrt{X_0^2+Y_0^2};
\end{equation}

\begin{equation}
\phi_{\rm uv} = \arctan \frac{Y_0}{X_0}.
\end{equation}

\begin{table*}
\centering
\caption{Fitted parameters derived from the first scan of Jupiter: main beam pointing directions defined with respect to the nominal telescope line of sight, FWHM, ellipticity, and orientation.}
\begin{tabular}{c r@{.}l@{ $\pm$ }r@{.}l r@{.}l@{ $\pm$ }r@{.}l r@{.}l@{ $\pm$ }r@{.}l r@{.}l@{ $\pm$ }r@{.}l c@{ $\pm$ }c}
\hline
\hline
\noalign{\vskip 2pt}
{\bf Beam} & \multicolumn{4}{c}{\bf X$_0$} & \multicolumn{4}{c}{\bf Y$_0$} & \multicolumn{4}{c}{\bf FWHM} & \multicolumn{4}{c}{\bf e} & \multicolumn{2}{c}{\bf $\psi_{{\rm ell}}$} \\
\noalign{\vskip 2pt}
\noalign{\vskip 2pt}
& \multicolumn{4}{c}{} & \multicolumn{4}{c}{} & \multicolumn{4}{c}{(arcmin)} & \multicolumn{4}{c}{} & \multicolumn{2}{c}{(deg)} \\
\noalign{\vskip 2pt}
\hline
\noalign{\vskip 2pt}
\multicolumn{19}{l}{\bf 70 GHz} \\
\noalign{\vskip 4pt}
\texttt{18M} & --0&03879 &  0&00002 & --0&04334 &  0&00002 & 13&44 &  0&13 &  1&23 &  0&02 &   85 &    2 \\ 
\texttt{18S} & --0&03878 &  0&00002 & --0&04335 &  0&00002 & 13&50 &  0&12 &  1&27 &  0&02 &   86 &    2 \\ 
\texttt{19M} & --0&04873 &  0&00002 & --0&02759 &  0&00002 & 13&14 &  0&13 &  1&25 &  0&02 &   78 &    2 \\ 
\texttt{19S} & --0&04874 &  0&00002 & --0&02758 &  0&00002 & 13&07 &  0&13 &  1&28 &  0&02 &   79 &    2 \\ 
\texttt{20M} & --0&05438 &  0&00002 & --0&01138 &  0&00002 & 12&84 &  0&12 &  1&27 &  0&02 &   71 &    2 \\ 
\texttt{20S} & --0&05438 &  0&00002 & --0&01137 &  0&00002 & 12&84 &  0&13 &  1&29 &  0&02 &   72 &    2 \\ 
\texttt{21M} & --0&05460 &  0&00002 &  0&01034 &  0&00002 & 12&77 &  0&10 &  1&28 &  0&02 &  107 &    1 \\ 
\texttt{21S} & --0&05459 &  0&00002 &  0&01035 &  0&00002 & 12&87 &  0&12 &  1&29 &  0&02 &  106 &    2 \\ 
\texttt{22M} & --0&04860 &  0&00002 &  0&02654 &  0&00002 & 12&92 &  0&11 &  1&27 &  0&01 &  101 &    2 \\ 
\texttt{22S} & --0&04861 &  0&00002 &  0&02653 &  0&00002 & 12&97 &  0&12 &  1&28 &  0&02 &  101 &    2 \\ 
\texttt{23M} & --0&03849 &  0&00002 &  0&04237 &  0&00002 & 13&35 &  0&12 &  1&23 &  0&02 &   92 &    2 \\ 
\texttt{23S} & --0&03850 &  0&00002 &  0&04235 &  0&00002 & 13&36 &  0&13 &  1&28 &  0&02 &   92 &    2 \\ 
\hline
\noalign{\vskip 2pt}
\multicolumn{19}{l}{\bf 44 GHz} \\
\noalign{\vskip 4pt}
\texttt{24M} & --0&07102 &  0&00007 & --0&00058 &  0&00009 & 23&18 &  0&51 &  1&39 &  0&06 &   89 &    3 \\ 
\texttt{24S} & --0&07101 &  0&00006 & --0&00060 &  0&00008 & 23&04 &  0&45 &  1&34 &  0&05 &   89 &    3 \\ 
\texttt{25M} &  0&04199 &  0&00014 &  0&07605 &  0&00012 & 30&23 &  0&94 &  1&19 &  0&07 &  114 &    9 \\ 
\texttt{25S} &  0&04193 &  0&00015 &  0&07607 &  0&00012 & 30&94 &  0&95 &  1&19 &  0&07 &  117 &    9 \\ 
\texttt{26M} &  0&04165 &  0&00016 & --0&07727 &  0&00013 & 30&29 &  1&06 &  1&19 &  0&08 &   62 &   10 \\ 
\texttt{26S} &  0&04163 &  0&00015 & --0&07728 &  0&00012 & 30&64 &  0&97 &  1&19 &  0&07 &   61 &    9 \\ 
\hline
\noalign{\vskip 2pt}
\multicolumn{19}{l}{\bf 30 GHz} \\
\noalign{\vskip 4pt}
\texttt{27M} & --0&06810 &  0&00014 &  0&03326 &  0&00019 & 33&02 &  1&09 &  1&37 &  0&05 &  101 &    5 \\ 
\texttt{27S} & --0&06811 &  0&00014 &  0&03326 &  0&00019 & 33&11 &  1&13 &  1&38 &  0&05 &  101 &    5 \\ 
\texttt{28M} & --0&06823 &  0&00015 & --0&03412 &  0&00020 & 33&10 &  1&18 &  1&37 &  0&05 &   78 &    5 \\ 
\texttt{28S} & --0&06825 &  0&00014 & --0&03412 &  0&00018 & 33&09 &  1&08 &  1&37 &  0&05 &   78 &    5 \\ 
\hline
\end{tabular}
\label{Jupiter 1}
\end{table*}

\begin{table*}
\centering
\caption{Fitted parameters derived from the second scan of Jupiter: main beam pointing directions defined with respect to the nominal telescope line of sight, FWHM, ellipticity, and orientation.}
\begin{tabular}{c r@{.}l@{ $\pm$ }r@{.}l r@{.}l@{ $\pm$ }r@{.}l r@{.}l@{ $\pm$ }r@{.}l r@{.}l@{ $\pm$ }r@{.}l c@{ $\pm$ }c}
\hline
\hline
\noalign{\vskip 2pt}
{\bf Beam} & \multicolumn{4}{c}{\bf X$_0$} & \multicolumn{4}{c}{\bf Y$_0$} & \multicolumn{4}{c}{\bf FWHM} & \multicolumn{4}{c}{\bf e} & \multicolumn{2}{c}{\bf $\psi_{{\rm ell}}$} \\
\noalign{\vskip 2pt}
\noalign{\vskip 2pt}
& \multicolumn{4}{c}{} & \multicolumn{4}{c}{} & \multicolumn{4}{c}{(arcmin)} & \multicolumn{4}{c}{} & \multicolumn{2}{c}{(deg)} \\
\noalign{\vskip 2pt}
\hline
\noalign{\vskip 2pt}
\multicolumn{19}{l}{\bf 70 GHz} \\
\noalign{\vskip 4pt}
\texttt{18M} & --0&03879 &  0&00002 & --0&04334 &  0&00002 & 13&40 &  0&11 &  1&23 &  0&02 &   85 &    2 \\ 
\texttt{18S} & --0&03879 &  0&00002 & --0&04334 &  0&00002 & 13&45 &  0&10 &  1&28 &  0&02 &   86 &    1 \\ 
\texttt{19M} & --0&04872 &  0&00002 & --0&02758 &  0&00002 & 13&13 &  0&11 &  1&25 &  0&02 &   78 &    2 \\ 
\texttt{19S} & --0&04873 &  0&00002 & --0&02758 &  0&00002 & 13&08 &  0&11 &  1&28 &  0&02 &   79 &    1 \\ 
\texttt{20M} & --0&05438 &  0&00002 & --0&01137 &  0&00002 & 12&81 &  0&10 &  1&27 &  0&02 &   71 &    1 \\ 
\texttt{20S} & --0&05438 &  0&00002 & --0&01136 &  0&00002 & 12&83 &  0&11 &  1&29 &  0&02 &   72 &    1 \\ 
\texttt{21M} & --0&05461 &  0&00001 &  0&01034 &  0&00002 & 12&74 &  0&09 &  1&28 &  0&02 &  108 &    1 \\ 
\texttt{21S} & --0&05459 &  0&00002 &  0&01035 &  0&00002 & 12&86 &  0&10 &  1&29 &  0&02 &  106 &    1 \\ 
\texttt{22M} & --0&04859 &  0&00001 &  0&02654 &  0&00002 & 12&89 &  0&09 &  1&26 &  0&02 &  101 &    1 \\ 
\texttt{22S} & --0&04859 &  0&00001 &  0&02653 &  0&00002 & 12&95 &  0&10 &  1&28 &  0&02 &  101 &    1 \\ 
\texttt{23M} & --0&03849 &  0&00002 &  0&04237 &  0&00002 & 13&32 &  0&11 &  1&24 &  0&02 &   93 &    2 \\ 
\texttt{23S} & --0&03850 &  0&00002 &  0&04235 &  0&00002 & 13&32 &  0&11 &  1&28 &  0&02 &   93 &    1 \\ 
\hline
\noalign{\vskip 2pt}
\multicolumn{19}{l}{\bf 44 GHz} \\
\noalign{\vskip 4pt}
\texttt{24M} & --0&07102 &  0&00006 & --0&00057 &  0&00008 & 23&18 &  0&44 &  1&39 &  0&05 &   90 &    3 \\ 
\texttt{24S} & --0&07100 &  0&00005 & --0&00062 &  0&00007 & 23&04 &  0&40 &  1&34 &  0&04 &   90 &    3 \\ 
\texttt{25M} &  0&04201 &  0&00012 &  0&07605 &  0&00010 & 30&16 &  0&80 &  1&19 &  0&06 &  115 &    7 \\ 
\texttt{25S} &  0&04196 &  0&00013 &  0&07607 &  0&00011 & 30&88 &  0&81 &  1&19 &  0&06 &  116 &    7 \\ 
\texttt{26M} &  0&04166 &  0&00014 & --0&07727 &  0&00011 & 30&16 &  0&91 &  1&19 &  0&07 &   62 &    8 \\ 
\texttt{26S} &  0&04165 &  0&00013 & --0&07728 &  0&00011 & 30&50 &  0&83 &  1&19 &  0&06 &   61 &    8 \\ 
\hline
\noalign{\vskip 2pt}
\multicolumn{19}{l}{\bf 30 GHz} \\
\noalign{\vskip 4pt}
\texttt{27M} & --0&06810 &  0&00012 &  0&03323 &  0&00016 & 33&03 &  0&93 &  1&36 &  0&04 &  101 &    4 \\ 
\texttt{27S} & --0&06810 &  0&00012 &  0&03324 &  0&00017 & 33&25 &  0&99 &  1&38 &  0&04 &  101 &    4 \\ 
\texttt{28M} & --0&06825 &  0&00013 & --0&03413 &  0&00017 & 33&16 &  1&02 &  1&37 &  0&04 &   78 &    5 \\ 
\texttt{28S} & --0&06823 &  0&00012 & --0&03415 &  0&00016 & 33&20 &  0&93 &  1&37 &  0&04 &   78 &    4 \\ 
\hline
\end{tabular}
\label{Jupiter 2}
\end{table*}

\begin{table*}
\centering
\caption{Fitted parameters derived from the third scan of Jupiter: main beam pointing directions defined with respect to the nominal telescope line of sight, FWHM, ellipticity, and orientation.}
\begin{tabular}{c r@{.}l@{ $\pm$ }r@{.}l r@{.}l@{ $\pm$ }r@{.}l r@{.}l@{ $\pm$ }r@{.}l r@{.}l@{ $\pm$ }r@{.}l c@{ $\pm$ }c}
\hline
\hline
\noalign{\vskip 2pt}
{\bf Beam} & \multicolumn{4}{c}{\bf X$_0$} & \multicolumn{4}{c}{\bf Y$_0$} & \multicolumn{4}{c}{\bf FWHM} & \multicolumn{4}{c}{\bf e} & \multicolumn{2}{c}{\bf $\psi_{{\rm ell}}$} \\
\noalign{\vskip 2pt}
\noalign{\vskip 2pt}
& \multicolumn{4}{c}{} & \multicolumn{4}{c}{} & \multicolumn{4}{c}{(arcmin)} & \multicolumn{4}{c}{} & \multicolumn{2}{c}{(deg)} \\
\noalign{\vskip 2pt}
\hline
\noalign{\vskip 2pt}
\multicolumn{19}{l}{\bf 70 GHz} \\
\noalign{\vskip 4pt}
\texttt{18M} & --0&03878 &  0&00002 & --0&04334 &  0&00002 & 13&40 &  0&11 &  1&24 &  0&02 &   85 &    2 \\ 
\texttt{18S} & --0&03878 &  0&00002 & --0&04334 &  0&00002 & 13&46 &  0&10 &  1&28 &  0&02 &   86 &    1 \\ 
\texttt{19M} & --0&04871 &  0&00002 & --0&02758 &  0&00002 & 13&13 &  0&11 &  1&25 &  0&02 &   79 &    2 \\ 
\texttt{19S} & --0&04872 &  0&00002 & --0&02758 &  0&00002 & 13&10 &  0&11 &  1&28 &  0&02 &   79 &    1 \\ 
\texttt{20M} & --0&05437 &  0&00002 & --0&01138 &  0&00002 & 12&83 &  0&11 &  1&27 &  0&02 &   71 &    1 \\ 
\texttt{20S} & --0&05437 &  0&00002 & --0&01137 &  0&00002 & 12&84 &  0&11 &  1&29 &  0&02 &   72 &    1 \\ 
\texttt{21M} & --0&05460 &  0&00001 &  0&01034 &  0&00002 & 12&77 &  0&09 &  1&28 &  0&02 &  107 &    1 \\ 
\texttt{21S} & --0&05458 &  0&00002 &  0&01035 &  0&00002 & 12&87 &  0&10 &  1&30 &  0&02 &  106 &    1 \\ 
\texttt{22M} & --0&04858 &  0&00001 &  0&02654 &  0&00002 & 12&92 &  0&09 &  1&27 &  0&02 &  102 &    1 \\ 
\texttt{22S} & --0&04859 &  0&00001 &  0&02653 &  0&00002 & 12&99 &  0&10 &  1&28 &  0&02 &  101 &    1 \\ 
\texttt{23M} & --0&03849 &  0&00002 &  0&04236 &  0&00002 & 13&32 &  0&11 &  1&24 &  0&02 &   93 &    2 \\ 
\texttt{23S} & --0&03850 &  0&00002 &  0&04235 &  0&00002 & 13&33 &  0&11 &  1&28 &  0&02 &   93 &    1 \\ 
\hline
\noalign{\vskip 2pt}
\multicolumn{19}{l}{\bf 44 GHz} \\
\noalign{\vskip 4pt}
\texttt{24M} & --0&07101 &  0&00006 & --0&00057 &  0&00008 & 23&20 &  0&44 &  1&39 &  0&05 &   89 &    3 \\ 
\texttt{24S} & --0&07099 &  0&00005 & --0&00062 &  0&00007 & 23&16 &  0&40 &  1&34 &  0&04 &   89 &    3 \\ 
\texttt{25M} &  0&04198 &  0&00012 &  0&07605 &  0&00010 & 30&15 &  0&76 &  1&19 &  0&06 &  115 &    7 \\ 
\texttt{25S} &  0&04194 &  0&00012 &  0&07606 &  0&00010 & 30&91 &  0&77 &  1&19 &  0&05 &  117 &    7 \\ 
\texttt{26M} &  0&04165 &  0&00013 & --0&07726 &  0&00011 & 30&26 &  0&87 &  1&19 &  0&06 &   61 &    8 \\ 
\texttt{26S} &  0&04166 &  0&00012 & --0&07726 &  0&00010 & 30&62 &  0&79 &  1&19 &  0&06 &   62 &    7 \\ 
\hline
\noalign{\vskip 2pt}
\multicolumn{19}{l}{\bf 30 GHz} \\
\noalign{\vskip 4pt}
\texttt{27M} & --0&06810 &  0&00012 &  0&03322 &  0&00016 & 33&13 &  0&92 &  1&37 &  0&04 &  101 &    4 \\ 
\texttt{27S} & --0&06810 &  0&00012 &  0&03322 &  0&00016 & 33&30 &  0&97 &  1&37 &  0&04 &  101 &    4 \\ 
\texttt{28M} & --0&06823 &  0&00013 & --0&03414 &  0&00017 & 33&32 &  1&03 &  1&37 &  0&08 &   78 &    5 \\ 
\texttt{28S} & --0&06822 &  0&00012 & --0&03414 &  0&00016 & 33&24 &  0&93 &  1&36 &  0&04 &   78 &    4 \\ 
\hline
\end{tabular}
\label{Jupiter 3}
\end{table*}

\begin{table*}
\centering
\caption{Fitted parameters derived from the fourth scan of Jupiter: main beam pointing directions defined with respect to the nominal telescope line of sight, FWHM, ellipticity, and orientation.}
\begin{tabular}{c r@{.}l@{ $\pm$ }r@{.}l r@{.}l@{ $\pm$ }r@{.}l r@{.}l@{ $\pm$ }r@{.}l r@{.}l@{ $\pm$ }r@{.}l c@{ $\pm$ }c}
\hline
\hline
\noalign{\vskip 2pt}
{\bf Beam} & \multicolumn{4}{c}{\bf X$_0$} & \multicolumn{4}{c}{\bf Y$_0$} & \multicolumn{4}{c}{\bf FWHM} & \multicolumn{4}{c}{\bf e} & \multicolumn{2}{c}{\bf $\psi_{{\rm ell}}$} \\
\noalign{\vskip 2pt}
\noalign{\vskip 2pt}
& \multicolumn{4}{c}{} & \multicolumn{4}{c}{} & \multicolumn{4}{c}{(arcmin)} & \multicolumn{4}{c}{} & \multicolumn{2}{c}{(deg)} \\
\noalign{\vskip 2pt}
\hline
\noalign{\vskip 2pt}
\multicolumn{19}{l}{\bf 70 GHz} \\
\noalign{\vskip 4pt}
\texttt{18M} & --0&03878 &  0&00002 & --0&04333 &  0&00003 & 13&39 &  0&14 &  1&24 &  0&03 &   85 &    2 \\ 
\texttt{18S} & --0&03878 &  0&00002 & --0&04334 &  0&00003 & 13&46 &  0&14 &  1&28 &  0&03 &   86 &    2 \\ 
\texttt{19M} & --0&04872 &  0&00002 & --0&02758 &  0&00003 & 13&13 &  0&15 &  1&25 &  0&03 &   79 &    2 \\ 
\texttt{19S} & --0&04872 &  0&00002 & --0&02757 &  0&00003 & 13&07 &  0&14 &  1&28 &  0&03 &   79 &    2 \\ 
\texttt{20M} & --0&05437 &  0&00002 & --0&01137 &  0&00003 & 12&84 &  0&14 &  1&27 &  0&03 &   71 &    2 \\ 
\texttt{20S} & --0&05438 &  0&00002 & --0&01136 &  0&00003 & 12&83 &  0&15 &  1&29 &  0&03 &   72 &    2 \\ 
\texttt{21M} & --0&05460 &  0&00002 &  0&01035 &  0&00002 & 12&75 &  0&12 &  1&28 &  0&02 &  108 &    2 \\ 
\texttt{21S} & --0&05459 &  0&00002 &  0&01035 &  0&00003 & 12&87 &  0&14 &  1&29 &  0&03 &  106 &    2 \\ 
\texttt{22M} & --0&04858 &  0&00002 &  0&02654 &  0&00002 & 12&92 &  0&13 &  1&26 &  0&02 &  101 &    2 \\ 
\texttt{22S} & --0&04859 &  0&00002 &  0&02654 &  0&00003 & 12&99 &  0&13 &  1&28 &  0&03 &  101 &    2 \\ 
\texttt{23M} & --0&03849 &  0&00002 &  0&04238 &  0&00003 & 13&32 &  0&14 &  1&24 &  0&03 &   93 &    2 \\ 
\texttt{23S} & --0&03850 &  0&00002 &  0&04236 &  0&00003 & 13&33 &  0&15 &  1&28 &  0&03 &   93 &    2 \\ 
\hline
\noalign{\vskip 2pt}
\multicolumn{19}{l}{\bf 44 GHz} \\
\noalign{\vskip 4pt}
\texttt{24M} & --0&07101 &  0&00008 & --0&00056 &  0&00011 & 23&16 &  0&60 &  1&39 &  0&07 &   90 &    4 \\ 
\texttt{24S} & --0&07099 &  0&00007 & --0&00061 &  0&00010 & 23&02 &  0&54 &  1&34 &  0&06 &   90 &    4 \\ 
\texttt{25M} &  0&04200 &  0&00016 &  0&07604 &  0&00014 & 30&24 &  1&06 &  1&20 &  0&08 &  116 &   10 \\ 
\texttt{25S} &  0&04194 &  0&00016 &  0&07606 &  0&00014 & 31&01 &  1&09 &  1&19 &  0&08 &  117 &   10 \\ 
\texttt{26M} &  0&04167 &  0&00018 & --0&07726 &  0&00015 & 30&23 &  1&22 &  1&20 &  0&09 &   60 &   11 \\ 
\texttt{26S} &  0&04167 &  0&00017 & --0&07728 &  0&00014 & 30&69 &  1&12 &  1&19 &  0&08 &   61 &   10 \\ 
\hline
\noalign{\vskip 2pt}
\multicolumn{19}{l}{\bf 30 GHz} \\
\noalign{\vskip 4pt}
\texttt{27M} & --0&06810 &  0&00016 &  0&03323 &  0&00022 & 33&00 &  1&28 &  1&36 &  0&05 &  101 &    6 \\ 
\texttt{27S} & --0&06810 &  0&00017 &  0&03324 &  0&00023 & 33&12 &  1&36 &  1&38 &  0&06 &  100 &    6 \\ 
\texttt{28M} & --0&06822 &  0&00018 & --0&03414 &  0&00024 & 33&19 &  1&45 &  1&37 &  0&11 &   78 &    7 \\ 
\texttt{28S} & --0&06821 &  0&00017 & --0&03413 &  0&00022 & 33&19 &  1&32 &  1&37 &  0&10 &   78 &    6 \\ 
\hline
\end{tabular}
\label{Jupiter 4}
\end{table*}

\begin{table*}
\centering
\caption{Fitted parameters derived from the fifth scan of Jupiter: main beam pointing directions defined with respect to the nominal telescope line of sight, FWHM, ellipticity, and orientation. Data at 30 and 44 GHz are missing due to spacecraft manoeuvrements during the observations.}
\begin{tabular}{c r@{.}l@{ $\pm$ }r@{.}l r@{.}l@{ $\pm$ }r@{.}l r@{.}l@{ $\pm$ }r@{.}l r@{.}l@{ $\pm$ }r@{.}l c@{ $\pm$ }c}
\hline
\hline
\noalign{\vskip 2pt}
{\bf Beam} & \multicolumn{4}{c}{\bf X$_0$} & \multicolumn{4}{c}{\bf Y$_0$} & \multicolumn{4}{c}{\bf FWHM} & \multicolumn{4}{c}{\bf e} & \multicolumn{2}{c}{\bf $\psi_{{\rm ell}}$} \\
\noalign{\vskip 2pt}
\noalign{\vskip 2pt}
& \multicolumn{4}{c}{} & \multicolumn{4}{c}{} & \multicolumn{4}{c}{(arcmin)} & \multicolumn{4}{c}{} & \multicolumn{2}{c}{(deg)} \\
\noalign{\vskip 2pt}
\hline
\noalign{\vskip 2pt}
\multicolumn{19}{l}{\bf 70 GHz} \\
\noalign{\vskip 4pt}
\texttt{18M} & --0&03878 &  0&00002 & --0&04334 &  0&00002 & 13&39 &  0&12 &  1&24 &  0&02 &   85 &    2 \\ 
\texttt{18S} & --0&03878 &  0&00002 & --0&04334 &  0&00002 & 13&46 &  0&12 &  1&28 &  0&02 &   86 &    2 \\ 
\texttt{19M} & --0&04872 &  0&00001 & --0&02759 &  0&00001 & 13&13 &  0&06 &  1&25 &  0&01 &   78 &    1 \\ 
\texttt{19S} & --0&04872 &  0&00001 & --0&02758 &  0&00001 & 13&08 &  0&05 &  1&28 &  0&01 &   79 &    1 \\ 
\texttt{20M} & --0&05437 &  0&00001 & --0&01138 &  0&00001 & 12&82 &  0&05 &  1&27 &  0&01 &   71 &    1 \\ 
\texttt{20S} & --0&05438 &  0&00001 & --0&01137 &  0&00001 & 12&83 &  0&05 &  1&29 &  0&01 &   72 &    1 \\ 
\texttt{21M} & --0&05460 &  0&00001 &  0&01034 &  0&00001 & 12&74 &  0&04 &  1&28 &  0&01 &  108 &    1 \\ 
\texttt{21S} & --0&05459 &  0&00001 &  0&01035 &  0&00001 & 12&86 &  0&05 &  1&30 &  0&01 &  106 &    1 \\ 
\texttt{22M} & --0&04859 &  0&00001 &  0&02654 &  0&00001 & 12&92 &  0&05 &  1&26 &  0&01 &  101 &    1 \\ 
\texttt{22S} & --0&04860 &  0&00001 &  0&02653 &  0&00001 & 13&00 &  0&05 &  1&28 &  0&01 &  101 &    1 \\ 
\texttt{23M} & --0&03849 &  0&00002 &  0&04236 &  0&00002 & 13&30 &  0&12 &  1&24 &  0&02 &   93 &    2 \\ 
\texttt{23S} & --0&03850 &  0&00002 &  0&04235 &  0&00002 & 13&33 &  0&13 &  1&28 &  0&02 &   93 &    2 \\ 
\hline
\end{tabular}
\label{Jupiter 5}
\end{table*}

\begin{table*}
\centering
\caption{Fitted parameters derived from the sixth scan of Jupiter: main beam pointing directions defined with respect to the nominal telescope line of sight, FWHM, ellipticity, and orientation.}
\begin{tabular}{c r@{.}l@{ $\pm$ }r@{.}l r@{.}l@{ $\pm$ }r@{.}l r@{.}l@{ $\pm$ }r@{.}l r@{.}l@{ $\pm$ }r@{.}l c@{ $\pm$ }c}
\hline
\hline
\noalign{\vskip 2pt}
{\bf Beam} & \multicolumn{4}{c}{\bf X$_0$} & \multicolumn{4}{c}{\bf Y$_0$} & \multicolumn{4}{c}{\bf FWHM} & \multicolumn{4}{c}{\bf e} & \multicolumn{2}{c}{\bf $\psi_{{\rm ell}}$} \\
\noalign{\vskip 2pt}
\noalign{\vskip 2pt}
& \multicolumn{4}{c}{} & \multicolumn{4}{c}{} & \multicolumn{4}{c}{(arcmin)} & \multicolumn{4}{c}{} & \multicolumn{2}{c}{(deg)} \\
\noalign{\vskip 2pt}
\hline
\noalign{\vskip 2pt}
\multicolumn{19}{l}{\bf 70 GHz} \\
\noalign{\vskip 4pt}
\texttt{18M} & --0&03879 &  0&00001 & --0&04335 &  0&00001 & 13&40 &  0&14 &  1&24 &  0&03 &   85 &    2 \\ 
\texttt{18S} & --0&03878 &  0&00001 & --0&04336 &  0&00001 & 13&45 &  0&14 &  1&28 &  0&03 &   86 &    2 \\ 
\texttt{19M} & --0&04872 &  0&00001 & --0&02760 &  0&00001 & 13&13 &  0&15 &  1&25 &  0&03 &   78 &    2 \\ 
\texttt{19S} & --0&04872 &  0&00001 & --0&02759 &  0&00001 & 13&09 &  0&14 &  1&28 &  0&03 &   79 &    2 \\ 
\texttt{20M} & --0&05437 &  0&00001 & --0&01139 &  0&00001 & 12&83 &  0&14 &  1&27 &  0&03 &   71 &    2 \\ 
\texttt{20S} & --0&05438 &  0&00001 & --0&01138 &  0&00002 & 12&82 &  0&15 &  1&29 &  0&03 &   72 &    2 \\ 
\texttt{21M} & --0&05461 &  0&00001 &  0&01033 &  0&00001 & 12&73 &  0&12 &  1&28 &  0&02 &  108 &    2 \\ 
\texttt{21S} & --0&05460 &  0&00001 &  0&01034 &  0&00001 & 12&85 &  0&14 &  1&29 &  0&03 &  106 &    2 \\ 
\texttt{22M} & --0&04859 &  0&00001 &  0&02652 &  0&00001 & 12&92 &  0&13 &  1&26 &  0&02 &  101 &    2 \\ 
\texttt{22S} & --0&04860 &  0&00001 &  0&02652 &  0&00001 & 12&99 &  0&13 &  1&28 &  0&03 &  101 &    2 \\ 
\texttt{23M} & --0&03849 &  0&00001 &  0&04235 &  0&00001 & 13&32 &  0&14 &  1&24 &  0&03 &   93 &    2 \\ 
\texttt{23S} & --0&03851 &  0&00001 &  0&04234 &  0&00001 & 13&32 &  0&15 &  1&28 &  0&03 &   93 &    2 \\ 
\hline
\noalign{\vskip 2pt}
\multicolumn{19}{l}{\bf 44 GHz} \\
\noalign{\vskip 4pt}
\texttt{24M} & --0&07103 &  0&00007 & --0&00058 &  0&00010 & 23&22 &  0&60 &  1&39 &  0&07 &   90 &    4 \\ 
\texttt{24S} & --0&07100 &  0&00004 & --0&00062 &  0&00006 & 22&91 &  0&54 &  1&34 &  0&06 &   90 &    4 \\ 
\texttt{25M} &  0&04199 &  0&00012 &  0&07604 &  0&00010 & 30&14 &  1&06 &  1&20 &  0&08 &  116 &   10 \\ 
\texttt{25S} &  0&04194 &  0&00012 &  0&07605 &  0&00010 & 31&00 &  1&09 &  1&19 &  0&08 &  117 &   10 \\ 
\texttt{26M} &  0&04166 &  0&00013 & --0&07727 &  0&00011 & 30&22 &  1&22 &  1&20 &  0&09 &   60 &   11 \\ 
\texttt{26S} &  0&04166 &  0&00012 & --0&07729 &  0&00011 & 30&70 &  1&12 &  1&19 &  0&08 &   61 &   10 \\ 
\hline
\noalign{\vskip 2pt}
\multicolumn{19}{l}{\bf 30 GHz} \\
\noalign{\vskip 4pt}
\texttt{27M} & --0&06811 &  0&00009 &  0&03322 &  0&00012 & 32&68 &  1&28 &  1&36 &  0&05 &  101 &    6 \\ 
\texttt{27S} & --0&06810 &  0&00009 &  0&03323 &  0&00012 & 33&02 &  1&36 &  1&38 &  0&06 &  100 &    6 \\ 
\texttt{28M} & --0&06824 &  0&00010 & --0&03414 &  0&00013 & 32&99 &  1&45 &  1&37 &  0&11 &   78 &    7 \\ 
\texttt{28S} & --0&06823 &  0&00009 & --0&03415 &  0&00012 & 32&89 &  1&32 &  1&37 &  0&10 &   78 &    6 \\ 
\hline
\end{tabular}
\label{Jupiter 6}
\end{table*}

\begin{table*}
\centering
\caption{Fitted parameters derived from the seventh scan of Jupiter: main beam pointing directions defined with respect to the nominal telescope line of sight, FWHM, ellipticity, and orientation.}
\begin{tabular}{c r@{.}l@{ $\pm$ }r@{.}l r@{.}l@{ $\pm$ }r@{.}l r@{.}l@{ $\pm$ }r@{.}l r@{.}l@{ $\pm$ }r@{.}l c@{ $\pm$ }c}
\hline
\hline
\noalign{\vskip 2pt}
{\bf Beam} & \multicolumn{4}{c}{\bf X$_0$} & \multicolumn{4}{c}{\bf Y$_0$} & \multicolumn{4}{c}{\bf FWHM} & \multicolumn{4}{c}{\bf e} & \multicolumn{2}{c}{\bf $\psi_{{\rm ell}}$} \\
\noalign{\vskip 2pt}
\noalign{\vskip 2pt}
& \multicolumn{4}{c}{} & \multicolumn{4}{c}{} & \multicolumn{4}{c}{(arcmin)} & \multicolumn{4}{c}{} & \multicolumn{2}{c}{(deg)} \\
\noalign{\vskip 2pt}
\hline
\noalign{\vskip 2pt}
\multicolumn{19}{l}{\bf 70 GHz} \\
\noalign{\vskip 4pt}
\texttt{18M} & --0&03879 &  0&00001 & --0&04333 &  0&00001 & 13&40 &  0&07 &  1&24 &  0&01 &   85 &    1 \\ 
\texttt{18S} & --0&03878 &  0&00001 & --0&04334 &  0&00001 & 13&46 &  0&06 &  1&28 &  0&01 &   86 &    1 \\ 
\texttt{19M} & --0&04872 &  0&00001 & --0&02758 &  0&00001 & 13&13 &  0&07 &  1&25 &  0&01 &   78 &    1 \\ 
\texttt{19S} & --0&04872 &  0&00001 & --0&02757 &  0&00001 & 13&09 &  0&07 &  1&28 &  0&01 &   79 &    1 \\ 
\texttt{20M} & --0&05438 &  0&00001 & --0&01137 &  0&00001 & 12&84 &  0&06 &  1&27 &  0&01 &   71 &    1 \\ 
\texttt{20S} & --0&05438 &  0&00001 & --0&01136 &  0&00001 & 12&84 &  0&07 &  1&29 &  0&01 &   72 &    1 \\ 
\texttt{21M} & --0&05461 &  0&00001 &  0&01034 &  0&00001 & 12&77 &  0&05 &  1&28 &  0&01 &  107 &    1 \\ 
\texttt{21S} & --0&05460 &  0&00001 &  0&01035 &  0&00001 & 12&87 &  0&06 &  1&29 &  0&01 &  106 &    1 \\ 
\texttt{22M} & --0&04859 &  0&00001 &  0&02654 &  0&00001 & 12&93 &  0&06 &  1&26 &  0&01 &  101 &    1 \\ 
\texttt{22S} & --0&04860 &  0&00001 &  0&02653 &  0&00001 & 12&98 &  0&06 &  1&28 &  0&01 &  101 &    1 \\ 
\texttt{23M} & --0&03850 &  0&00001 &  0&04237 &  0&00001 & 13&32 &  0&06 &  1&24 &  0&01 &   93 &    1 \\ 
\texttt{23S} & --0&03851 &  0&00001 &  0&04236 &  0&00001 & 13&34 &  0&07 &  1&28 &  0&01 &   93 &    1 \\ 
\hline
\noalign{\vskip 2pt}
\multicolumn{19}{l}{\bf 44 GHz} \\
\noalign{\vskip 4pt}
\texttt{24M} & --0&07103 &  0&00004 & --0&00056 &  0&00005 & 23&28 &  0&27 &  1&39 &  0&03 &   89 &    1 \\ 
\texttt{24S} & --0&07101 &  0&00003 & --0&00061 &  0&00004 & 23&08 &  0&24 &  1&34 &  0&03 &   90 &    1 \\ 
\texttt{25M} &  0&04198 &  0&00010 &  0&07606 &  0&00009 & 29&64 &  0&71 &  1&19 &  0&05 &  117 &    7 \\ 
\texttt{25S} &  0&04192 &  0&00011 &  0&07606 &  0&00009 & 30&40 &  0&73 &  1&18 &  0&05 &  119 &    7 \\ 
\texttt{26M} &  0&04165 &  0&00011 & --0&07726 &  0&00009 & 29&87 &  0&75 &  1&19 &  0&05 &   61 &    7 \\ 
\texttt{26S} &  0&04165 &  0&00010 & --0&07727 &  0&00009 & 30&24 &  0&71 &  1&18 &  0&05 &   60 &    7 \\ 
\hline
\noalign{\vskip 2pt}
\multicolumn{19}{l}{\bf 30 GHz} \\
\noalign{\vskip 4pt}
\texttt{27M} & --0&06811 &  0&00007 &  0&03323 &  0&00010 & 33&05 &  0&57 &  1&36 &  0&02 &  101 &    2 \\ 
\texttt{27S} & --0&06811 &  0&00008 &  0&03324 &  0&00010 & 33&21 &  0&60 &  1&38 &  0&02 &  101 &    2 \\ 
\texttt{28M} & --0&06824 &  0&00008 & --0&03413 &  0&00011 & 33&28 &  0&64 &  1&36 &  0&05 &   78 &    3 \\ 
\texttt{28S} & --0&06823 &  0&00007 & --0&03414 &  0&00010 & 33&22 &  0&58 &  1&37 &  0&02 &   78 &    2 \\ 
\hline
\end{tabular}
\label{Jupiter 7}
\end{table*}

\label{appendix}


\bibliographystyle{aat}
\bibliography{Planck_bib,LFI_beams_bib,conviqt_bib}
\raggedright 
\end{document}